\documentclass[pre,twocolumn,amssymb,floatfix,superscriptaddress,footinbib,A4paper,longbibliography]{revtex4-1}

\usepackage{footmisc}
\usepackage{enumerate}
\usepackage{graphicx,placeins}
\usepackage{xcolor}
\usepackage{ragged2e}  
\usepackage[applemac]{inputenc} 
\usepackage{fontenc}  
\usepackage{amsmath}
\usepackage{soul}
\usepackage{hyperref}

\usepackage[normalem]{ulem}
\definecolor{darkgreen}{rgb}{0.07, 0.53, 0.07}
\definecolor{purple}{rgb}{0.5, 0, 0.7}
\definecolor{orange}{rgb}{1,0.5,0}
\definecolor{cyan}{rgb}{0,0.8,0.68}
\definecolor{validate}{rgb}{0,0.5,0}
\newcommand{\red}[1]{{\color{red}{#1}}}

\newcommand{\mf}[1]{{\color{red}{#1}}} 
 
\renewcommand{\red}[1]{{\color{black}{#1}}}
\renewcommand{\mf}[1]{{\color{black}{#1}}}

\newcommand{\perror}{p_{\rm err}}
\newcommand{\pthr}{p_{\rm thr}}
\newcommand{\upi}{\mathrm{i}}
\newcommand{\upe}{\mathrm{e}}
\newcommand{\upd}{\mathrm{d}}
\newcommand{\IFone}{\mathrm{IF}_\mathrm{1qb}}

\begin{document}
\title{
Optimizing resource efficiencies for scalable full-stack quantum computers
}

\author{Marco \surname{Fellous-Asiani}}
\email{fellous.asiani.marco@gmail.com}
\affiliation{Centre for Quantum Optical Technologies, Centre of New Technologies, University of Warsaw, Banacha 2c, 02-097 Warsaw, Poland}
\affiliation{Universit\'e Grenoble Alpes, CNRS, Grenoble INP, Institut N\'eel, 38000 Grenoble, France}

\author{Jing Hao \surname{Chai}}
\affiliation{Universit\'e Grenoble Alpes, CNRS, Grenoble INP, Institut N\'eel, 38000 Grenoble, France}
\affiliation{Centre for Quantum Technologies, National University of Singapore, Singapore}

\author{Yvain \surname{Thonnart}}
\affiliation{Universit\'e Grenoble Alpes, CEA-LIST, F-38000 Grenoble, France}

\author{Hui Khoon \surname{Ng}}
\email{huikhoon.ng@yale-nus.edu.sg}
\affiliation{Yale-NUS College, Singapore}
\affiliation{Centre for Quantum Technologies, National University of Singapore, Singapore}
\affiliation{MajuLab, International Joint Research Unit UMI 3654, CNRS, Universit{\'e} C{\^o}te d'Azur,
Sorbonne Universit{\'e}, National University of Singapore, Nanyang Technological University, 
Singapore}

\author{Robert S.~\surname{Whitney}}
\email{robert.whitney@grenoble.cnrs.fr}
\affiliation{Universit\'e Grenoble Alpes, CNRS, LPMMC, 38000 Grenoble, France.}

\author{Alexia \surname{Auff{\`e}ves}}
\email{alexia.auffeves@cnrs.fr}
\affiliation{MajuLab, CNRS-UCA-SU-NUS-NTU International Joint Research Laboratory}
\affiliation{Centre for Quantum Technologies, National University of Singapore, 117543 Singapore, Singapore}

\begin{abstract}
In the race to build scalable quantum computers, minimizing the resource consumption of their full stack to achieve a target performance becomes crucial. It mandates a synergy of fundamental physics and engineering: the former for the microscopic aspects of computing performance, and the latter for the macroscopic resource consumption. For this we propose a holistic methodology dubbed Metric-Noise-Resource (MNR) able to quantify and optimize all aspects of the full-stack quantum computer, bringing together concepts from quantum physics (e.g., noise on the qubits), quantum information (e.g., computing architecture and type of error correction), and enabling technologies (e.g., cryogenics, control electronics, and wiring).  This holistic approach allows us to define and study resource efficiencies as ratios between performance and resource cost. As a proof of concept, we use MNR to minimize the power consumption of a full-stack quantum computer, performing noisy or fault-tolerant computing with a target performance for the task of interest. Comparing this with a classical processor performing the same task, we identify a quantum energy advantage in regimes of parameters distinct from the commonly considered quantum computational advantage. This provides a previously overlooked practical argument for building quantum computers. 
While our illustration uses highly idealized parameters inspired by superconducting qubits with concatenated error correction, the methodology is universal---it applies to other qubits and error-correcting codes---and  provides experimenters with guidelines to build energy-efficient quantum processors. In some regimes of high energy consumption, it can reduce this consumption by orders of magnitudes. Overall, our methodology lays the theoretical foundation for resource-efficient quantum technologies. 
\end{abstract}
\maketitle

\section{Introduction}

There is a lot of excitement and hope that quantum information processing will help us solve problems of importance for society. Potential applications are numerous, ranging from optimisation \cite{Farhi2014Nov,Amaro2022Feb}, cryptography \cite{Shor2006Jul,Haner2020Apr} to finance \cite{chakrabarti2021threshold,rebentrost2018quantum}. The simulation of quantum systems \cite{Bauer2020Nov,Cao2019Oct,Ma2020Jul} for quantum chemistry and material science holds the promise to understand fundamental phenomena and design new materials and new drugs \cite{Cao2018Nov, Zinner2021Jul}. Different experimental platforms are currently investigated, including photonics \cite{Slussarenko2019Dec}, ion traps \cite{Bruzewicz2019Jun}, spin qubits \cite{Burkard2021Dec}, superconducting qubits \cite{kjaergaard2020superconducting}, among others \cite{Childress2013Feb,Laird2013Aug,lahtinen2017short}. Owing to impressive experimental efforts, qubit fidelities are starting to approach the fault-tolerance thresholds for scalable quantum computers. Quantum computational advantages have been claimed \cite{Google2019,zhong2020quantum}, even as the concept is still being discussed \cite{Pednault2019Oct,Zhou2020Nov}. 

Making quantum computers a concrete reality has a physical resource cost, especially when large-scale processors are targeted. In this spirit, the number of physical qubits required to implement large-scale computations has started to be investigated in various fault-tolerant schemes, including those that employ concatenated codes \cite{Chamberland2016Sep,suchara2013qure,Suchara2013Dec}, surface codes \cite{Gidney2021Apr,suchara2013qure,Suchara2013Dec}, and bosonic qubits \cite{Guillaud2020Sep,Chamberland2020Dec,noh2022low}. The total number of logical gates and qubits required by many algorithms have also been estimated, e.g., for decryption tasks \cite{Gidney2021Apr,grassl2016applying,pavlidis2012fast,haner2020improved} and material \cite{campbell2021early, kivlichan2020improved,kim2022fault,delgado2022simulate,lemieux2021resource} or electromagnetic \cite{scherer2017concrete} simulations. \red{These studies play an important role in identifying strategies for scaling up quantum processors.}

\red{The question of energy consumption was mentioned in the seminal experimental demonstration reported in \cite{Arute}: a 5-orders-of-magnitude difference between the power consumption of the quantum processor and the one of the classical supercomputer performing the same task was announced. However, the study was on a 50-qubit quantum processor and at the present time it remains unclear how the energy consumption of future quantum processors will scale.  On one hand, some studies anticipate energy savings, thanks to the complexity gains provided by quantum logic, see e.g. \cite{jaschke2022quantum}. They rely on subtle algorithmic details but then assume very simple models for the hardware. On the other hand, studies based on precise hardware details (but agnostic on algorithmic details) \cite{martin2022energy,krinner2019engineering}, usually conclude that the overheads needed to control the physical qubits could be so large that they will be to be an issue for scalability \cite{Almudever2017Mar,mcdermott2018quantum,bertels2020quantum,donati2021look,martin2022energy,frank2022cryo,Jokar2022Feb,Bandic2022Apr} especially in achieving large-scale, fault-tolerant quantum computers. 
This lack of consensus reveals the need for a holistic  methodology coupling data from the hardware and the algorithmic frameworks. }

\begin{figure}
\includegraphics[width=0.90\columnwidth]{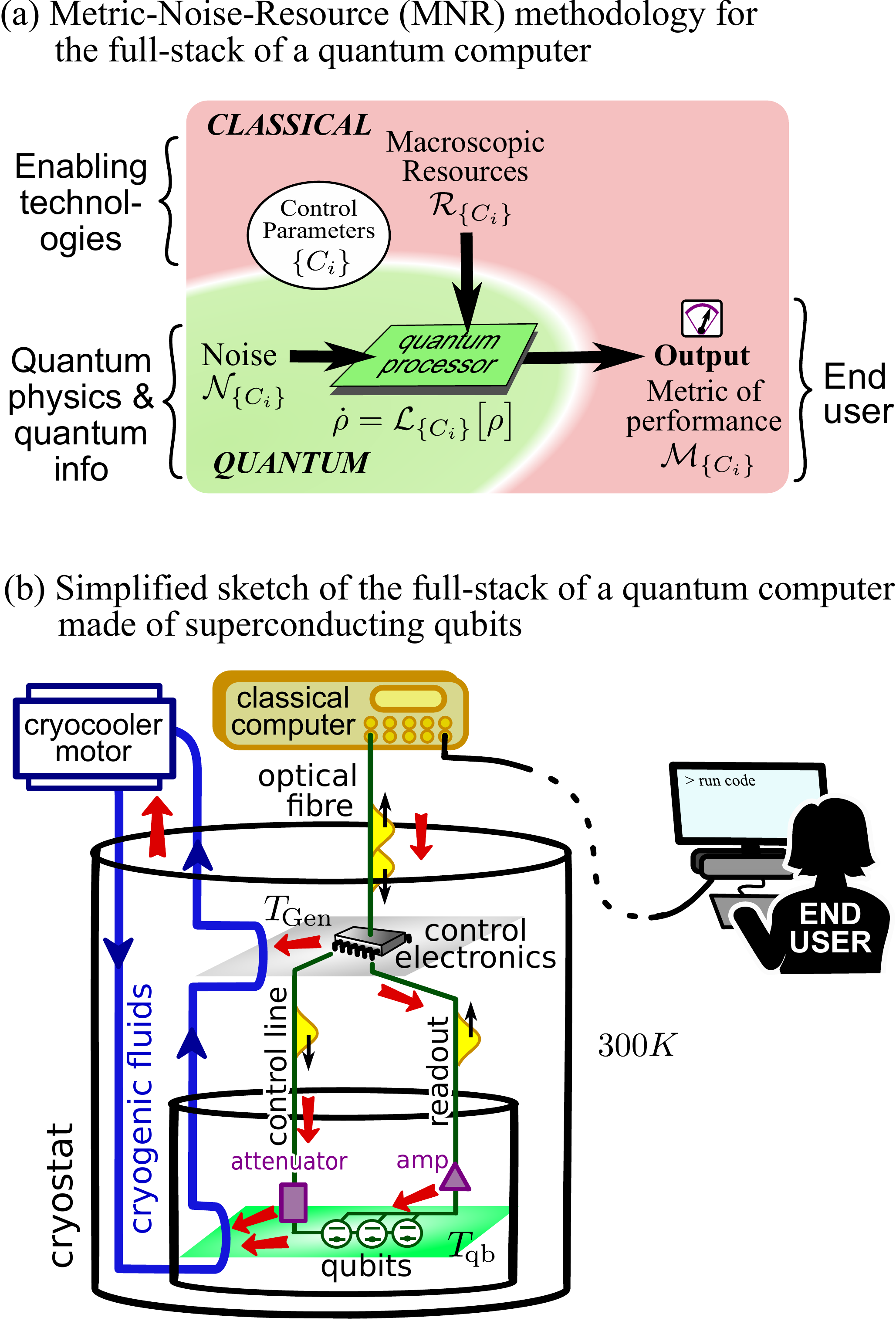}
\caption{(a)  Schematic of our  Metric-Noise-Resource (MNR) methodology, which models the battle of control against noise.
The control parameters affect the metric of performance, the noise, and the resource consumption. Such parameters include qubit temperature, the amount of error correction, etc.  
The metric of performance can be improved by using resources to reduce the noise (e.g., cooling the qubits), or spending resources to make the metric less sensitive to existing noise (e.g., better error correction).
(b)  A sketch of the physical elements in a typical full-stack superconducting quantum computer, with qubits at
temperature $T_{\rm qb}$, and classical control electronics at temperature $T_{\rm gen}$. The classical computer at room temperature compiles the user-specified algorithm and code into a sequence of physical gate operations, interprets detected errors in real time, and can modify the gate sequence to correct them. Black arrows indicate information flows. Red arrows indicate heat flows that bring noise that can cause gate errors (heat conducted by wiring, heat generated by attenuators and amplifiers, etc.); more details are shown in Fig.~\ref{Fig:full-stack-stages}. We model the full stack by considering each physical element in (b) in terms of its effect on the metric, noise, and resource in (a).
\label{Fig1}
}
\end{figure}

Setting up a holistic methodology is highly challenging as it requires modeling the full stack of the quantum computer \cite{Almudever2017Mar,bertels2020quantum,donati2021look,Bandic2022Apr,almudever2021structured,murali2019full,rodrigo2021double,gokhale2020full}, and coordinating inputs from currently separated areas of expertise as sketched in Fig.~\ref{Fig1}.
Improving computational performance requires programming the processor to implement a given algorithm, and to reach a satisfactory level of control over noise while the algorithm is being executed. These optimizations are performed at the quantum level and rely on detailed knowledge of quantum hardware and software, e.g.,quantum control, noise modelling, environmental engineering, quantum error correction, quantum algorithms, qubit fabrication, etc. However, achieving that in a physical device requires the use of macroscopic resources provided by enabling technologies at the classical level (e.g., cryogenic systems \cite{martin2022energy,krinner2019engineering}, classical computers for control \cite{mcdermott2018quantum,ishida2018towards,debenedictis2020adiabatic,kang2022cryo,park202113,Bardin2019Oct}, lasers, detectors \cite{You2020Sep,Walsh2021Apr}, amplifiers \cite{Malnou2021Oct,Planat2020Apr,krantz2019quantum}, etc.). Hence, understanding and managing the resource bill of future quantum processors cannot be restricted to the quantum level, as it provides no access to the macroscopic costs. Reciprocally, a solely macroscopic approach is blind to the computing performance---we do not know what we are paying for. Resource-cost assessments and optimizations must be jointly conducted by an interdisciplinary combination of expertise. In the energetic context, this was dubbed the Quantum Energy Initiative \cite{QEI-perspective}.

\red{In this paper, we present a methodology that allows us to optimize the resource cost for the full-stack of a quantum computer, under the constraint of a certain performance. Such an optimization under constraint
is only possible if the methodology can treat all elements of the quantum computer in a single framework, including both the hardware (such as attenuators and cryogenics) and the quantum software (such as different quantum gates that implement the same quantum algorithm).  We provide such a holistic methodology, and show how it relates the microscopic description of the quantum computation to the macroscopic resources consumed by the cryogeny, wiring, classical control electronics, etc.  We apply it to examples of quantum tasks with}
increasing complexity, from the operation of a single-qubit gate, a noisy circuit, to a fault-tolerant algorithm. Each requires the specification of a relevant metric of performance and a detailed description of the physical processes at play. In each example, we use our approach to show how resource costs scale with the size of the computational task. This reveals concrete instances of extreme sensitivity to both hardware characteristics (e.g., qubit quality or efficiency of the electronics) and software characteristics (architecture of the circuits, or type of error-correcting code), and the necessity to treat both aspects in a coordinated manner.

As an important outcome, we \red{analyze} quantum resource efficiencies as ratios between the metric of computing performance and the resource cost. Different efficiencies can be defined, depending on the metric and the resource(s) of interest. Some permit the benchmarking of different qubit technologies or computing architectures. Others allow a comparison between quantum and classical processors, and to define a quantum advantage from the resource perspective, which can further boost practical interest in quantum computers. Focusing on the example of Shor's algorithm to break RSA encryption, our calculations single out regimes of quantum energy advantage over classical supercomputers for problem sizes still accessible by such supercomputers, including in cases where the quantum computer takes more time to do the job than the supercomputer. These results show that the energy advantage is reached in different regimes than the computational advantage, providing a new and so far overlooked potential practical reason to build quantum computers.

Throughout this paper, we take parameters inspired by the superconducting platform and existing technologies for the control electronics, wiring, and cryogeny. However, our approach is generic and versatile, capable of providing general behaviors and typical orders of magnitude for a wide variety of settings. \red{In particular, it shows how diverse the parameters that can affect power consumption are, with a crucial one being qubit quality. It also singles out surprising effects:
while it is often optimal to make the qubits cold enough to minimize error correction, our approach shows that there are regimes where the opposite is true; regimes where it is more energy efficient to have warmer qubits with more error correction.}

For illustrative simplicity, we base our examples on the concatenated 7-qubit code, which is well documented and allows for straightforward analytical expressions, but can be demanding in terms of physical quantum resources. This choice leads us to use parameter values sometimes beyond the current state of the art.
Nevertheless, we invite the reader to appreciate our results as proofs of concept of our methodology. It can provide on-demand practical guidelines to experimentalists and engineers looking to build resource-efficient quantum processors, allowing them to clearly identify the sequence of challenges to be met. Ultimately, systematic applications of the methodology can help avoid ecologically unacceptable outcomes, such as the current rapid increase in energy consumption of servers for consumer electronics \cite{Puebla} and artificial intelligence \cite{AI}. Thus, throughout the paper, we keep the methodology as apparent as possible, so as it can be applied to different qubit platforms and enabling technologies, as well as other error-correcting codes.

Our article is organized as follows. We present our general optimization methodology in Sec.~\ref{Sect:MNR} for any kind of resource and any kind of quantum computing platform. In Secs.~\ref{Sect:noisy-gate} to \ref{Sect:fault-tolerant}, we apply it to the special case of a superconducting quantum computer, focusing on energy and power, to illustrate the use of our methodology and of the kinds of conclusions one can draw from such an analysis. Section~\ref{Sect:noisy-gate} describes a simple example for a noisy single-qubit gate, establishing the basic connection between microscopic qubit parameters, and macroscopic power consumption. Section~\ref{Sect:NISQ} focuses on a noisy circuit, revealing the close interplay between inputs from the software and the physics of the hardware. Sections \ref{Sect:noisy-gate} and \ref{Sect:NISQ} are largely pedagogical in their aims, to shed light on how the performance defined at the quantum level can impact the resource consumption at the macroscopic level. Section~\ref{Sect:fault-tolerant} considers a full fault-tolerant quantum computer, using concatenated codes for error correction, and performing a calculation of difficulty similar to breaking the well-known RSA encryption. Estimates for fault-tolerant quantum computing based on the currently popular surface codes are given at the end of Sec.~\ref{Sect:fault-tolerant}. We summarize our findings in Sec.~\ref{Sect:discussion}.  
\section{Metric-Noise-Resource Methodology}
\label{Sect:MNR}

A quantum computer is a programmable machine whose job is to perform a well-defined sequence of operations on an ensemble of qubits: After the circuit is programmed, the qubits are prepared in a reference state, unitary operations implementing a computational task are then applied, and the qubits are finally measured to extract the result of the calculation. Quantum noise perturbs this sequence, giving rise to errors that decrease the computing performance. This has to be countered by an increase in noise mitigation measures, in an attempt to reduce the occurrence of errors and to remove their effects on the computation. Such increased noise mitigation is usually associated with increased resource costs. In some cases, the increased resource cost can itself result in more noise. For instance, more error correction requires more physical qubits, and that can result in additional sources of crosstalk, increasing the noise affecting the quantum processor \cite{PRXQuantum}. Hence, finding the minimal resource cost to reach a target performance requires one to explore non-trivial sweet spots. Such an investigation involves coordinated inputs from the software and hardware, at the quantum and classical levels of description. 
\red{Here we present a holistic methodology to treat the whole range of inputs. For reasons that become clear below, we have dubbed it the Metric-Noise-Resource (MNR) methodology, or more simply MNR \footnote{\red{MNR was briefly summarized for non-experts in A. Auff\`eves' perspective article \cite{QEI-perspective}, citing this work as the place its would be presented as a complete scientific methodology, used to make quantitative predictions.}}.}

The basics of MNR are sketched in Fig.\ref{Fig1}(a). The first step consists of identifying the set of parameters---dubbed ``control parameters" ${\cal C}_i$s--- that allows us to execute a quantum algorithm with a given target performance. It is with respect to these parameters that the resource cost shall be minimized. Control parameters can be of various kinds. Some characterize the quantum processor or the hardware controlling it. Typical examples include the temperature of the qubits, or the strength of the attenuators on the control lines. Some are of software nature, reflecting the fact that the same algorithm can be executed by different circuits. Examples include the degree of compression of the circuit, or any other quantity capturing the circuit architecture. A crucial control parameter is the size of the quantum error-correcting code, i.e., the number of physical qubits per logical qubit in a fault-tolerant quantum computation. 

Once this identification has been done, we can turn to the first element of MNR: the metric  of performance ${\cal M}$ (later dubbed ``metric," for brevity). It is a number measuring the quality of the computation, for which a larger number means a better computation. Naturally, there is some flexibility in the choice of the metric. Some are defined at a low level, focusing on the precision with which states can be prepared while executing the programmed sequence of gates. A natural example is the fidelity, which quantifies the distance between the ideal and the real processor states before the extraction of the result. Other metrics are user-oriented, such as the Q-score \cite{Martiel2021Jun} or the quantum volume \cite{Cross2019Sep} which estimates the maximal size of the problems that can be solved on a quantum computing platform. Some user-oriented metrics aim to benchmark classical and quantum processors, and to identify quantum computational advantages. Whichever the chosen metric, it directly depends on the level of control over physical processes in the quantum computer.

This brings us naturally to the second element of MNR: the noise in the physical processes. It is taken into account by modeling the dynamics of the noisy quantum processor executing the algorithm. This involves a given time-dependent Hamiltonian, together with a noise model, in the form of a master equation whose expression depends on the control parameters. Many parameters can enter this noise model, such as the temperature of the qubits, and the temperature of the external control electronics. The time-dependent Hamiltonian corresponds to the sequence of gates applied to the qubits, which is set by the circuit architecture. Hence, such
 a model allows us to derive a quantitative expression for  the metric, as a function of the ${C}_i$s.

The third ingredient of MNR is the resource ${\cal R}$ of interest we wish to minimize. Formally, a resource can be any cost-function which depends on the set of control parameters. While MNR is general and could tackle economic costs, here we shall focus on physical resource costs. They can be extremely diverse in nature, e.g., the physical volume occupied by the quantum processor, the total frequency bandwidth allocated to the qubits, the duration of the algorithm, the amount of classical information processed to perform error correction, or the energy consumption. 
In this paper we will address the last of these, by considering the electrical power consumed while a quantum computation is being performed. 

Once these steps are completed, MNR basically reduces to a constrained optimization. Fixing a target metric ${\cal M}_0$  boils down to setting a tolerable level of control over noise for a properly programmed processor: It provides a first constraint that the control parameters have to satisfy. The resource cost ${\cal R}$ is then minimized as a function of the control parameters under this implicit constraint. An optimal set of control parameters gives the minimal resource consumption ${\cal R}^\text{min}({\cal M}_0)$ needed to reach the metric ${\cal M}_0$ \footnote{In our models, 
to minimize the resource consumption, we should take the \textit{smallest} metric that allows us to perform the task of interest. Put differently, the minimum resource required to achieve  $\mathcal{M} \geq \mathcal{M}_0$ is reached when $\mathcal{M} = \mathcal{M}_0$. 
We have noticed that this is true whenever the resources and metrics grow monotonically with at least one control parameter.
This applies in our case, since the power consumption and metric always increase monotonically as the qubit temperature is reduced, or as the attenuation is increased. 
In contrast, if one had a case where the resources or metrics were non-monotonic functions of \textit{all} the control parameters, then one might be able to achieve lower power consumption by going to a higher metric than the target necessary for the task of interest, taking $\mathcal{M} > \mathcal{M}_0$.}. For instance, if the qubit temperature is a control parameter, MNR provides an optimal working temperature for the qubits to reach a target metric ${\cal M}_0$ with a minimal resource cost and can lead to non-trivial values as shown below. It thus provides practical inputs to designing resource-efficient quantum computations. 

\red{MNR relates a metric to its macroscopic resource cost. This makes it drastically different from the common point of view to-date, which has been to target the largest metrics, whatever the resource cost. It was inspired by our earlier work \cite{PRXQuantum}, which pointed out that a constraint on resources has a profound effect on fault-tolerant quantum computing. However unlike here, we only considered quantum-level resources.} 

\subsection*{Resource efficiencies}
 \label{Sect:efficiency}

In general, efficiencies characterise the balance between a performance and the resource consumed in achieving it. MNR provides systematic relations between performance and resource costs. Hence, it naturally leads one to define and optimize resource efficiencies for quantum computing.  For classical supercomputers, the target performance is computing power, expressed in FLoating-point Operations Per Second (FLOPS). The energy efficiency is built as the ratio of computing power over the power consumption (the resource)  of the processor. It is measured in FLOPS/W, has the dimension of the inverse of an energy, and gives rise to the Green 500 ranking of the most energy-efficient supercomputers \cite{Green500}. In this paper we shall explore quantum equivalents of this energy efficiency. Within the MNR methodology, the resource efficiency is generically defined as $\eta = {\cal M}/{\cal R}$,
where ${\cal M}$ is the metric, and ${\cal R}$ the resource cost. As mentioned above, two kinds of metrics can be considered: low-level metrics and user-oriented metrics. The resource cost can be defined at the quantum level, or at the macroscopic level, giving rise to bare and dressed efficiencies, respectively. 
The quantum level is crucial for understanding the physics of qubits, while the macroscopic level will be what matters for large-scale applications.

Sections~\ref{Sect:noisy-gate} and \ref{Sect:NISQ} respectively involve noisy gates and circuits. A low-level metric, the fidelity, is natural in both cases. Modeling the processor at the quantum level provides an implicit relation between the noise, the control, and the chosen metric. Applying the MNR methodology to minimize the power consumption ${\cal R}^{\text{min}}({\cal M}_0)$ for the target metric ${\cal M}_0$ unambiguously sets the maximal efficiency of the task at the target ${\cal M}_0$. Such an efficiency is well suited for benchmarking different technologies of qubits, or different computing architectures implementing the same algorithm, i.e., to compare different quantum computing platforms.

Sections~\ref{Sect:NISQ} and \ref{Sect:fault-tolerant} involve algorithms. We thus introduce user-oriented metrics there to explore another kind of resource efficiency. As a typical example, in Sec.~\ref{Sect:fault-tolerant} we consider the breaking of RSA-encryption. There the relevant metric is the maximal key size that can be broken with a well-defined probability of success. We estimate the energy consumed by full-stack quantum and classical processors as a function of this size. Beyond a typical size, we show that quantum processors are more energy-efficient than classical ones, highlighting a new and essential practical advantage of quantum computing.

\section{Noisy single-qubit gate}
\label{Sect:noisy-gate}

We start by applying the MNR methodology to the simplest component of a quantum computer: a resonant, noisy single-qubit gate. This allows us to introduce the generic type of qubits we will be working with throughout the paper, which takes values inspired by the superconducting platform \cite{kjaergaard2020superconducting,krantz2019quantum,Google-Sycamore-datasheet}. We consider only errors due to spontaneous emission and thermal noise, both  unavoidable as soon as the qubit is driven by control lines bringing pulses from the signal generation stage to the processor (see below). All other sources of noise are neglected.

\subsection{Quantum level} 
\label{Sect:quantum-level}

Let us first focus on the characteristics of the gate at the quantum level. The ground and first excited states of the superconducting qubit are denoted $|0\rangle$ and $|1\rangle$, respectively, with a transition frequency set to $\omega_0= 2\pi \times 6$\,GHz.
The gate is implemented by driving the qubit with resonant microwave pulses at a frequency $\omega_0$ and an amplitude that induces a classical Rabi frequency $\Omega$.  We consider square pulses of duration $\tau_\mathrm{1qb}$, giving rise to the qubit Hamiltonian $H = -\frac{1}{2}\hbar \omega_0 \sigma_z + \frac{1}{2}\hbar \Omega (\upe^{\upi\omega t/2}\sigma_- + \upe^{-\upi\omega t/2}\sigma_+)$, with $\sigma_\pm\equiv\frac{1}{2}(\sigma_x\mp\upi\sigma_y)$. For model simplicity, all single-qubit gates are taken as $X$ gates, i.e., each gate is a $\pi$-pulse of duration $\tau_{1\text{qb}} = \pi/\Omega$. For driving pulses propagating in control lines, $\Omega$ and the spontaneous emission rate $\gamma$ are not independent, with $\Omega= \sqrt{(4 \gamma)/(\hbar \omega_0)} \sqrt{P}$, where $P$ is the average power of the microwave pulse \cite{PRXQuantum,cottet2017observing}. In the present section, dedicated to the study at the quantum level, the gate duration $\tau_\mathrm{1qb}$ is taken as the control parameter. The resource cost is defined as the power $P_\pi$ consumed to bring the qubit from $|0\rangle$ to $|1\rangle$: 
\begin{align}
   P_\pi=\frac{\hbar \omega_0 \pi^2}{4 \gamma \tau_\text{1qb}^2}.
  \label{P_pi}
\end{align}
The spontaneous emission rate $\gamma$ is set by the specific qubit technology; we use $\gamma^{-1}=1\,{\rm ms}$ in Sec.~\ref{Sect:noisy-gate} and \mf{$\gamma^{-1}=10 \,{\rm ms}$ in Sec.~}\ref{Sect:NISQ}. The action of the noise alone is described by a map ${\cal N}$, obtained by integrating the Lindblad equation over a time interval $\tau_{1{\rm qb}}$,
\begin{equation}\label{Eq:Lindblad}
    \frac{\upd\rho}{\upd t} =  \gamma n_\mathrm{noise} {\cal L}[\sigma_-^\dagger](\rho) + \gamma (n_\mathrm{noise}+1) {\cal L}[\sigma_-](\rho),
\end{equation}
with ${\cal L}[A](\cdot)\equiv A\cdot A^\dagger-\frac{1}{2}\{A^\dagger A,\cdot\}$,  $\{\cdot,\cdot\}$ is the anti-commutator, and $n_\text{noise}$ denotes the number of thermal photons. We assume this same noise map $\cal N$ for every single-qubit gate, and write \mf{${\cal G}= G {\cal N}$}, where $\cal G$ and $G$ are the maps for the noisy and ideal gates, respectively. 

We define our ``low-level" metric to quantify the performance of the noisy single-qubit gate as the worst-case gate fidelity,
\begin{equation}\label{eq:F1}
{\cal M}_\mathrm{1qb}\equiv \min_\rho{\cal F}_{\cal G}(\rho).
\end{equation}
where $\cal F_{\cal G}(\rho)$ is the (square of the) fidelity 
\red{between the output of the ideal gate and the output of the noisy gate.
\mf{Then ${\cal F}_{\cal G}(\rho)\equiv \big[\textrm{Tr}\sqrt{(G\rho G^{\dagger})^{1/2}{\cal G}(\rho)(G\rho G^{\dagger})^{1/2}}\big]^2$.}}
The concavity of the fidelity ensures that the worst-case fidelity is attained on a pure state. The minimization in Eq.~\eqref{eq:F1} can thus be restricted to over pure states only, and ${\cal F}_{\cal G}(\rho)$ simplifies to \mf{$\langle \psi| G^{\dagger} {\cal G}(\psi) G|\psi\rangle=\langle \psi| {\cal N}(\psi) |\psi\rangle$} for $\rho\equiv |\psi\rangle\langle\psi|\equiv \psi$. It is useful to rewrite the metric as ${\cal M}_\mathrm{1qb}\equiv 1-\IFone$ where $\IFone$ is now the worst-case (i.e., maximum) gate \emph{in}fidelity. Straightforward algebra yields an expression for $\IFone$ in terms of the gate and noise parameters:
\begin{align}
\IFone&=\gamma \tau_\text{1qb} \big(1+n_\text{noise}
    \big).
    \label{maxIF_1qb}
\end{align}
$\IFone$ scales like $\gamma \tau_\text{1qb}$, the number of spontaneous events during the gate, and increases with the number of thermal photons $n_\text{noise}$. Equation \eqref{maxIF_1qb} provides us with an implicit relation between the noise ($\gamma$ and $n_\text{noise}$), the control ($\tau_{1\text{qb}}$), and the metric (${\cal M}_{1\mathrm{qb}}=1-\IFone$). 
The metric can be increased by reducing the time to perform the gate operation, $\tau_\text{1qb}$. However, Eq.~(\ref{P_pi}) tells us that this comes at the cost of increased power consumption. \\

\noindent{\it Bare efficiency.---}
We now define the bare efficiency $\eta_0$, with ``bare" meaning that the resource cost $P_{\pi}$ is defined at the quantum level:
\begin{eqnarray}
\eta_{0} \ =\ \frac{{\cal M}_{1_\text{qb}}}{P_\pi}.
\label{Eq:eta_Rabi}
\end{eqnarray}
In the MNR methodology, we impose the metric to be equal to a given target value, ${\cal M}_{1_\text{qb}}={\cal M}_0$.  Let us consider the case where the thermal noise is negligible compared to spontaneous decay, i.e., $n_\text{noise} \ll 1$,  yielding ${\cal M}_0 = {\cal M}_{1_\text{qb}}= 1-\gamma \tau_{1\text{qb}}$. Now, we wish to minimize the resource cost $P_{\pi}$ for the desired ${\cal M}_0$. Such minimization is performed as a function of the control parameter $\tau_\mathrm{1qb}$ and gives rise to the maximal efficiency $\eta^{\max}_\text{0}$. From Eqs.~\eqref{P_pi} and \eqref{maxIF_1qb}, we can see that $\tau_\mathrm{1qb}$ affects both the resource and the metric. This allows us to write $\eta^{\max}_\text{0}$ solely as a function of the target metric,
\begin{eqnarray}
\eta^{\max}_\text{0}({\cal M}_0)=\frac{{\cal M}_0}{P^\text{min}_{\pi}({\cal M}_0)}=  \frac{4}{\pi^2}\frac{{\cal M}_0 (1-{\cal M}_0)^2}{\gamma \hbar \omega_0}. \qquad
\label{Eq:eta_Rabi-low_noise}
\end{eqnarray}
Equation \eqref{Eq:eta_Rabi-low_noise} tells us that the bigger the target metric of performance, the smaller the efficiency. In other words, increasing the target by one digit (e.g.\ to take ${\cal M}_0$ from 0.9 to 0.99) costs more and more power---we will see this general trend in all our examples below.
It also reveals the natural unit of power to be $\gamma \hbar\omega_0$, which is the power dissipated into the environment through spontaneous decay events.  The larger the noise rate $\gamma$, the larger the power dissipated, as the gate has to be performed more quickly to maintain the equality ${\cal M}_{1_\text{qb}}=\mathcal{M}_0$. Hence, $P^{\min}_{\pi}(\mathcal{M}_0)$ increases \footnote{More precisely, ${\cal M}_{1_\text{qb}}=\mathcal{M}_0$ implies that $\tau_\text{1qb}=(1-\mathcal{M}_0)/\gamma$. Replacing it in the expression of $P_{\pi}$ (see \eqref{P_pi}) shows that the latter increases with $\gamma$.}, decreasing the efficiency. 
Hence, at the level of single gates, good qubits characterized by small $\gamma$ are typically more energy efficient than bad ones. This observation will carry through to the macroscopic level in all examples in this work.
\\

\subsection{Macroscopic level} 
\label{Sect:single-qubit-macro-level}

\begin{figure}
     \centering
     \includegraphics[width=\columnwidth]{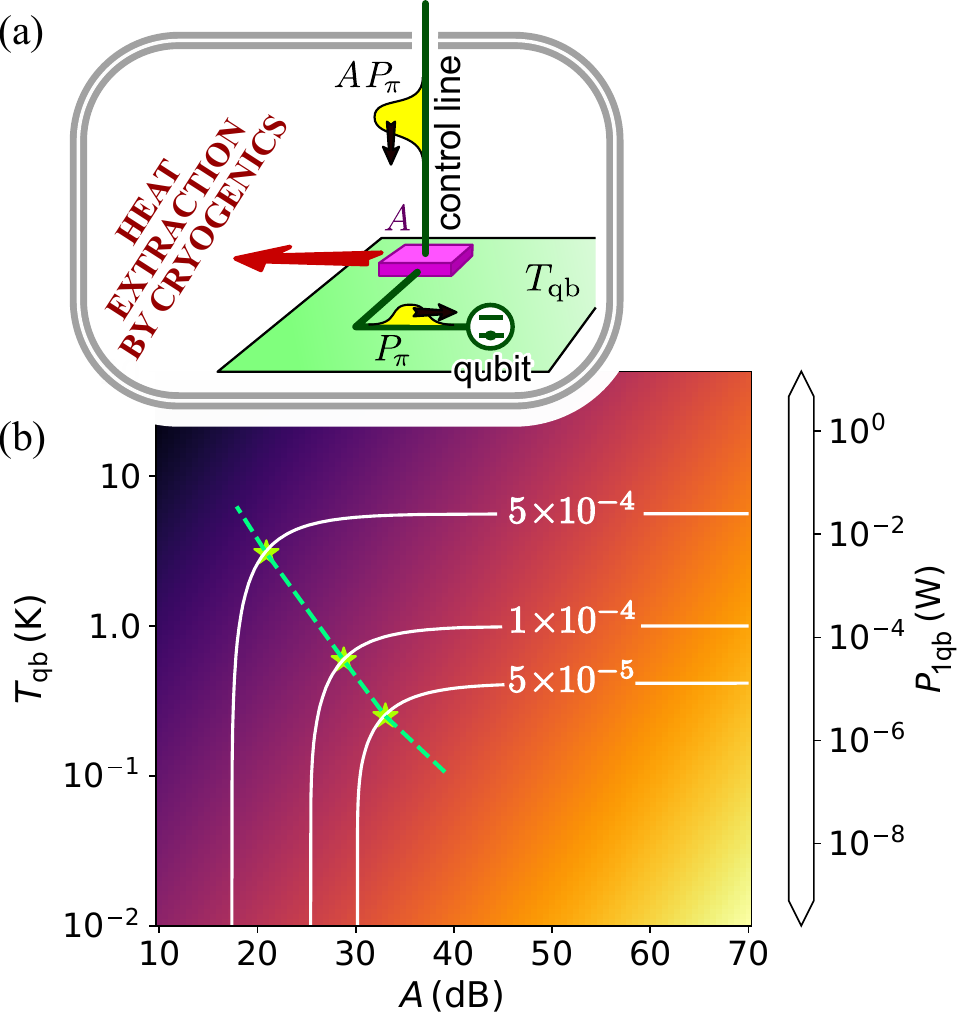}
\caption{(a) Sketch of our simplified model of a single qubit in a cryostat. (b) The color scale is the power consumption (in  Watts) of a single-qubit gate as a function of the qubit temperature $T_{\text{qb}}$ and attenuation $A$. Parameter values are $\gamma^{-1} = 1\,\mathrm{ms}$ and $\tau_{\rm 1qb}=25\,\mathrm{ns}$. Each contour is associated with the target metric (worst-case infidelity) indicated in white. Green stars mark the optimal parameters (those that minimize the power consumption) for each value of the target metric.}
\label{Fig:single-qubit}
\end{figure}

We now model the macroscopic chain of control to take into account its resource cost. Note that from now on, only such macroscopic resource costs will be considered, for which ``dressed" efficiencies are appropriate. \red{Here we present the basic approach within a simplified model, which will be made more realistic in Sec.~\ref{Sect:fault-tolerant}.
This simple example is limited to a control line funneling driving pulses from outside the cryostat onto the qubit through a single attenuator, with the cryogenics evacuating the heat dissipated by that attenuator. The implementation of the gate is depicted in Fig.~\ref{Fig:single-qubit}(a). The qubit is put in a cryostat and cooled down to the temperature $T_\text{qb}$ (typically below a Kelvin). The driving signal is generated at room temperature $T_\mathrm{ext}$ and sent into the cryostat through a control line. Alongside the signal, the line also brings in unwanted room-temperature thermal noise, unavoidable whenever we require external control.}

To mitigate the noise, the signal is first generated with a high amplitude for a strong signal-to-noise ratio. An attenuator is then placed on the line \footnote{To keep this example pedagogical, we consider a single attenuator. A more realistic chain of attenuators is addressed in Sec.~\ref{Sect:fault-tolerant}.} (at the qubit level at temperature $T_\text{qb}$), which lowers the input pulse power by an amount $A$.  Thus $A$ and $T_\text{qb}$ are the two control parameters optimized in the present section. For simplicity, we fix the gate duration to be $\tau_\text{1qb} = 25$ns~\cite{kjaergaard2020superconducting,krantz2019quantum,Google-Sycamore-datasheet}, chosen to avoid leakage errors \cite{Werninghaus2021Jan} not modeled here.
The two control parameters,  $A$ and $T_\text{qb}$, impact the gate noise in the following manner (see, e.g., Eq.~(10.13) in Ref.~\cite{pozar2011microwave}):
\begin{align}
    n_\text{noise} 
    \,=\,\frac{A-1}{A}n_\text{BE}(T_{\text{qb}})+\frac{1}{A} n_\text{BE}(T_{\text{ext}}),
    \label{Eq:n_noise-simple}
\end{align}
where 
$n_\text{BE}(T)= 1\big/[\upe^{\hbar\omega_0/(k_\text{B}T)}-1]$ is the Bose-Einstein photon distribution at temperature $T$.
Here, $A$ is expressed in natural units: $A=10^{A_{\rm dB}/10}$, where $A_{\rm dB}$ is the attenuation expressed in dB \cite{attenuation}. The noise model is now entirely defined by Eqs.~\eqref{Eq:Lindblad} and \eqref{Eq:n_noise-simple}. Keeping the fidelity in Eq.~\eqref{eq:F1} as the metric, increasing it boils down to increasing the level of attenuation $A$, or decreasing the qubit temperature $T_{\text{qb}}$. 
 
We finally define the macroscopic resource of interest. To get a signal of power $P_\pi$ on the qubit, a power $A\,P_\pi$ is injected into the cryostat, giving $\dot{Q} \approx A P_{\pi}$ as the rate of heat generation from the attenuator at $T_\mathrm{qb}$ \footnote{The signal and its reflection after interaction with the qubit are dissipated in the attenuator, which we take to have $A\gg 1$. Then, it is a reasonable approximation to take the attenuator's heat dissipation as equal to the injected signal power.}. We assume Carnot-efficient heat extraction, 
\red{as it already gives the right order of magnitude for large-scale cooling to cryogenic temperatures that can be done at 10\% to 30\% of Carnot efficiency, like the cooling capabilities at CERN \cite{Parma2014}}. Then the cryogenic electrical power consumption (dubbed ``cryo-power" below) needed to run the gate is
\begin{equation}
    P_{1\mathrm{qb}}(T_{\text{qb}},A) = \frac{T_\text{ext}-T_{\text{qb}}}{T_{\text{qb}}} A P_\pi.
    \label{Eq:power_singlequbit}
\end{equation}
This is the resource we consider in the present section. Putting together Eqs.~\eqref{maxIF_1qb}--\eqref{Eq:power_singlequbit}, we can see that increasing the metric by reducing $T_{\text{qb}}$ or increasing $A$ (taking $A\gg1$ as in typical experiments), increases the resource cost $P_{1\mathrm{qb}}(T_{\text{qb}},A)$. This behavior is apparent in Fig.~\ref{Fig:single-qubit}(b), where cryo-power is plotted as a function of $A$ and $T_\text{qb}$. If we target a specific value ${\cal M}_0$ for the metric, i.e., we require ${\cal M}_{1\text{qb}}={\cal M}_0$, this sets an implicit relation between $A$ and $T_\text{qb}$, giving rise to the contours marked in the figure.

In the MNR methodology, ${\cal M}_{1\text{qb}}={\cal M}_0$ is the constraint under which $P_{1\mathrm{qb}}(T_{\text{qb}},A) $ is minimized. Using Eqs.~\eqref{maxIF_1qb} and \eqref{Eq:n_noise-simple}, this constraint can be explicitly written as
\begin{align}
    1-\gamma \tau_{\text{1qb}}\left(1+\frac{A-1}{A}n_\text{BE}(T_{\text{qb}})+\frac{1}{A} n_\text{BE}(T_{\text{ext}})\right)={\cal M}_0.
\end{align} 
In Fig.~\ref{Fig:single-qubit}(b), this is indicated by the white contours, while the points with minimum power consumption are marked with green stars. This provides our first example of a non-trivial sweet spot, where the metric defined at the quantum level impacts the macroscopic resource cost, and is an explicit illustration of the necessity to coordinate inputs from both levels of description. \\

\begin{figure}
    \centering
    \includegraphics[width=0.96\columnwidth]{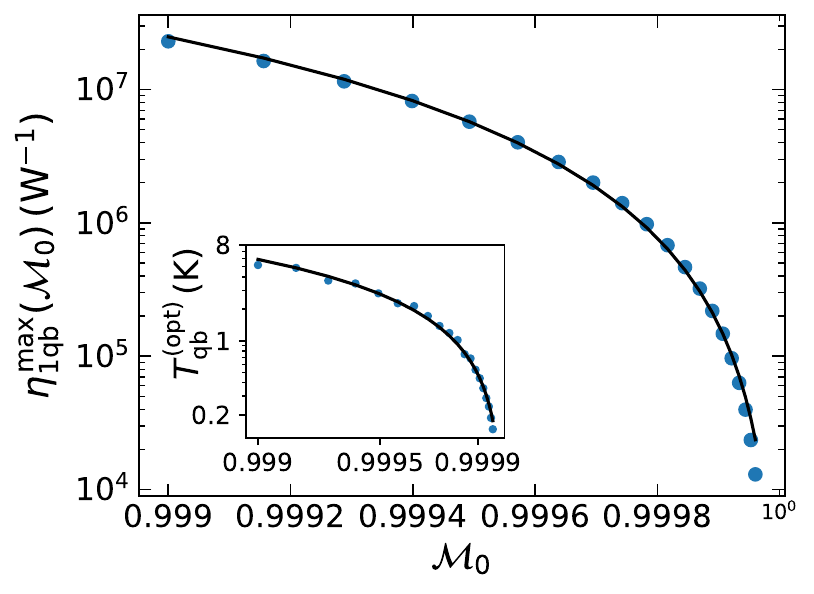}
    \caption{Maximal dressed efficiency $\eta_{\text{1qb}}^{\text{max}}({\cal M}_0)$ (in $W^{-1}$) for a single-qubit gate, as defined in Eq.~(\ref{Eq:eta_fullstack}). Inset: Optimal qubit temperature as a function of the target metric ${\cal M}_0$. Here, $\gamma^{-1} = 1\,\mathrm{ms}$.}  
    \label{Fig:Pow_mtarget}
\end{figure}

\noindent{\it Dressed efficiency.---} 
We now minimize the cryogenic power consumption as a function of the two control parameters $A$ and $T_\mathrm{qb}$, under the constraint ${\cal M}_{1_\text{qb}}={\cal M}_0$.
We denote this minimum by $P^\text{min}_{1\text{qb}}({\cal M}_0)$. This defines the maximal dressed efficiency of the single-qubit gate:
\begin{eqnarray}
\eta_{1\text{qb}}^{\text{max}}({\cal M}_0) =\frac{{\cal M}_0}{P^\text{min}_{1\text{qb}}({\cal M}_0)}\ . 
\label{Eq:eta_fullstack}
\end{eqnarray}
$\eta_{1\text{qb}}^{\text{max}}$ is plotted in Fig.~\ref{Fig:Pow_mtarget} as a function of ${\cal M}_0$, revealing the same behavior as $\eta^{\max}_0$: The larger the target metric, the smaller the efficiency. The inset gives the qubit temperature that achieves the minimal power consumption, as a function of ${\cal M}_0$. The maximal dressed efficiency is much lower than the maximal bare efficiency $\eta^{\max}_0$, with a typical reduction by 3 orders of magnitude.  For example, $\eta^{{\rm max}}_{{\rm1qb}}(0.99965) = 3\times 10^{6}\,\mathrm{W}^{-1}$, while $\eta^{\max}_0(0.99965) \sim 10^{10} W^{-1}$. While these two examples are not strictly comparable (the gate duration was optimized for the microscopic efficiency, but fixed at $\tau_{\rm 1qb}=25\,{\rm ns}$ for the macroscopic case), 
the main difference is that the cryogenic power consumption is larger than the microscopic power $P_\pi$ by a magnification factor $A T_\text{ext}/T_\text{qb}$ which can be very large ($\approx 2\times 10^4$ for ${\cal M}_{\text{1qb}} =0.99965$). 
This illustrates the reduction of efficiency when going from the microscopic to the macroscopic level.

\section{Example of noisy computation}
\label{Sect:NISQ}

Noisy computations are currently considered in the search for use-cases with a quantum computational advantage in the Noisy Intermediate Scale Quantum (NISQ) \cite{Preskill2018Aug} setting, as opposed to fault-tolerant quantum computing (FTQC) which we discuss in Sec.~\ref{Sect:fault-tolerant}. 

\red{Here we consider a simplified model of noisy computation (chosen for pedagogy rather than realism), performed with the simplified qubit model in the previous section. 
We use this simplified model to introduce how details of the algorithmic implementation enter MNR as control parameters that can be adjusted to minimize the resource consumption. 

Readers who want a more realistic indication of the minimal power consumption of a noise computation should look at the corner of Fig.~\ref{Fig:P_C-map}b marked $k=0$ (recalling that $k=0$ means that there is no error correction). It is based on our complete full-stack model in Sec.~\ref{Sect:fault-tolerant}, rather than the simplified model presented here. 
Although its assumptions are for large-scale fault-tolerant calculations not 
NISQ ones (it assumes large-scale cryogenics and certain approximations mentioned in \cite{footnote-noise-approx}), we expect its conclusion of a few milliwatts per physical qubit
to be reasonable for an optimistic estimate of the NISQ regime.}

\begin{figure}
    \centering
    \includegraphics[width=0.9\columnwidth]{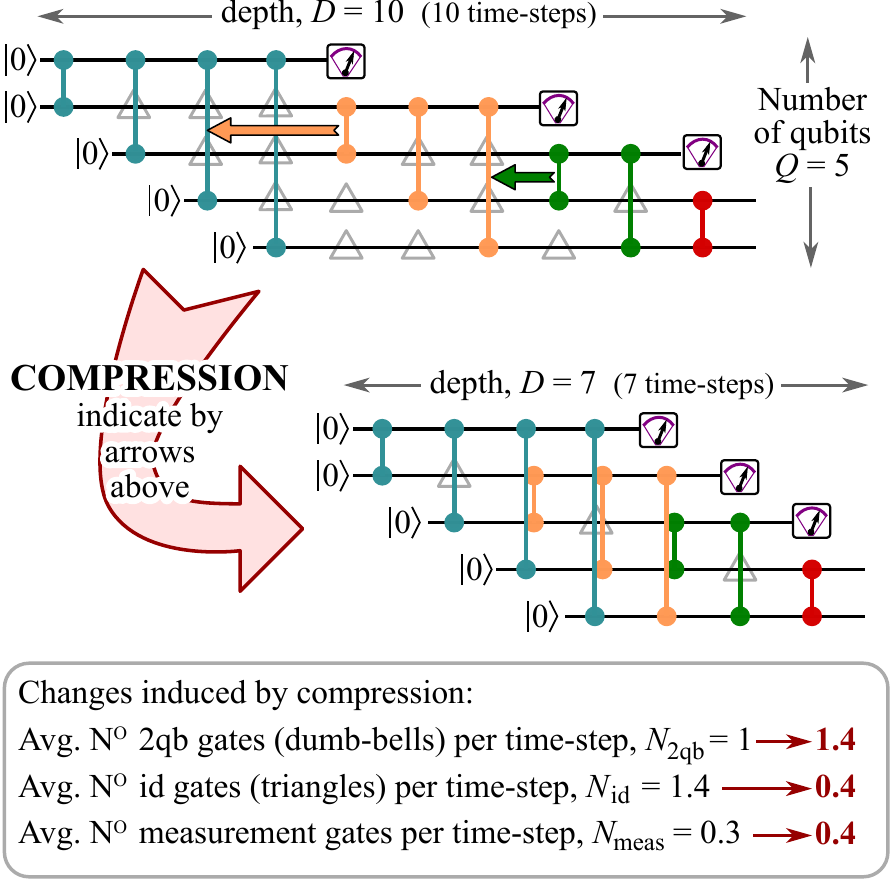}
    \caption{A hypothetical circuit made of two-qubit (2qb) gates, each indicated by a colored dumbbell. Each qubit has a 2qb gate with each of the qubits below it. Thus, if the circuit has $Q$ qubits, then it contains $\tfrac{1}{2}Q(Q-1)$ 2qb gates. In the upper circuit, no two 2qb gates are performed in parallel, so its depth is $D=\tfrac{1}{2}Q(Q-1)$, and there are many time-steps in which qubits are simply storing information for a later time-step. These are indicated by gray triangles, corresponding to noisy identity (id) gates.  One can ``compress'' the circuit step-by-step, by moving 2qb gates in the direction of the arrows shown on the upper circuit. 
    This increases the average number of 2qb gates per time-step, reducing the number of id gates and thereby reducing $D$.
    The lower circuit is the fully compressed version of the upper circuit.}
    \label{Fig:NISQ_circuit}
\end{figure}

To understand how the algorithm is taken into account by MNR,
it is important to note that the same algorithm can be implemented using different circuits. Here,
the {\it algorithm} refers to the overall operation we want to perform on the qubits, while a {\it circuit} is an instruction set specifying the sequence of gates on the qubits to carry out the algorithm. In the MNR methodology, the architecture of the circuit can be viewed as a control parameter for the algorithmic task. 
Our simple example here is the algorithm implemented by the circuit in Fig.~\ref{Fig:NISQ_circuit}, which bears structural similarities with a quantum Fourier transform circuit ~\cite{Nielsen2011Jan}. It comprises sequences of two-qubit (2qb) gates grouped into sub-circuits, marked with different colors in Fig.~\ref{Fig:NISQ_circuit}). For $Q$ qubits, this circuit has $Q-1$ sub-circuits, and $\frac{1}{2}Q(Q-1)$ 2qb gates. It can be ``compressed" by having the sub-circuits overlap, noting that there are idling qubits within each sub-circuit \footnote{We take the uncompressed circuit and label sub-circuits  from left to right (from 1 to $Q-1$).
Then at compression step $i$, we compress all sub-circuits to the left of the sub-circuit $i$,
and leave uncompressed all sub-circuits to the right of sub-circuit $i$.
Here ``compress'' means moving sub-circuits into each other, as in Fig.~\ref{Fig:NISQ_circuit},
such that as many gates are performed in parallel as possible.
The compression factor is then defined as $\epsilon=(i-1)/(Q-2)$, so  $\epsilon=0$  (meaning $i=1$) is the uncompressed circuit,
and  $\epsilon=1$  (meaning $i=Q-1$) is the fully compressed circuit.}. We define a compression parameter $\epsilon$, set to be 0 for the scenario where all sub-circuits are performed in sequence with no overlap (top circuit of Fig.~\ref{Fig:NISQ_circuit}). We can then make a succession of compression where some sub-circuits are partially performed in parallel with their preceding sub-circuits. Maximum compression occurs when $(Q-3)/(Q-1)$ sub-circuits are partially parallelized in this manner (bottom circuit of Fig.~\ref{Fig:NISQ_circuit}). 

In this section, the compression $\epsilon$ plays the role of a control parameter of software nature. This comes in addition to the hardware parameters used for the single-qubit gate, namely, the processor temperature $T_{\rm qb}$ and the control line attenuation $A$ (here taken to be identical for all lines). We will thus minimize the resource cost as a function of the triplet $(T_{\rm qb},A,\epsilon)$.

\subsection{Noise model and low-level metric}
In this section, we consider circuits built from a typical minimal gate set consisting of identity (id), single-qubit (1qb) and two-qubit (2qb) gates acting on $Q$ qubits. We assume the 2qb gates are implemented with a cross-resonance scheme \cite{chow2011simple,Sheldon2016Jun} in which the two qubits interact by sending a microwave signal to one qubit at the frequency of the other qubit. Such 2qb gates rely on resonant excitations similar to those employed in the 1qb gates. We thus assume 1qb and 2qb gates to have similar costs, and take that cost to be $P_\pi$ of Eq.~(\ref{P_pi}). 
2qb gates are, however, slower than 1qb gates, and we set $\tau_{\rm 2qb}=100\,$ns \cite{chow2011simple,Sheldon2016Jun}. Finally, the quantum computer runs at a clock frequency determined by its slowest gate. We thus set the clock period, or the time-step for gate applications, to be $\tau_{\rm step}=\tau_{\rm 2qb}= 100\,$ns. 

We first establish the relation between the local noise afflicting individual gate operations and the global metric characterizing the overall circuit performance. 

\mf{We follow the previous section in assuming that the only noise felt by the qubits is the unavoidable noise coming from the control lines. This is modelled as simple probabilistic noise in which each qubit has a probability to have an error during one timestep
equal to the worst-case infidelity  $\IFone$ of the process at each timestep, determined from \eqref{Eq:Lindblad}. Here $\IFone$ is defined as in \eqref{maxIF_1qb}, but $\tau_\text{1qb}$ is replaced by $\tau_{\rm step}=100ns$. 
Then a two-qubit gate has twice the infidelity of a single-qubit gate, because two qubits participate in a two-qubit gate.
So each id gate and each 2qb gate respectively have probabilities equal to $\IFone$ and $2\,\IFone$ of generating an error in the computation.

To quantify the algorithmic performance, we choose a low-level metric, ${\cal M}_\mathrm{algo}=1-P^{\text{error}}_\epsilon$, where $P^{\text{error}}_\epsilon$ is the probability that at least one error occurred within the circuit with compression $\epsilon$. For the algorithm to have a reasonable chance of success, $\IFone$ should be small, because of that, the error probability of the circuit can be approximated by: $P^{\text{error}}_\epsilon={\cal N}_g(\epsilon)\IFone$, 
where ${\cal N}_g(\epsilon)\equiv\big({\cal N}_{\rm id}(\epsilon)+{\cal N}_{\rm 1qb}(\epsilon)+2{\cal N}_{\rm 2qb}(\epsilon)\big)$. Here, ${\cal N}_i(\epsilon)$ is the total number of gates of type $i$ in the circuit with compression $\epsilon$.
}
\mf{From this, we deduce:}
\begin{align}
{\cal M}_{\text{algo}}(\epsilon,A,T_{{\rm qb}}) \, =\, 1- {\cal N}_g (\epsilon)\,\IFone (A, T_{{\rm qb}}).
\label{Eq:M_algo}
\end{align}
Eq.~\eqref{Eq:M_algo} makes explicit the influence of control parameters of software ($\epsilon$) and hardware ($A,T_{\rm qb}$) natures on the global performance of the algorithm.

\subsection{Resource cost}

\red{Whenever the calculation time is a parameter
(as it is when we introduce the circuit compression shown in Fig.~\ref{Fig:NISQ_circuit}), minimizing the average power consumption during the calculation is {\it not} the same as minimizing the energy cost of the calculation (since that energy cost is average power $\times$ calculation time).
So should we minimized power or energy?

We argue that power consumption should be minimized whenever that power consumption is large enough to be the principal engineering challenge. For example, there are many engineering reasons why it is much harder to consume 1\,GW for 1 minute, rather than 250\,kW for 3 days, even though the two have similar total energy costs.
Our main full-stack calculation, in section~\ref{Sect:fault-tolerant}, is in the regime where the power consumption is so high that it will be a huge engineering challenge. Thus it is critical to minimize this power. 
Hence, for simplicity, we also minimize power consumption for the pedagogical examples in the work, including for the simplified model of a NISQ calculation considered in this section.

In many cases, we believe that both minimizations will give similar results.  Minimizing energy cost will tend to promote shorter calculation times than minimizing power consumption alone (since it corresponds to minimizing power $\times$ calculation time). However, we observe that minimizing power consumption already tends to favour solutions with fairly short calculation times (see e.g., Sec.~\ref{Sect:logical-depth}).
So the parameters that minimize power consumption may not be far from those that minimize energy cost.}

We take the resource cost to be the total cryo-power averaged over a specified circuit that implements the algorithm. This is given by
\begin{equation}
    P_\epsilon(A,T_{{\rm qb}})=P_{\text{1qb}}(A,T_{{\rm qb}}) N_{\text{1qb},\epsilon} + P_{\text{2qb}}(A,T_{{\rm qb}}) N_{\text{2qb},\epsilon}.
    \label{Eq:power_generic_physical_1}
\end{equation}
\red{Here the cryo-powers supplied to perform a 1qb and 2qb gate are 
$P_{\text{1qb}}(A,T_{{\rm qb}})$ and $P_{\text{2qb}}(A,T_{{\rm qb}})$, respectively,
and it is assumed that identity (id) gates require no power.}
$N_{\text{1qb},\epsilon}$  and $N_{\text{2qb},\epsilon}$ are the average number of 1qb and 2qb gates, respectively, run in parallel per time-step of the circuit with compression $\epsilon$. 
Since we consider the power consumption during the execution of the algorithm, we shall only consider one run. Energy considerations of specific NISQ algorithms such as VQE or QAOA may require one to take into account the number of runs needed to reach a result with a certain accuracy.

\begin{figure}
    \centering
    \includegraphics[width=0.98\columnwidth]{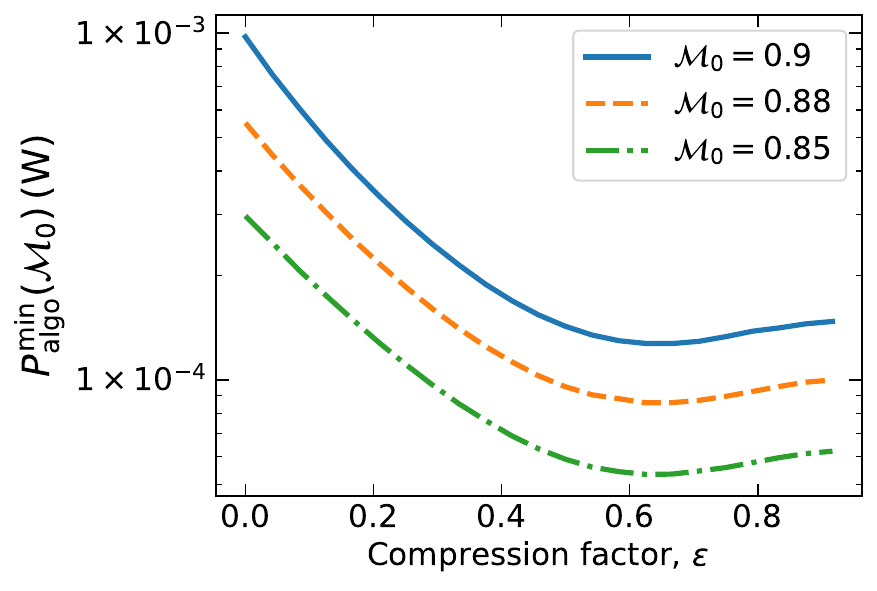}
    \caption{Minimum power as a function of the compression $\epsilon$ of the circuit (see text). The circuit is a 25-qubit version of the one shown in Fig.~\ref{Fig:NISQ_circuit}. Here, $\gamma^{-1} = 10\,\mathrm{ms}$.}
    \label{Fig:NISQ_depthChanging}
\end{figure}

\red{In our plots, Fig.~\ref{Fig:NISQ_depthChanging} and onwards, we have to choose certain parameters. There we assume each time-step is 100\,ns, where this is the time for a 2qb gate,
when 1qb gates take only $25\,$ns.
We assume that it takes about the same power to drive 1qb and 2qb gates, given by Eqs.~(\ref{P_pi},\ref{Eq:power_singlequbit}). However, as a 1qb gate is completed in a quarter of a time-step, its power consumption averaged over the time-step is 
a quarter that in Eq.~(\ref{Eq:power_singlequbit}), so $P_{\text{1qb}}(A,T_{{\rm qb}})=P_{\text{2qb}}(A,T_{{\rm qb}})/4$.}

Eqs.~\eqref{Eq:M_algo} and \eqref{Eq:power_generic_physical_1} allow us to optimize the total cryo-power as a function of the three control parameters $(A, T_{{\rm qb}}, \epsilon)$, under the constraint of a target metric ${\cal M}_\text{algo} = {\cal M}_0$. 
To make the impact of the circuit compression obvious, we first minimize the cryo-power with respect to $A$ and $T_\text{qb}$, for various values of the compression $\epsilon$. We performed this optimization for a circuit with $Q=25$ qubits; see Fig.~\ref{Fig:NISQ_depthChanging}. The plot shows that the sweet spot of minimal power consumption occurs when the circuit is partially compressed, revealing a competition between two mechanisms. On the one hand, compressed circuits correspond to a reduced total number of gates ${\cal N}$ (including id gates), with a reduced risk of error according to Eq.~\eqref{Eq:M_algo}, but to a larger number of gates run in parallel, leading to a larger power consumption. On the other hand, uncompressed circuits, with more idling qubits, increase the total error probability, which has to be compensated for by lowering the error rate per gate, achieved by lowering $T_{\rm qb}$ which also increases the cryo-power per gate (see Fig.~\ref{Fig:single-qubit}). This demonstrates that resource optimizations require coordinated inputs from the hardware and the software.\\

\red{Fig.~\ref{Fig:NISQ_depthChanging} illustrates that minimizing the average power consumption is not equivalent to minimizing the energy cost of the calculation.  Multiplying this power by the calculation time (with its relation to the compression factor explained above), we see that one can get a lower total energy cost for a higher compression factor (shorter calculation time) than that which minimizes the power consumption. However, one also sees that the difference is not huge (less than a factor of two  difference in energy consumption for the simple model in Fig.~\ref{Fig:NISQ_depthChanging}), so a circuit optimised for minimum power consumption will not be far from one optimized for minimum energy consumption.}

\begin{figure}
    \centering
    \includegraphics[width=0.98\columnwidth]{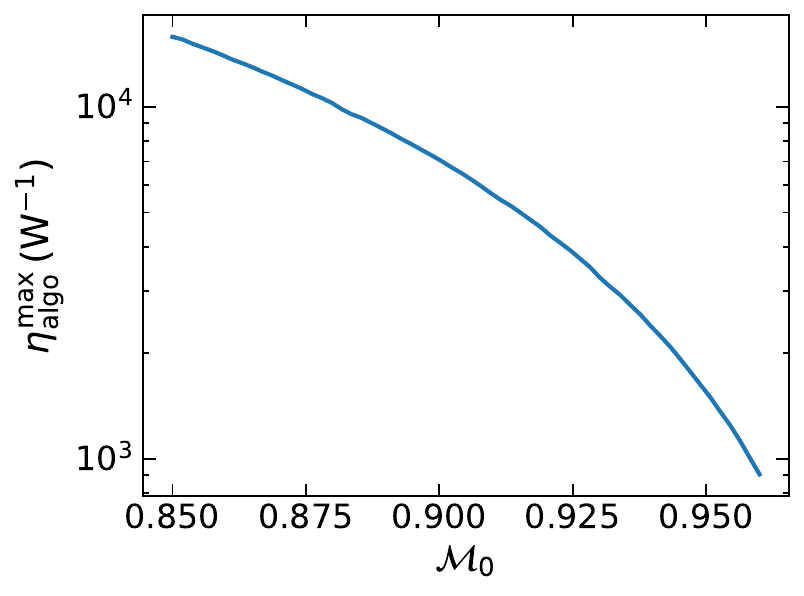}
    \caption{Optimal efficiency $\eta_\text{algo}^\text{max}(\mathcal{M}_0)$ as a function of the target algorithmic fidelity $\mathcal{M}_0$. The circuit is a 25-qubit version of the one shown in Fig.~\ref{Fig:NISQ_circuit}. Here, $\gamma^{-1} = 10\,\mathrm{ms}$.}
    \label{Fig:NISQ_circuitEfficiency_metric}
\end{figure}

\begin{figure}
    \centering
    \includegraphics[width=0.98\columnwidth]{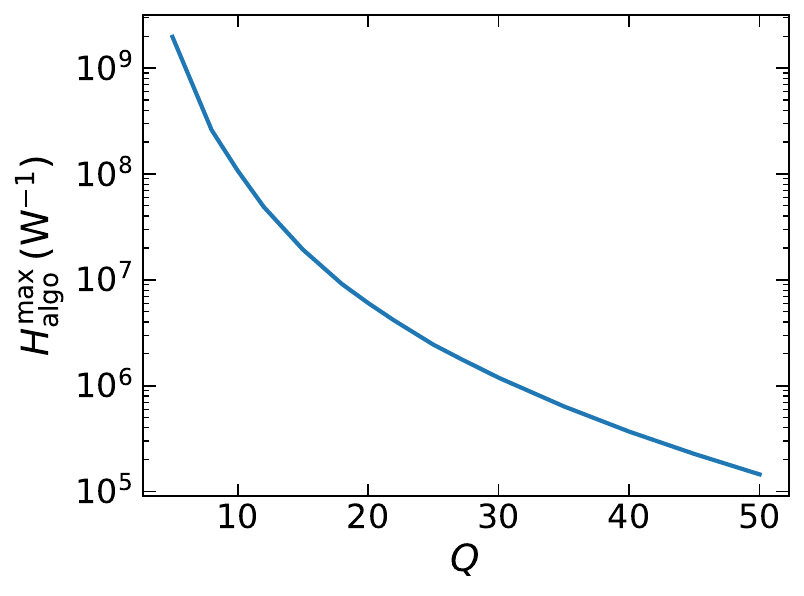}
    \caption{Optimal efficiency $H_{\text{algo}}^{\text{max}}(Q)$ as a function of the target metric $Q$, the number of qubits needed for an algorithmic success probability of $2/3$ (see text). Here, $\gamma^{-1} = 10\,\mathrm{ms}$.}
    \label{Fig:NISQ_circuitEfficiency_qubits}
\end{figure}

\subsection{Resource efficiencies}
Resource efficiencies for quantum algorithms executed on noisy circuits can be defined in two ways, depending on the target performance. First, one may adopt a low-level metric, such as the fidelity used above. This invites us to define an efficiency $\eta_\mathrm{algo}\equiv {\cal M}_\mathrm{algo}/P_{\epsilon}(A,T_{{\rm qb}})$ for a specified circuit implementation of the algorithm. For a given target metric ${\cal M}_0$, minimizing $P_{\epsilon}(A,T_{{\rm qb}})$ as a function of $A$, $T_{{\rm qb}}$, and the circuit compression $\epsilon$ defines the maximal algorithmic efficiency $\eta^{{\rm max}}_{{\rm algo}}({\cal M}_0)$. This is plotted on Fig.~\ref{Fig:NISQ_circuitEfficiency_metric} as a function of ${\cal M}_\mathrm{algo}\equiv{\cal M}_0$, for circuits with $Q=25$ qubits. While it follows the same behavior as the single-qubit gate efficiency, a direct comparison is difficult, because of the complexity of the relationship between the fidelity of a single gate and the fidelity of a whole circuit.

Second, the performance of an algorithm can be quantified by user-oriented metrics. A typical such metric is the size of the problem which can be solved with a given probability of success. In this example, it can be measured by the size of the data register $Q$ that carries out the algorithm, assuming the algorithm is executed with a $2/3$ success probability.
The maximal user-oriented efficiency is given by ${\rm H}^{\rm{max}}_{\rm{algo}}=Q/P^{\rm{min}}_{Q,\epsilon}(A,T_{{\rm qb}})$. 
Here, we have introduced $P^{\rm{min}}_{Q,\epsilon}(A,T_{{\rm qb}})$, the minimal cryo-power for a success probability of $2/3$ for a circuit of size $Q$, optimized with respect to the compression $\epsilon$, attenuation $A$ and processor temperature $T_{{\rm qb}}$. The maximal efficiency is plotted in Fig.~\ref{Fig:NISQ_circuitEfficiency_qubits} as a function of the target metric $Q$.

This section provides a pedagogical example to understand the impacts of the hardware and software choices on the energy consumption of a quantum computation. It also allows us to play with two different kinds of metrics and efficiencies, either low level or user oriented. To be truly informative, the user-oriented efficiency should be compared to a classical value quantifying the efficiency of a classical processor performing the same algorithm. A larger efficiency reached by the quantum processor provides a signature of a quantum energy advantage. This regime and the potential to reach it is studied in Sec.~\ref{Sect:fault-tolerant} in the context of fault-tolerant computation.


\section{Fault-tolerant computation}
\label{Sect:fault-tolerant}

We now turn to the macroscopic power consumption and energy efficiency of fault-tolerant quantum computation, currently the only known route to useful large-scale quantum computers. Fault-tolerant quantum computation is built upon the technique of quantum error correction. The basic idea of quantum error correction is to distribute each qubit of information over many physical qubits to form what is known as a logical qubit. This gives the logical qubit some resilience against noise that usually affects physical qubits individually. It requires, however, the use of several physical qubits to carry one logical qubit, and the addition of regular error correction operations, namely, syndrome measurements to diagnose what errors occurred, and recovery gates to remove the effects of those errors on the logical qubit.
This means a significant increase in the number of qubits, measurements, and gates, each of which has imperfections, and can potentially add noise to the computer. Computing with such logical qubits is said to be done in a fault-tolerant manner if the error correction operations, as well as the computational operations on the logical qubits (i.e., the logical gates), are designed so that the addition of so many more noisy physical components for the error correction still has the net effect of removing more errors than it introduces. 
This turns out to be possible only if the physical error rate is below some threshold level, often referred to as the fault-tolerance threshold.

From the user perspective, the fault-tolerant nature of the quantum computer is invisible. The user states the problem to be solved, and the algorithm to solve it, in terms of an ideal (noise-free) operation performed on a given input, and specifies a target metric (e.g., probability of success). This is then converted by the compiler to physical noisy qubits, gates, and measurements, using a prescribed fault-tolerant quantum computing scheme. The user-given input is represented by the logical qubits, encoded into the physical qubits carrying the information. The logical gates between those logical qubits that carry out the steps of the user-specified algorithm are converted into a sequence of physical gates between the physical qubits that make up each logical qubit.

For our simulation, we shall consider fault-tolerant quantum computing built from concatenating a 7-qubit code \cite{Steane1996Jul,Gottesman1997May,Steane1997Mar,Aliferis2006,Nielsen2011Jan}. This is a very well-studied scheme, and has the advantage over more recent proposals (e.g., those based on topological codes) in that it has fairly complete and well-documented analyses, allowing us to be sure we do not overlook any resource requirements. 
However, it is widely believed that fault-tolerant proposals based on surface codes require vastly less resources than the 7-qubit code.
In Sec.~\ref{sub:surfacecode}, we extend our results to such surface codes, confirming this but pointing out open questions there that make our estimates possibly unreliable.
The advantage of our complete analysis of the 7-qubit allows the reader to clearly see what questions they need to answer before doing similar estimates with their favorite fault-tolerant scheme.

\subsection{MNR on a fault-tolerant algorithm}
\label{sub:FTMNR}
Before coming to the specifics of our model we provide the reader with a general view on the approach that is valid for any quantum error-correcting code. We let $\perror$ denote the error probability of a physical qubit \footnote{As in Sec.~\ref{Sect:NISQ}, we attribute all noise in the physical gates and measurements as arising from the noise in the individual qubits that the gates and measurements are operating on. Additional control noise occurs in a realistic device, but this can be easily incorporated into our description by regarding $p_\mathrm{err}$ as the maximum over the physical qubit error probability and the error probability associated with the gate/measurement control.}, which is provided by a microscopic model of the noise. If the error correction is successful, the error probability of a logical qubit is reduced to $p_{\rm err;L}=f(k,\perror)$, where $f$ is a function and $k$ quantifies the amount of error correction---the concatenation level in the case of  our concatenated 7-qubit code example. The price to pay for this reduction of errors is that the number of physical qubits per logical qubit grows with $k$; we denote this number by $g(k)$.

Throughout this section, we consider a simple ``rectangular" circuit, with the goal of preserving $Q_L$ (logical) qubits of quantum information, for a total of $D_L$ (logical) time-steps. Such a rectangular circuit approximates well many fault-tolerant quantum algorithms based on $Q_L$ qubits and having a circuit with depth $D_L$, and still yields similar orders of magnitude for the power consumption and the metric (see Sec.~\ref{Sect:fullstackP}). As $(Q_L,D_L)$ is set by the choice of algorithm and circuit, $k$ is the only software parameter we are left with to perform our optimizations. 

The metric $\mathcal{M}_{\text{FT}}$ we will consider is the probability of success of the rectangular circuit. Denoting the number of locations where logical errors can happen as ${\cal N}_L=Q_L\times D_L$, we find 
\begin{equation} \label{eq:MFT}
\mathcal{M}_{\text{FT}}=(1-p_{\rm err;L})^{{\cal N}_L}.
\end{equation} 
Targeting a total success probability of  $\mathcal{M}_{\text{FT}}=2/3$, it translates into a maximal allowed value for $p_{\rm err;L}$. This maximal allowed value shrinks as the size ${\cal N}_L$ of the circuit grows. Hence, performing bigger computations while maintaining the same target metric mandates more error correction, and hence the consumption of more physical resources.

Estimating the physical resource cost requires the use of a full-stack model. Elaborations of the simple cases studied earlier to give a full-stack model that incorporates more experimental details  are presented in Sec.~\ref{sub:hardware}, leading to a larger set of hardware control parameters. We also need to specify the physical circuit that carries out the quantum algorithm, which depends on the parameter $k$. Altogether, we can establish the generic expression of the full-stack power consumption $P_{\text{FT}}$:
\begin{eqnarray}
    P_{\text{FT}}= P_{\text{1qb}} N_{\text{1qb}} 
        + P_{\text{2qb}} N_{\text{2qb}}
        + P_{\text{meas}} N_{\text{meas}} 
        + P_\text{Q} Q . \quad\quad
    \label{Eq:power_generic_physical}
\end{eqnarray}
The first three terms capture the dynamical power consumption: they are nonzero only when a computation is running and involve active gates and measurements. Measurements must be modeled since syndrome measurements for error correction take place all along the fault-tolerant quantum computation. As in Eq.~(\ref{Eq:power_generic_physical_1}), $N_{\rm 1qb}$ and $N_{\rm 2qb}$ are the average numbers of physical 1qb gates and 2qb gates, respectively, performed in parallel, while $N_{\rm meas}$ is the average number of physical qubit measurements performed in parallel. These three quantities are determined solely by the software, i.e., the algorithm, the choice of error-correcting code, and the parameter $k$. Conversely, $P_{\text{1qb}}$, $P_{\text{2qb}}$ and $P_{\text{meas}}$ are, respectively, the full-stack power consumption of 1qb gates, 2qb gates, and measurements, including all cryogenic and electronic costs; these depend solely on the hardware parameters. Finally, the fourth term in Eq.~(\ref{Eq:power_generic_physical}) captures the static power consumption, which we will take to be proportional to the number of physical qubits, $Q = g(k)\times Q_L$. Its expression depends both on software and hardware parameters.

The MNR methodology then simply consists of the following steps: (i) Consider an algorithm characterized by $(Q_L,D_L)$ and a target probability of success equal to $2/3$. As in Sec.~\ref{Sect:NISQ}, this $2/3$ is a common choice for the success probability for a single run of an algorithm, with an exponential chance of yielding the correct answer with a constant number of re-runs. Owing to Eq.~\eqref{eq:MFT}, this sets an implicit relation between the hardware control parameters and $k$. (ii) Minimize the power consumption $P_{\text{FT}}$ as a function of the control parameters under the constraint of reaching a probability of success $\mathcal{M}_{\text{FT}}=2/3$.

\begin{figure}
    \includegraphics[width=0.9\columnwidth]{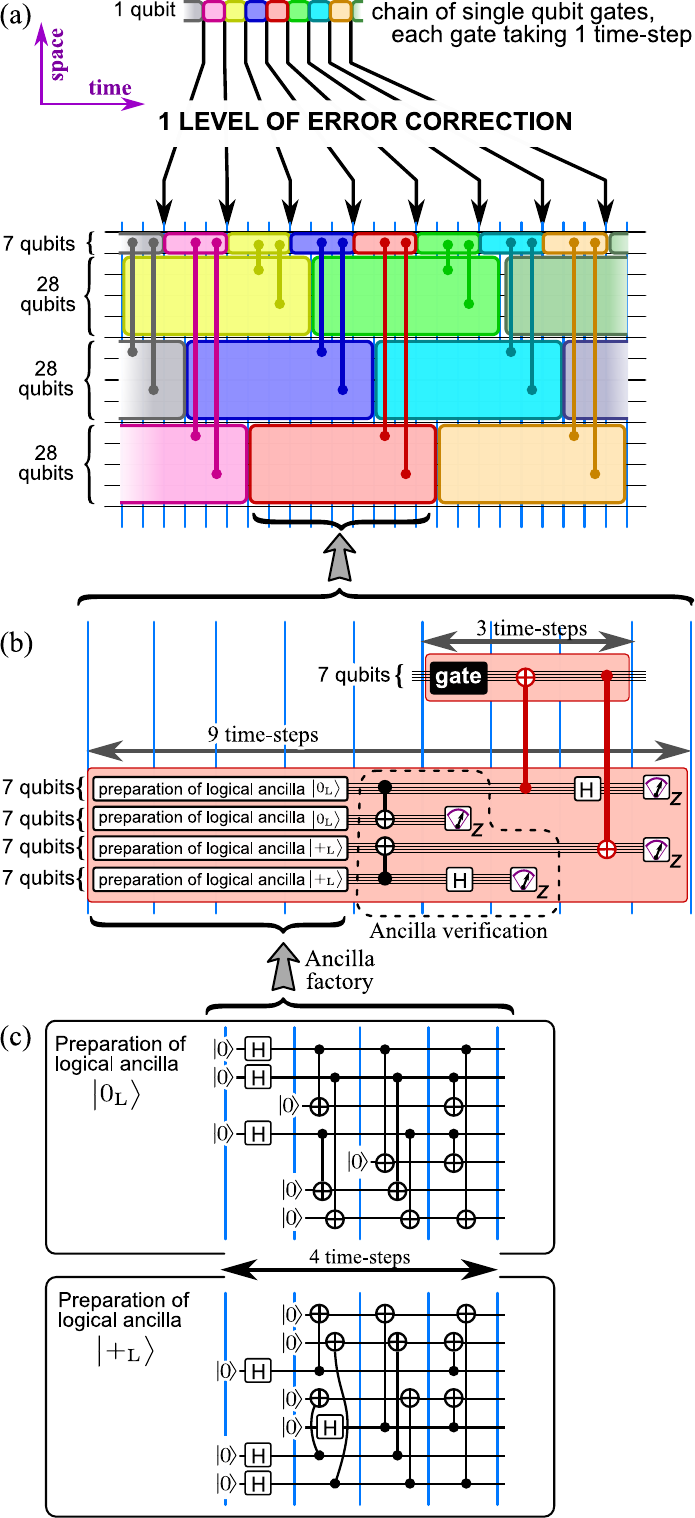}
    \caption{Circuit for one level of the 7-qubit error correction code on a series of single-qubit gates \cite{Steane1997Mar,Aliferis2006}.  Gates acting on groups of 7 qubits are transversal. The {\it ancilla factory} prepares the state of the logical ancillas \cite{ancilla-always-good}.  Vertical blue lines indicate time-steps in the algorithm. See the main text for a brief description of this circuit, with additional technical details given in App.~\ref{Sect:Appendix-logical_to_physical}.
    }
    \label{Fig:steane_method}
\end{figure}

\subsection{Noise and metric for the 7-qubit code}
\label{sub:noise-metric}
Let us first consider the noise at the level of a single physical qubit. Instead of the infidelity, we consider the qubit's error probability $\perror$. We employ the same noise model as that of Sec.~\ref{Sect:noisy-gate}. We write 
 $\perror = \tfrac{1}{2}\gamma \tau_\text{step}  \,\left(\tfrac{1}{2}+ n_\text{noise}\right)$, see footnote \cite{footnote-noise-approx}.
Here,
$\tau_\text{step}$ is the time-step of the quantum computer, taken equal to the time taken to perform the slowest qubit gate, i.e., the 2qb gate in our model.

From now on we focus on the 7-qubit code. The basic components of the fault-tolerance scheme are illustrated in Fig.~\ref{Fig:steane_method}, starting at the top with the logical circuit to be implemented (drawn here, for simplicity, for just single-logical-qubit gates). Each logical gate is broken down into the physical qubits and gate operations that are needed to implement it in a fault-tolerant manner, with qubits that carry the actual logical information, as well as ancillary qubits (or just ``ancillas") that permit the syndrome measurements for error correction. Shown also in Fig.~\ref{Fig:steane_method} are the details of the preparation of the state of the ancilla for the error correction to work in a fault-tolerant manner. These details are critical inputs to our power-consumption calculations below (see App.~\ref{Sect:Appendix-logical_to_physical}).

The power of the code can be increased, thereby acquiring the ability to remove more errors, by concatenating the basic 7-qubit code: At the first level of concatenation, the logical qubit is encoded into 7 physical qubits; at the next level of concatenation, the logical qubits of the previous level are treated like physical qubits, and the logical qubit at this level is encoded into 7 logical qubits of the previous level, thus employing $7^2$ physical qubits in all; and so on in a recursive manner. Error correction is done at every level of the concatenation.
After $k$ levels of concatenation, the error probability per logical qubit per (logical) time-step can be shown to be \cite{Aliferis2006}
\begin{eqnarray}
p_{\rm err;L} = \pthr  \ \left(\perror\big/\pthr\right)^{2^k} ,
\label{Eq:logical-gate-error-prob}
\end{eqnarray}
where $\pthr\approx 2 \times 10^{-5}$. 
Here $1/\pthr$ is an integer that counts the number of ways the extra physical elements (qubits, gates, and measurements) added to correct errors can have faults; see, for example, Ref.~\cite{Nielsen2011Jan} for a fuller explanation. $\pthr$ is the aforementioned fault-tolerance threshold: The error per logical qubit decreases as $k$ increases only if the qubit error probability $\perror$ is less than $\pthr$. This is an important constraint on the physical qubits that we will consider for our simulations, requiring fidelities which can be significantly beyond the state of the art. 

Increasing $k$ increases the ability to remove errors, and hence compute more accurately. The price to pay, however, is a large increase in the number of physical qubits. For a computation with $Q_{\rm L}$ logical qubits, the fault-tolerant scheme requires $Q$ physical qubits where
\begin{eqnarray}
Q\ \equiv\ g(k)\times  Q_{\rm L} \ =\ (91)^k Q_{\rm L},
\label{Eq:physical_vs_logical-qubits}
\end{eqnarray}
This formula can be understood from Fig.~\ref{Fig:steane_method}, which illustrates how fault-tolerant quantum computation with the 7-qubit code works. We focus on the first level of concatenation, where one logical qubit is encoded into 7 physical qubits. In addition to the 7 physical qubits, one needs 28 physical qubits as ancillas to facilitate syndrome measurements for the code. The 28
ancillas are explained in Refs.~\cite{Steane1997Mar,Aliferis2006}.
In short, they should be understood as two groups of 14 ancillas each, with one group for each of the two kinds ($X$ or $Z$) of syndrome measurements needed for the 7-qubit code. Each group of 14 should again be thought of as two groups of 7 ancillas;  one of these groups is used to verify the quality of the ancillas in the other group, necessary to guarantee fault tolerance. Furthermore, the ancillas have to be prepared in specific states for the syndrome measurement. As the ancilla preparation takes a certain number of time-steps to complete (four time steps, as shown in Fig.~\ref{Fig:steane_method}), in order for all 28 ancillas to be ready at the time they are needed in the syndrome measurement, we find that, at any one time-step, there must be 3 groups of 28 ancillas each in various stages of preparation; see Fig.~\ref{Fig:steane_method}. This then gives the $91 = 7 + 3 \times 28$ in Eq.~\eqref{Eq:physical_vs_logical-qubits}, for $k=1$. Then, the recursive structure of the concatenation, treating each logical qubit as if it were a physical qubit at the next level, gives the $(91)^k$ factor for $k$ levels of concatenation in Eq.~\eqref{Eq:physical_vs_logical-qubits}. 

This systematic analysis allows us to derive the number of 1qb gates, 2qb gates, and measurements running in parallel as needed in Eq.~\eqref{Eq:power_generic_physical} to estimate the power consumption; the details are given in App.~\ref{Sect:Appendix-logical_to_physical}. Finally, the overall metric introduced in Sec.~\ref{sub:FTMNR} for our generic algorithm is given by
\begin{align}
    \mathcal{M}_{\text{FT}}=\left[1-\pthr \,\big(\perror\big/\pthr\big)^{2^k}\right]^{{\cal N}_L}.
    \label{Eq:metric_logical}
\end{align}
For simplicity, we use the linear approximation, 
\begin{align}
    \mathcal{M}_{\text{FT}}=1- {{\cal N}_L} \,\pthr \,\big(\perror\big/\pthr\big)^{2^k},
    \label{Eq:metric_logical-linear}
\end{align}
which slightly overestimates the effect of the errors (i.e., slightly underestimates the metric).

\subsection{Full stack hardware model} 
\label{sub:hardware}
\begin{figure*} 
    \includegraphics[width=0.78\textwidth]{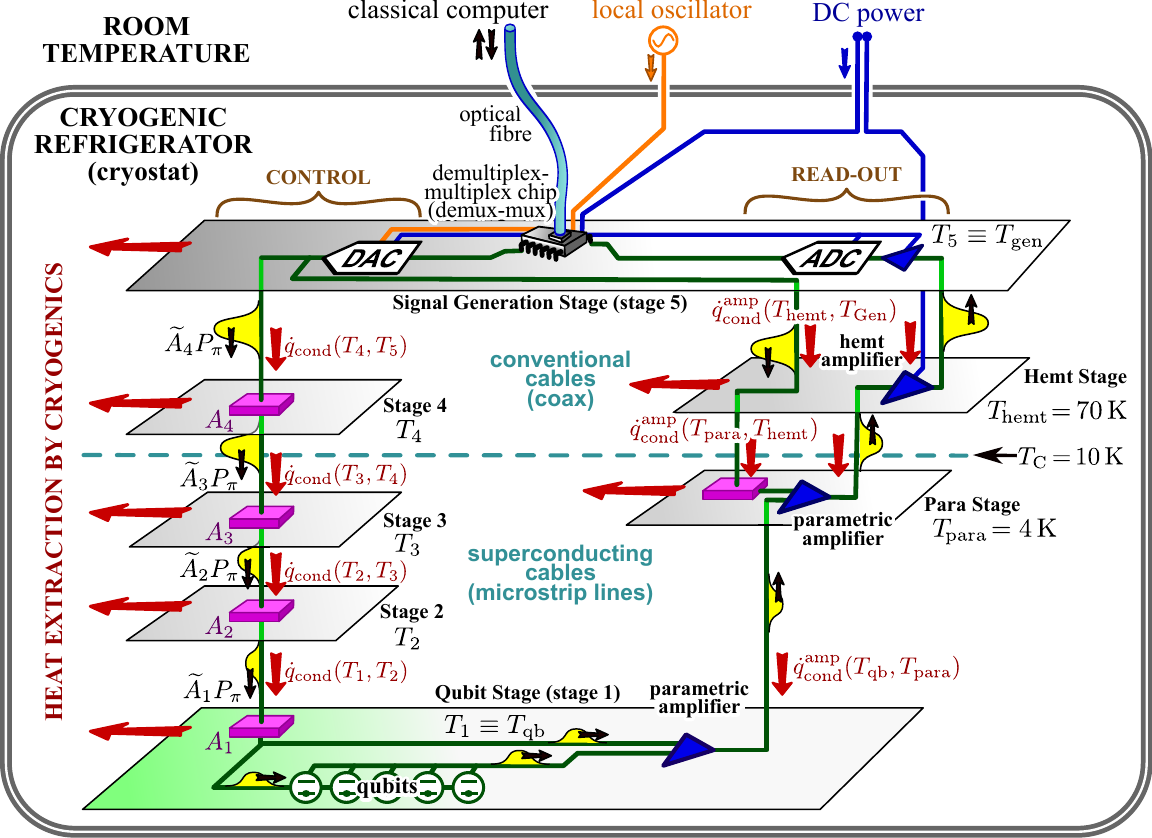}
    \vskip 2mm
    {\scriptsize
    \begin{tabular}{|c|c|} 
    \hline
    {\bf Component} & {\bf Ratio to qubits} \\
    \hline
      optical fiber $+$ demux-mux & 1 for 270 qubits \\
      local oscillator & 1 wire (mm $\varnothing$) for $10^6$ qubits \\
      \ dc (demux-mux$+$DAC$+$ADC) \ & 1 wire (mm $\varnothing$) for $10^6$ qubits \\
      dc (hemt amplifiers) & 1 wire  (mm  $\varnothing$) for $10^8$ qubits \\
      Qubit control (1 \& 2qb gates) &  1 DAC$+$cable (with attenuators) for 25 qubits \\
      Qubit read-out (measurement) & \ 1 cable$+$paramps$+$hemt\,amp$+$ADC  for 100 qubits \  \\
      \hline
    \end{tabular}}
    \caption{
    \label{Fig:full-stack-stages} 
   Sketch of our model of the multi-stage cryogenics with all components. It is important to maximize the number of physical qubits per other component (using multiplexing, etc), and the table gives reasonable values for the ratio of the number of components to the number of physical qubits.
    The qubit control lines are particularly crucial to the energy consumption; we model them with 4 stages of attenuation (attenuators in purple), with conventional coaxial cables down 10\,K and superconducting microstrip lines below that. The read-out is less crucial to the energy consumption, if one uses superconducting parametric amplifiers at $T_{\rm qb}$ and $T_{\rm amp}=4$\,K, with a third amplification stage using High Electron Mobility Transistors (hemt) at $T_{\rm hemt}=70$\,K.  Black arrows indicate the flow of information/signals, while red arrows indicate heat conduction.  The demux-mux, DAC, ADC, attenuators and amplifiers all also generate heat. 
    }
\end{figure*}

We now present our full-stack model, which goes significantly beyond the pedagogical model used in earlier sections. In short, we replace the simplified set-up of Fig.~\ref{Fig:single-qubit}(a) by the full set-up of Fig.~\ref{Fig:full-stack-stages}. This involves key improvements over the  simplified set-up of Fig.~\ref{Fig:single-qubit}(a) that brings us closer to experimental reality. These improvements dealing with the control electronics and the cryogeny are presented below, with more details in the appendices. We take inspiration from current technologies for the improved model. Nevertheless, our interest is in understanding general trends that will provide guidelines for ongoing and future research, and this leads us to consider values that are beyond the current state of the art.

\subsubsection{Cryogenic model}
The first improvement to bring us closer to experimental reality is that we spread the attenuation on the microwave control lines over multiple temperatures stages (see the left side of Fig.~\ref{Fig:full-stack-stages}).  This is known to be much more energy efficient than placing all the attenuation at $T_{\rm qb}$, as we had done in Fig.~\ref{Fig:single-qubit}.
Much of the heat generated by attenuators is thus dissipated at higher temperatures, where it costs much less power to extract it.  Adding more temperature stages always reduces power consumption, but it is often technically challenging. We observe that the benefits of adding another stage becomes small once there are about five stages, so we take five temperature stages here. Appendix \ref{sect:Attenuation-multi-stage} gives the detailed specifications of these five stages of attenuation.
The heat conducted by the control lines turns out to be significant, and to minimize this heat conduction, we assume all wiring to be superconducting below 10\,K, and thus conduct vastly less heat than normal metal wires. The heat conduction properties of these control lines are given in App.~\ref{Sect:conduction}.
As above, we assume that the cryogenics has Carnot efficiency, and thus use the minimal possible power to evacuate heat as allowed by the laws of thermodynamics.  We take this for simplicity as it already gives the right order of magnitude for large-scale cryogenics, where the state-of-the-art is 10\% to 30\% of Carnot efficiency \cite{Parma2014}.  Evidently, results change if one considers small-scale cryostats which operate far below Carnot efficiency, as shown in the example in App.~\ref{App:varying_power_per_physical}.

\subsubsection{Control electronics}
A second improvement to bring us closer to experimental reality is that we now add the circuitry to read out the qubits (see the right side of Fig.~\ref{Fig:full-stack-stages}). The signal from the qubits has to be amplified significantly above the thermal noise level at the temperature stage that the signal is being sent to. As amplifiers generate heat, it is again much more energy efficient to have a chain of amplifiers at different temperature stages, than to have all the amplification occur at the qubit temperature.  Superconducting parametric amplifiers generate much less heat than conventional amplifiers, but they can only operate at temperatures below 10\,K. At higher temperatures, the best option is amplifiers based on high-electron mobility transistors (hemt). Here, we take the amplification chain inspired by recent experiments \cite{Malnou2021Oct}: We assume one superconducting parametric amplifier at the qubit temperature, which sends the signal to another superconducting parametric amplifier at 4\,K. This then sends the signal to a hemt amplifier at 70\,K, which finally sends it to the chip that reads out the signal. Appendix \ref{Sect:sig_gen_amps} gives the detailed specifications for this chain of amplifiers. The readout lines are the same materials as the control lines, so their heat conduction properties are those described in App.~\ref{Sect:conduction}.
 
The third and final improvement is that we assume there is a signal generation stage at temperature $T_{\rm gen}$, with chips that carry out the signal generation and readout. Below, we want to find the optimal value of $T_{\rm gen}$ that minimizes the power consumption. For this, we need to know the heat dissipated by the signal generation stage, which requires a specification of what it contains. Our model assumes that the signal generation stage receives digitized instructions of the waveform to generate for each gate operation down an optical fiber from a conventional (classical) room-temperature computer. The signal generation stage contains a chip (demux) that de-multiplexes the photonic signal in the optical fiber, and turns it into digital electrical signals. These digital signals are turned into analogue signals in the DACs, and are then superimposed on the local-oscillator signal (at 6 GHz) to make the microwave signal that performs the desired gates on the target qubits. At the same time, the signal generation stage takes the microwave waveform coming from the measurement of the qubit through the amplifiers, and digitizes it in the ADC. This is then turned into a multiplexed photonic signal, which is sent through the optical fiber to the conventional room-temperature computer. This (classical) computer demultiplex it and digitally demodulates the waveform, allowing it to deduce the state of the qubit in question \cite{krantz2019quantum}. It also decodes (i.e., interprets) the syndromes coming from the error correction procedure and manages the algorithm at the logical level. Further details of this are provided in App.~\ref{Sect:Appendix-classical-computer}, which argues that this classical computer will not be a significant contribution to the power consumption, and so can be neglected at our level of approximation.

\subsubsection{Control parameters}
We can now summarize the four control parameters we will use for our optimizations, namely, $T_{\rm qb}$, $T_{\text{gen}}$, $A$ (the \textit{total} attenuation on the lines), and $k$, the concatenation level. The temperature of each stage 
and the amount of attenuation put on these stages, 
are taken to be functions of $T_{\rm qb}$, $T_{\text{gen}}$ and $A$ [see  Eq.~\eqref{Eq:A_i_and_T_i} in the appendix]. As explained around Eq.~\eqref{Eq:A_i_and_T_i}, we consider such constraints to lead to a relatively optimal distribution of attenuation and temperatures.

\subsection{Full-stack power cost for the 7-qubit code}
\label{Sect:fullstackP}

\subsubsection{Software assumptions}
We first write down Eq.~\eqref{Eq:power_generic_physical} describing the power consumption for the specific case of fault-tolerant quantum computing based on the 7-qubit code. 
\red{For this we need to look at the circuit for one level of error-correction for any Clifford logic gate, see Fig.~\ref{Fig:steane_method}. The circuit looks the same for such any logic gate, except for the contents of the black box marked ``gate'' (this ``gate'' in Fig.~\ref{Fig:steane_method}b is transversal, containing 7 physical gates corresponding to the logic gate). 
However, this black-box makes a very small contribution to the total number of gate operations in the circuit, so once the error correction is included (i.e. $k\geq 1$), all logical gates require about the same number of physical gate operations. Thus one expects that any logical gate will have a power consumption very similar to that of a (logical) identity gate that does nothing except preserve the quantum state of the logical qubit. This intuition is  confirmed and carefully quantified in  App.~\ref{Sect:Appendix-logical_to_physical}.
As a result, the power consumption of any given algorithm 
is almost independent of what the algorithm is actually doing at the logical level; it only depends on that algorithm's number of logical qubits $Q_{\rm L}$ and logical depth $D_{\rm L}$. We can thus take the power-consumption of any algorithm to be close to that of a logical memory whose only job is to preserve the state of $Q_{\rm L}$ logical qubits for $D_{\rm L}$ logical time-steps.}

The power consumption of such a circuit can be taken as proportional to  $Q_{\rm L}$ (see App.~\ref{Sect:Appendix-logical_to_physical}), with 
\begin{align}
    P_\text{FT}\simeq Q_{\rm L} \!\left( \frac{4(64)^k}{185} \big[16 P_{\text{2qb}} 
        +\! 7 P_{\text{1qb}} 
        +\! 7 P_{\text{meas}} \big]
        +\! (91)^k P_\text{Q} \!\right),
    \label{Eq:power_versus_QL}
\end{align}
using an approximation that gets better at higher $k$. Appendix \ref{Sect:Appendix-logical_to_physical} shows that this approximation gets the order of magnitude right for any circuit of Clifford gates at $k=1$, and is within a few percent of the correct result for $k\geq 2$.

\red{To keep the modelling here as compact as possible, we neglect the power consumption associated with fault-tolerant non-Clifford gates (such as $T$-gates). While a quantum computer without at least one type of non-Clifford gate is not universal (and can be efficiently simulated on a classical computer), the modelling of non-Clifford gates is very different than Clifford gates. App.~\ref{Sect:neglected_T_gates} discusses this modelling, and points out how rare non-Clifford gates are in the algorithms that we consider. It then argues that accounting for them would complicate the modelling without significantly changing the resulting power consumption.}

\subsubsection{Hardware assumptions}
What remains to be calculated is the contribution of each hardware component to $P_{\rm 1qb}$, $P_{\rm 2qb}$, $P_{\rm meas}$, and $P_\text{Q}$. We compute
$P_{\rm 1qb}$ and $P_{\rm 2qb}$ in the same manner as in the noisy quantum circuit in Sec.~\ref{Sect:NISQ}, except that we now account also for the chain of attenuators at different temperatures. The expressions for $P_\mathrm{1qb}$ and $P_\mathrm{2qb}$ are given in App.~\ref{sect:Attenuation-multi-stage}. For $P_{\rm meas}$, we use a similar approach to that for $P_{\rm 1qb}$, as the measurement in our model involves sending a microwave signal similar to that for a gate operation. Our estimations, however, show that in most cases $P_{\rm meas}$ is negligible compared to $P_{\rm 1qb}$, so we drop it. $P_\mathrm{1qb}$, $P_\mathrm{2qb}$, and $P_\mathrm{meas}$ grow whenever the qubit temperature is reduced to raise the physical qubit fidelity. A larger physical qubit fidelity hence requires a larger power consumption for gates and measurements. 

For our simulations, we make the qubit lifetime vary between 3.5\,ms and 1\,s. However, our main discussion will be based on a lifetime of 50\,ms which is about $100$ times better than the state of the art in transmon qubits \cite{Wang2022Jan}. This is necessary since, as mentioned above, a successful calculation using a fault-tolerant scheme built from the 7-qubit code requires an error probability smaller than the threshold of $\pthr= 2 \times 10^{-5}$, achievable only for qubits with a long enough lifetime. 

Next, $P_\text{Q}$ is the part of the power consumption that scales as the number of physical qubits, independent of whether gates or measurements are being performed. It has two different contributions. The first is the power consumption of the cryogenics to remove the heat conducted down the microwave lines that control and read out the qubits. Their thermal properties are given App.~\ref{Sect:conduction}. The second is the heat generated by all electronics that are always on. This includes the amplifiers at 4\,K and 70\,K, the electronics for control and readout at $T_{\rm gen}$, and the classical computer at ambient temperature. Their detailed specifications are given in App.~\ref{Sect:sig_gen_amps}, with the full list of parameters summarized in Table~\ref{table:parameters}. 

We consider three generic scenarios, labelled A, B, and C, for the control electronics.
Scenario A can be taken as a futuristic scenario for conventional CMOS technology where the control electronics typically dissipates $1$mW of heat per qubit at the temperature $T_{\rm{gen}}$. Current best estimates are closer to 15-30\,mW per qubit \cite{Park2021Feb,frank2022cryo,Kang2022Aug}, but these numbers are dropping as research progresses. Taking this optimistic value compared to current CMOS also reinforces the observation that we make in Sec.~\ref{Sect:impact_control_electronics} \footnote{When we started this work, it was suggested that 1\,mW per qubit was directly achievable \cite{Bardin2019Oct}, but very recent analyses \cite{Park2021Feb,frank2022cryo,Kang2022Aug}  accounted for more aspects of signal generation and argued that the current state-of-the-art is 15-30\,mW. Nonetheless, we believe reaching $1$ mW per physical qubit is an optimistic but reasonable target given the rapid progress in the field.}. 

Scenarios B and C respectively correspond to improvements by 2 and 4 orders of magnitude compared with scenario A in terms of heat dissipation per qubit. Scenario C can be taken as a futuristic projection for classical logic performance based on superconducting circuits known as single-flux quantum (SFQ) \cite{mcdermott2018quantum} which may potentially generate about 10\,000 times less heat than CMOS. However, our results should mainly be taken as an indication of the importance of research in this direction.

We conclude this summary of our full-stack model by noting that our simulations use generic numbers and orders of magnitudes. Our results should thus not be considered as precise estimates for a specific technology or platform. Instead, they enable us to observe general trends and thereby provide understanding that can guide future experiments.

\begin{figure*}
\includegraphics[width=\textwidth]{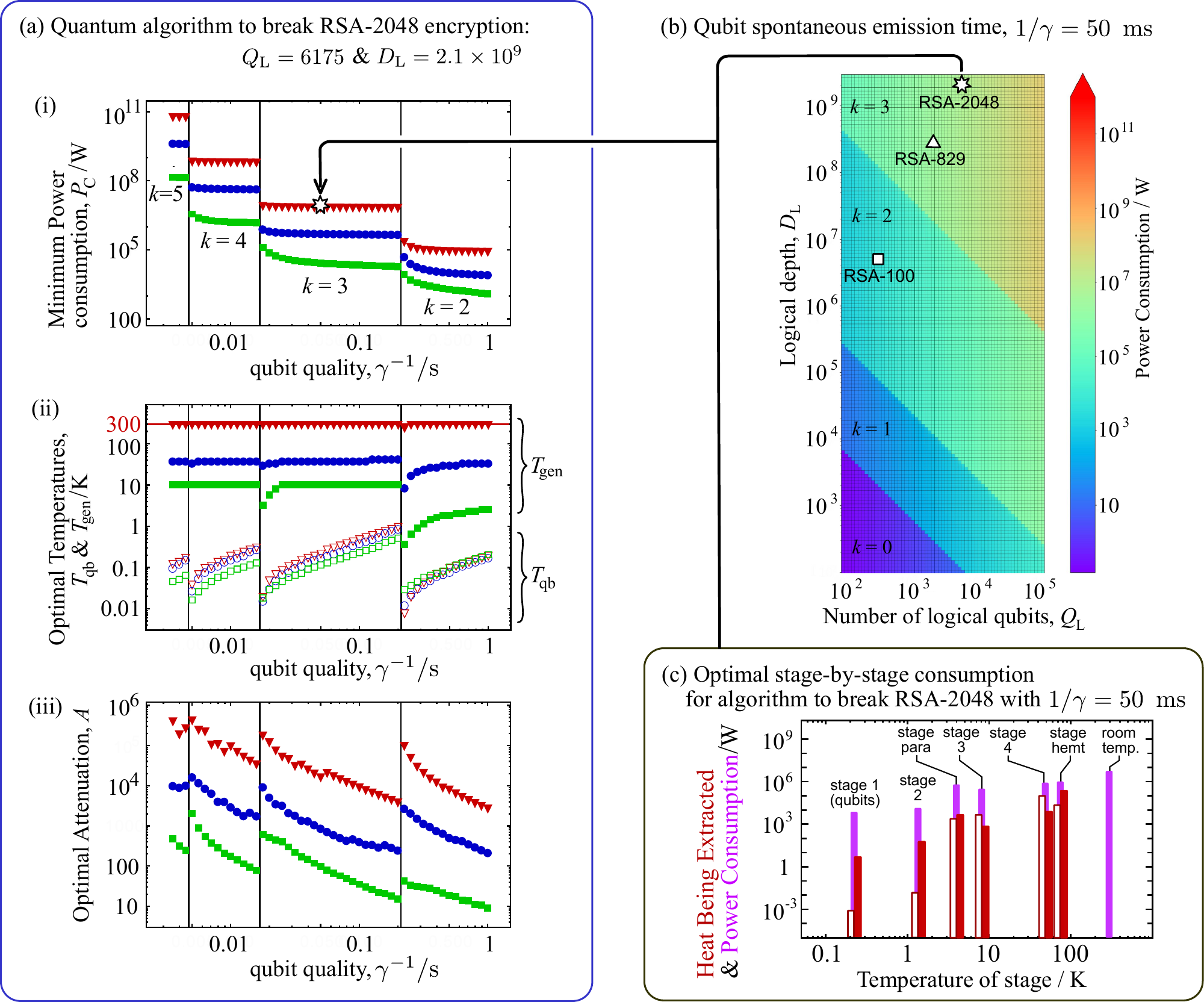}
\caption{(a) The upper plot is the minimum power consumption, $P_{\rm FT}$, as a function of the qubit quality (qubit lifetime $1/\gamma$) under the constraint that the computation successfully cracks the RSA-2048 encryption, a calculation that requires $Q_{\rm L}=6175$ logical qubits, and a logical depth of $D_{\rm L}=2.1\times10^9$. The different symbols are for the different types of control electronics at $T_{\rm gen}$.
Red triangles, blue circles, and green squares respectively correspond to our scenarios A, B and C; see main text. The lower two plots show the optimal temperatures and attenuations for which this minimum $P_{\rm FT}$ is achieved. Transitions between values of $k$ are first-order like, and we observe that their positions are fairly well estimated by a simple formula (see App.~\ref{Sect:Appendix-first-order}). 
(b) Minimum power consumption, $P_{\rm FT}$, as a function of the size of the calculation being carried out, for scenario A with $1/\gamma=50\,ms$, which corresponds to $\perror\simeq \pthr/40$ (see text). The star indicates values corresponding to an algorithm that breaks the RSA-2048 encryption, with a total power consumption of 7\,MW for $4.6\times 10^{9}$ physical qubits.  
(c) For every set of parameters, we have the heat extracted at each refrigeration stage after optimization. We show this here for the parameters corresponding to the point marked by the star in (a) and (b). The red/white bars are the heat extracted at each refrigeration stage, with white corresponding to the heat conduction down the cables, and red being the heat generated at each stage by attenuators or amplifiers. Each purple bar is the power consumption to extract this heat (assuming Carnot-efficient refrigeration). The purple bar at $T_{\rm gen}=$300\,K (room-temperature) is the power consumption of the control electronics.
\label{Fig:P_C-map}
}
\end{figure*}

\subsection{Minimization of Power Consumption} 
\label{Sect:optimizing-fault-tolerant}

We now use our model to minimize the macroscopic power consumption $P_\text{FT}$, under the constraint of a fixed target metric, ${\cal M}_{FT}=2/3$, following the MNR methodology presented in Sect.~\ref{sub:FTMNR}. Our results for $P_\text{FT}$ do not vary significantly for slight variations of the target metric from the specified 2/3 value \footnote{Note, however, that $P_\text{FT}$ diverges for ${\cal M}_{FT} \to 1$ (if this is allowed by the algorithm) because that corresponds to an algorithm that never gives the wrong answer; this would require an infinite amount of error correction, and hence would require infinite resources.}.
This target constrains the control parameters, since the metric depends on $p_{\text{err}}$, which in turn depends on the control parameters [see Eqs.~\eqref{Eq:metric_logical-linear} and \eqref{Eq:eta_vs_T}]. Under this constraint, we optimize the power consumption with respect to four control parameters: the temperature $T_{\rm gen}$ of the signal generation (top stage in Fig.~\ref{Fig:full-stack-stages}), the temperature $T_{\rm qb}$ of the qubits, the total attenuation $A$ between $T_{\text{gen}}$ and $T_{\text{qb}}$, and the level of concatenation $k$ for the fault-tolerant scheme. Figure \ref{Fig:P_C-map}(b) presents a two-dimensional map of our optimizations, i.e., the minimal power $P_{FT}$ consumed by our generic circuit of size $(Q_L,D_L)$. $P_{FT}$ increases with the number of logical qubits $Q_L$ and depth $D_L$, with the discontinuities corresponding to the change of concatenation level $k$. We considered scenario A, with high quality qubits characterized by $1/\gamma=50\rm{ms}$, corresponding to $\perror \sim p_{\rm{thr}}/40$ (i.e., $\perror = p_{\rm{thr}}/40$ when $T_{\rm qb}$ is small enough and $A$ large enough so that $\perror = \frac{1}{4}\gamma \tau_{\rm step}$).  

As mentioned at the beginning of this section, our chosen circuit provides a good approximation of the power consumption of any circuit involving the same values of $Q_L$ and $D_L$. We use this property to estimate the minimum power required to implement the set of quantum gates given in Ref.~\cite{Gidney2021Apr} for the Shor's algorithm that breaks the RSA encryption of an $n$-bit key. Fig.~\ref{Fig:P_C-map}a(i) shows $P_{FT}$ for as a function of the qubit quality $\gamma^{-1}$. The different curves are for different values of power dissipation for the electronics. Figs.~\ref{Fig:P_C-map}(a)(ii) and (iii) show the values of temperatures and attenuation that give the minimal power consumption. Finally, Fig.~\ref{Fig:P_C-map}(c) shows the heat evacuated (and the corresponding power consumption) at each temperature stage in the cryogenics. Our results allow us to make a number of observations likely to hold for a range of fault-tolerant quantum computing schemes, including those based on surface codes. These observations are detailed below, and deal with the respective impacts of the qubit fidelity, the control electronics, the cryogenics, and the logical depth.

\subsubsection{Impact of qubit fidelity} 
If the error probability is only slightly below the fault-tolerance threshold, we observe that the power consumption is unreasonably large. However, the power consumption drops very rapidly as the quality of the qubits increases. In the present model, increasing qubit quality means having the physical qubits couple more weakly to the microwave control line, and hence more microwave power to drive the qubits [see Eq.~\eqref{P_pi} for how the power to flip a qubit from $|0\rangle$ to $|1\rangle$ is proportional to the qubit quality, $\gamma^{-1}$].  Despite this, the gains from reducing the error rate (and hence reducing the necessary amount of error correction) greatly outweighs the costs of increasing the microwave power per physical qubit. Fig.~\ref{Fig:P_C-map} shows that a factor of 10 increase in the qubit quality (i.e., dividing $\gamma$ by 10) leads to a factor of 100 reduction in the overall power consumption for a given computational accuracy.  We believe that a large reduction in power consumption from improved qubit quality is likely to be a general trend in all parameter regimes, and indeed in all qubit technologies, placing a significant emphasis on developing qubits of the highest possible quality. 

It is worth noting that additional sources of noise (beyond the unavoidable noise in the lines that control the qubits) will always add to the resource cost. They will always increase the power consumption, as we are required to cool the qubits further, or provide additional error correction to achieve a given metric of performance.  Particularly dangerous are errors due to long-range cross-talk between qubits, since error correction can be of limited use against them \cite{PRXQuantum}.  

\subsubsection{Impact of control electronics}
\label{Sect:impact_control_electronics}
Once the cryogenics are optimized, we observe that
the control electronics are a dominant contribution for 
scenario A. This is clear from Fig.~\ref{Fig:P_C-map}(c), where we plot the heat to be extracted per stage in the cryogenics, and the corresponding power consumption per stage. The absolute magnitude of heat and power varies dramatically with the quality of the gates and with $Q_{\rm L}$ and $D_{\rm L}$, but we observe that the ratios between different stages do not vary very much. In all cases, we find that the total power consumption per physical qubit (given by $P_{\rm FT}/Q$)  is 1.3--2\,mW [it is 1.5\,mW for the star in Fig.~\ref{Fig:P_C-map}(a)]. Relatively little of that comes from the cryogenics below 4\,K; the dominant part ($\sim$1\,mW) comes from the control electronics at $T_\text{gen}$.

For this reason, our optimization in scenario A puts the control electronics at room temperature (i.e., the optimal $T_\text{gen}$ is ambient temperature), with the consequence that there are many millions of room-temperature cables (a few cables for every 25 physical qubit, in our model of multiplexing) going down into the cryostat. Placing the electronics at 4\,K will reduce the heat conduction into the cryostat, as there will then be almost no cables between 300\,K and 4\,K. However, the heat generated by these control electronics is vastly more than that brought in by the wires, and the resulting increased demand for cooling will increase the power consumption by a factor of about 75 (see Fig.~\ref{Fig:gain_power_4K_fct_gamma} in the appendix). 
It is only when the dissipation of the control electronics is orders of magnitude lower (such as 10\,$\mu$W in scenario B) that we observe a significant energetic advantage in placing these electronics at lower temperatures, as shown in the middle plot in Fig.~\ref{Fig:P_C-map}(a).

We recall that we have assumed Carnot-efficient cryogenics. If the cryogenics are only at 10-30\% of Carnot efficiency [see Sec.~\ref{Sect:cryo-efficiency}], then the optimal $T_{\rm gen}$ will remain at room temperature in scenario A (or for any CMOS technology consuming more power than scenario A). The power consumption of the cryogenics will be higher (e.g., ten times larger for cryogenics at 10\% Carnot efficiency), but the power consumption of the control electronics will remain a major cost. Hence, research to minimize this cost is crucial.

At the same time, we do not yet have a technology that can reliably install many millions of wires (with attenuators) between the room temperature stage and the qubits.
It may thus be necessary to put the control electronics at low temperatures, despite the cryogenic cooling costs. This makes it crucial to pursue ongoing research to improve the efficiency of cryo-CMOS, hand-in-hand with designing cryogenics that can efficiently evacuate large amounts of heat at the temperature chosen for the cyro-CMOS control electronics.

\subsubsection{Impact of cryogenics}
When we assume that the cryogenics are close to achieving Carnot efficiency, we observe that the cryo-power comes mainly from evacuating heat generated at temperatures above 4\,K.  An example of this is shown in Fig.~\ref{Fig:P_C-map}(c). This means its power consumption is almost independent of the qubit temperature, which is always significantly less than 4\,K. 
More precisely, the total power consumption has a large $T_{\rm qb}$-independent contribution, with the contribution for $T_{\rm qb}$ (which diverges at  $T_{\rm qb} \to 0$) dominating only for very small $T_{\rm qb}$. This causes the abrupt change in $P_{\rm FT}$ as $k \to (k-1)$ visible in Fig.~\ref{Fig:P_C-map}, although, if one were to magnify the curves sufficiently, one would see that the curves are continuous with a discontinuity in their derivative at the transition from $k$ to $k-1$ (see App.~\ref{Sect:Appendix-first-order}).

Notably, this observation comes with a caveat: It relies on the cryogenics being reasonably close to Carnot efficiency at low temperatures.
If this is not so, one can find cases where the power consumption depends largely on $T_{\rm qb}$. 
For example, many small-scale laboratory cryogenic systems have a
heat extraction at ultra-low temperature far from the Carnot efficiency, and some experimental qubits have significant additional sources of heat at $T_{\rm qb}$. In such situations, the overall power consumption may be dominated by the evacuation of heat at $T_{\rm qb}$.
Appendix \ref{App:varying_power_per_physical} gives an example of this in which the power consumption per physical qubit can vary by three orders of magnitude as $T_{\rm qb}$ changes. The minimal power consumption for a given metric of performance then depends more strongly on the qubit quality, and in a more subtle manner on many other hardware and software parameters.  Without the systematic optimization proposed here, one simply cannot know the optimal values of all the control parameters. Appendix \ref{App:varying_power_per_physical} has  examples where a poor choice of parameters can induce a power consumption of gigawatts, compared with megawatts when the optimal parameters are used. 
In contrast to conventional wisdom (quantified in App.~\ref{App:varying_power_per_physical}), power consumption is sometimes reduced by a strategy of raising the qubit temperature (and hence increasing the errors per physical qubit)
and compensating for it by having more error correction. Unexpectedly, whether this strategy is optimal or not, depends on parameters such as the power consumption of the control electronics.

The above caveat means that, for any given hardware, one has to carefully optimize the full-stack model to know whether the power consumption is dominated by effects above 4\,K, or by effects at $T_{\rm qb}$. 
Our analysis suggests that the most promising cases fall into the former category (dominated by effects above 4\,K),
and so this observation will apply to them.  Nevertheless, this has to be checked on a case-by-case basis.

\subsubsection{Impact of logical depth}
\label{Sect:logical-depth}
We observe that the power consumption increases with algorithmic depth. By reasoning in terms of power per qubit, it is easy to guess that the power consumption at each time-step in the calculation goes up with the number of qubits.  However, the power consumption at each time-step in the calculation also grows with the depth of the algorithm---it grows with the number of logical time-steps in the calculation, $D_{\rm L}$. This is because a longer computation requires lower error probability per gate-operation, in order to keep the quantum information error-free until the end of the computation. A longer calculation thus requires more noise mitigation by, for example, lowering the qubit temperature or performing more error correction.
This means more power consumption at every time-step in the calculation. This behaviour is quite different from deterministic computation in classical computing, where the power consumption of the computer depends only on the number of operations performed in parallel, usually not on the duration of the computation. 

This behavior is, however, not unique to quantum computers. An example of an analogous situation in classical computing is floating-point calculus for simulating chaotic systems  (e.g., a three-body problem \cite{Frolov2003}).
To achieve a given precision for a simulation of the chaotic system for $D$ time-steps, one is required 
to take a floating-point precision at each time step that grows with $D$, requiring more power consumption at each time-step. However, while power consumption increasing with the algorithmic depth, $D$, occurs in certain specific cases in classical computing, it is unavoidable in quantum computing.

\subsection{Quantum energy advantage} 
\label{Sect:quantum_energy_advantage}

Finally, we make use of our model to explore the potential of fault-tolerant quantum computers to achieve a quantum energy advantage. Usually, the concept of quantum advantage refers to a comparison between the computing powers of classical and quantum processors, with a quantum advantage being present if the quantum processor solves a problem in less time or space than the best in-class classical processor. Here, we bring the discussion to the energetic level, and ask when a fault-tolerant quantum computer can solve problems using less energy than a classical supercomputer, a feature we dub the quantum energy advantage. 

Intuitively, one expects such quantum energy advantage whenever the quantum computer solves a problem, intractable on a classical computer, in a reasonable time. Even if the quantum computer requires more power than the classical computer, the considerably shorter time taken for the quantum computer will give a lower energy cost. But the quantum energy advantage involves much more surprising regimes. In particular, it can appear for problems solvable on a classical computer, and even in cases where the quantum computation takes more time than the classical one. We see examples of these various cases below.

\begin{figure}
\includegraphics[width=0.9\columnwidth]{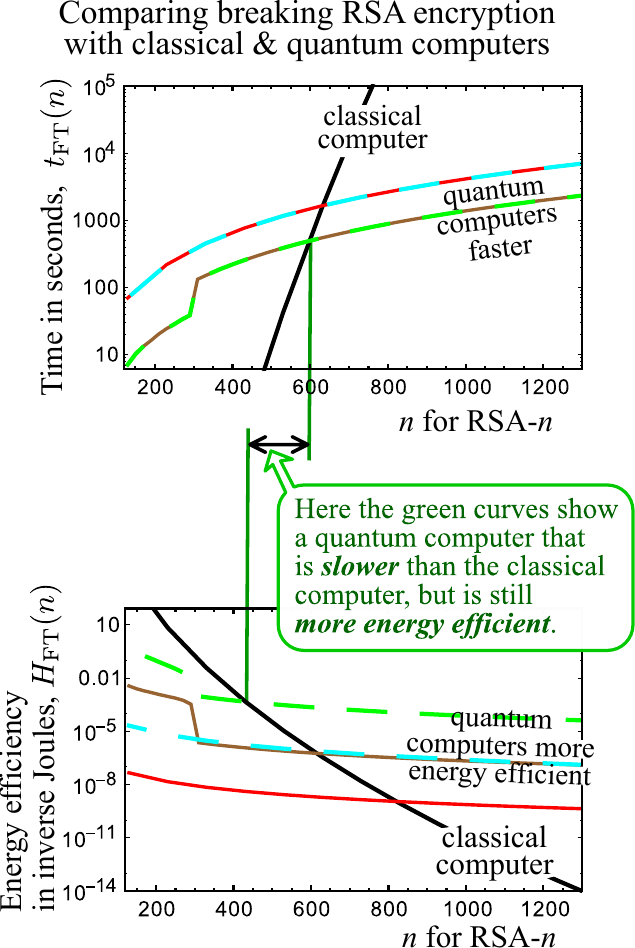}
\caption{ 
\label{Fig-RSA-quantum-versus-classical}
Comparison between classical and quantum computers performing the calculation to crack RSA-$n$ encryption as a function of the key-size $n$.  The plots show (a) the time $t_{FT}$ taken to perform the calculation and (b) the energy efficiency of the calculation. The time of the quantum computation is the number of time-steps in the algorithm multiplied by the time per step, $\tau_{\rm step}=100$\,ns. The efficiency is defined as ${\rm H}_{FT}(n) ={\rm E}_{FT}(n)$, where ${\rm E}_{FT}(n)={\rm P}_{FT}\times t_{FT}(n)$ for the quantum computer.
We plot this efficiency for the value of ${\rm P}_{FT}$ found from our minimization. The black curve is a classical supercomputer based on Ref.~\cite{Boudot2020Aug} (see App.~\ref{Sect:classical-RSA}). The colored curves are for quantum computers with various parameters: red for qubits with $1/\gamma=5\,$ms and the control electronics in scenario A,
brown for $1/\gamma=50\,$ms and scenario A,
cyan for $1/\gamma=5\,$ms and scenario C,
and green for $1/\gamma=50\,$ms and scenario C. 
In all three cases, the quantum computers become faster and more efficient than the classical computer as $n$ grows. The green curves shows a quantum computer that becomes more energy efficient than the classical computer {\it before} it becomes faster than it. 
}
\end{figure}

The manner in which a quantum computer solves a problem is so different from a classical computer that there is little sense in comparing the power consumption per gate-operation or per floating point operation; the energy efficiency measure of FLOPS/W used in classical computing is  not applicable to quantum computers. Instead, one must compare the power consumption of quantum and classical computers performing the same useful task.
As an example of such a useful task, we consider the cracking of an $n$-bit RSA key discussed earlier. There are well-documented quantum \cite{Gidney2021Apr} and classical \cite{Boudot2020Aug} algorithms for this task, allowing us to perform a fair energetic comparison. This time, we choose a user-oriented metric to quantify the performance of the quantum processor, namely, the size of the key $n$ which can be broken with success probability of $2/3$. In a similar spirit as the quantity defined in Sec.~\ref{Sect:NISQ}, a convenient user-oriented energy efficiency is ${\rm H}_{{FT}} = n/{\rm E}_{FT}(n)$, where ${\rm E}_{FT}(n) = {\rm P}_{FT}(n)\times t_{FT}(n)$ is the energy consumed by the fault-tolerant quantum computation, and $t_{FT}(n)$ is the duration of the computation. Such an efficiency corresponds to the inverse of the energy cost per bit of the key and has the dimensions of a bit/J.

We first consider a situation of quantum computational advantage for a calculation that can be completed on a classical computer in a week or two. This is the case for the RSA-830 encryption (i.e., public-key encryption with a 830-bit key), which has been cracked on a classical supercomputer \cite{Boudot2020Aug}, 
using the equivalent of 2700 core-years on Intel Xeon Gold 6130 CPUs. These consume about 12\,W per core, so cracking the encryption classically requires about a terajoule of energy. This yields a typical efficiency of $8\times 10^{-10}$\,bit/J.
If it were done using all the cores on a top-100 supercomputer, such as the JUWELS Module 1 \footnote{JUWELS Module 1 has 114\,000 cores with performance close to Xeon Gold 6130 CPUs, and is also in the top 100 for energy efficiency \cite{Green500}}, then we estimate  that its power consumption would be about 1.3\,MW (including cooling), and it would crack the encryption in 8-9 days (see App.~\ref{Sect:classical-RSA}). 

In contrast, a quantum computer with high-quality qubits ($1/\gamma=50$ms) and control electronics corresponding to scenario A, corresponding to the triangle in Fig.~\ref{Fig:P_C-map}b, can complete the calculation in about 16 minutes, with a power consumption of about 2.9\,MW after optimization. We again recall that we are optimistic on hardware parameters compared to the state of the art, and that we made some simplifying assumptions (see Apps.~\ref{Sect:neglected_T_gates} and \ref{sect:details-RSA}), but we also base our optimizations on an error-correcting code that is highly demanding in resource. 
This corresponds to a total energy consumption of 2.7\,GJ and an efficiency of  $3\times 10^{-7}$bit/J. 
This is more than two orders of magnitude less than the classical supercomputer, clearly pointing to a quantum energy advantage. 

For RSA-2048 encryption, considered un-crackable by classical supercomputers (even with an arbitrarily large power supply), Fig.~\ref{Fig:P_C-map}(b) gives a quantum computer power consumption of about 7\,MW after our optimization. If this could be achieved, it would be similar to the power consumption of some of the largest existing supercomputer clusters. The quantum computer would solve this RSA-2048 problem in 1.5 hours, for a total energy cost of about 38\,GJ, equivalent to the energy in about twenty car-tanks of gasoline \cite{gasoline} and an efficiency of $5\times 10^{-8}$bit/J.

In these two examples, the smaller energy cost of the quantum calculation arises because the quantum computation is much faster, while both the classical supercomputer and the quantum computer require megawatts of power. However, this is not the only route to a quantum energy advantage. To investigate this point more thoroughly, we computed the time $t_{FT}(n)$ needed for the quantum computer to break keys of increasing size $n$, and the corresponding energy efficiency ${\rm H}_{{FT}}(n)$. The comparison with classical computers allows us to define the respective regimes of computational and energy advantages. Both quantities, together with the estimations for their classical counterparts, are plotted in Fig.~\ref{Fig-RSA-quantum-versus-classical} for various qubit and control electronics technologies. As it appears on the figure, the quantum computational advantage solely depends on the qubit fidelity and remains insensitive to the dissipation by the control electronics---the better the qubits, the sooner the advantage. Conversely, the quantum energy advantage is sensitive to the qubit fidelity and other parameters such as the efficiency of the control electronics. Hence, both advantages are reached for sufficiently large keys, at the price of huge experimental and technological challenges. 

Interestingly, our study singles out cases where the quantum computer consumes less energy than a classical computer, even when the quantum calculation takes longer than the classical one. An example of this can be seen on the green curve in Fig.~\ref{Fig-RSA-quantum-versus-classical}, obtained for high-quality qubits with extremely efficient control electronics (scenario C). Fig.~\ref{Fig-RSA-quantum-versus-classical} illustrates that the computational advantage in the upper plot is of a different nature than the energy advantage in the lower plot, with the former being related to time, and the latter to energy. Hence, it should not come as a surprise that they occur in different parameter regimes. 

These observations suggest that an important future motivation for using quantum computers could be that they can solve problems in a more energetically efficient manner than classical computers, even when those problems are solvable on classical computers in an acceptable time (or even solvable faster on a classical computer). We believe that this may become a crucial source of applications for quantum computing, especially because an energy advantage without a computational one might ask for less demanding qubit quality and a quantum computer of much more moderate size. The search for quantum energy advantage should thus be an active goal of research, alongside the search for quantum computational advantage.

\subsection{The case of surface code}
\label{sub:surfacecode} 
It is likely that we can significantly improve the efficiency of quantum computers by taking more resource-efficient error-correcting codes than the concatenated 7-qubit code discussed in the previous sections. An example could be surface code, the current choice for fault-tolerant quantum computing being pursued in many experiments. Unlike the concatenated 7-qubit code, the literature lacks certain information about surface codes, which limits our capacity to make concrete claims. In particular, the error diagnosis from the measured syndromes is trivial for concatenated codes (see App.~\ref{Sect:Appendix-classical-computer}), while surface codes require complicated algorithms run on conventional computers to ``decode" the syndrome information to deduce what errors may have occurred, and to do this fast enough to correct them. These decoding algorithms remain a subject of investigation today and may carry a significant energy cost. We currently know the general scaling properties of some of the most well-known decoders (e.g., minimum-weight perfect matching \cite{Edmonds1,Edmonds2,Kolmogorov2009Jul,WangMWPMdecoder,fowler2015MPWM}, and union-find \cite{DelfosseUnionFind,Das2020Jan} decoders. However, we do not have access to the detailed pre-factors necessary for our resource estimates for a realistic classical computer that is powerful enough to decode the errors fast enough to correct them. 

Nevertheless, we can give estimates for some aspects of the full-stack fault-tolerant quantum computer with surface codes. We start from the proposal of Ref.~\cite{Gidney2021Apr} to crack the RSA-2048 encryption in about 8 hours, using 20 million superconducting qubits, with error probability per gate set to be $0.1\%$.
For scenario A, a suitably-optimized concatenated code has a total power consumption of $1-2$mW per physical qubit (including cryogenics, control and readout electronics, etc), whatever the qubit quality and whichever algorithm is being performed;  see Eqs.~\eqref{Eq:physical_vs_logical-qubits}-\eqref{Eq:power_versus_QL}.
We recall our caveat that scenario A calls for parameters beyond the current state of the art. Assuming similar numbers apply in the surface-code situation, the 20 million qubits needed to break RSA-2048 will then only require a power of 20 to 40\,kW \footnote{Considering a more conservative estimate for the consumption of the electronics generating the signals, i.e., 15-30\, mW per physical qubit \cite{Park2021Feb,frank2022cryo,Kang2022Aug} (instead of 1\,mW per physical qubit as used in our scenario A), the power consumption of the whole computer apart from the classical computer decoding the syndromes would be dominated by this electronics that should be put at room temperature. The power consumption would then be 300-600\,kW, which is less or comparable to a typical supercomputer (their consumption is usually in the range 1MW-10MW).}.
Then this part of the quantum computer will be truly green: It will take about 8 hours to do something no classical supercomputer could ever do, while using about the same amount of power as an electric car driving on an interstate highway. Of course, this estimate  excludes the power needed for running the decoding algorithm, which, as mentioned above, may be significant. However, as decoding algorithms get better, there is a reasonable hope that the quantum computer will require no more power than a classical supercomputer to do useful calculations that no supercomputer can ever do.

These considerations make it likely that such a quantum computer would have a much better quantum energy advantage than that discussed Sec.~\ref{Sect:quantum_energy_advantage}, including for tasks that could be done on a classical supercomputer in a reasonable time.

\section{Discussion and outlook}
\label{Sect:discussion}

This work presents and applies a new methodology, dubbed Metric-Noise-Resource (MNR), \red{that provides a holistic and quantitative model of the full-stack of a quantum computer. It identifies and quantifies the links between the quantum and macroscopic levels of such a quantum computer.} This provides the theoretical foundation for the minimization of the resource costs of quantum technologies.
We use it to arrive at a quantitative relation between the computing performance of a quantum processor and its macroscopic resource cost, and thus minimize the latter under a given constraint of performance.  The MNR methodology provides a common language to connect fundamental research and enabling technologies, fostering new synergies, such as those recently called for in the Quantum Energy Initiative \cite{QEI-perspective}. It allows us to define and optimize various quantum computing efficiencies, i.e., new versatile resource-based figures of merit that we expect to serve as a tool to benchmark a large range of hardware and software features, including qubit technologies, processor architectures, quantum error-correcting codes, etc. 

To illustrate our methodology, we applied it to the full-stack model of a superconducting quantum computer. There, we established the first estimates of energy costs in relation to the quality of the qubits, the type of quantum error-correcting codes employed for fault-tolerant computing, as well as the architecture and energetic performance of control electronics and cryogeny. We considered the case of RSA key breaking, a task with well-defined classical and quantum algorithms, and whose complexity is indicative of useful applications in physics research, chemical simulations, and various optimization tasks. 

Our model is based on highly optimistic assumptions for the qubits lifetimes and noise model (as needed for the 7-qubit code), and somewhat optimistic characteristics for other elements in the model. We sought to model all the main sources of energy consumption, while maintaining a reasonable simplicity in our model. Our results should hence not be interpreted as design blueprints for an energy-efficient quantum computer. Instead, they are a guide to what must be understood and improved in the physics and engineering of such a quantum computer.

We provided examples where the minimization of resources costs can reduce power consumption from gigawatts to megawatts. 
While this may arguably be an extreme case, our results suggest that such resource optimizations are likely to be critical in the success of both the current state of the art, and in futuristic scenarios. 
\red{Existing prototypes of quantum computers (for which energy optimization was not a priority) often consume hundreds of watts per physical qubit \cite{Arute}; our proposed optimization \mf{suggests that} future generations of quantum computers \mf{could} consume only a few milliwatts per physical qubit.
\red{In our main example, this includes the power cost of classical computing to decode the error-correction syndrome (this cost is negligible there).  In some other cases, this syndrome-decoding cost is not yet known (see Sec. \ref{sub:surfacecode}),
so it may be in addition to our prediction of a few milliwatts per physical qubit.}
}

In some of our examples, our minimization led to optimal parameters that could not have arisen from simpler estimates and are contrary to conventional wisdom. For example, it is sometimes energetically favourable to put the qubits at higher temperature, and compensate with more error correction, an unexpected strategy given how many more physical components are needed for the extra error correction. Only a systematic full-stack analysis using our MNR methodology can reliably determine these optimal parameters for a given quantum computing platform.

In addition, we observed evidence of a potential quantum energy advantage for RSA key breaking. While our optimization there involved somewhat futuristic scenarios with qubit quality and control electronics beyond the existing state of the art, it nevertheless suggests that quantum processors can consume less energy than classical ones even in the regime where the RSA keys are small enough for classical computers to break in a reasonable time. This is an extremely encouraging result, particularly as other error-correcting codes (such as surface code) could demand fewer resources than the simple one used in our model, while also working with much noisier qubits.

We also showed that a quantum energy advantage and a quantum computational advantage are different concepts that can correspond to different parameter regimes. Energy savings provided by the quantum logic thus emerge as a crucial practical interest of quantum computing, distinct from computational advantage.  This is a particularly interesting and open problem for the NISQ regime. From this perspective, it should be investigated whether the current generation (or the next generation) of noisy quantum computers could be better than classical computers for useful algorithms, not necessarily by showing a computational advantage, but by being more energy efficient instead. 

The MNR methodology presented in this work will help in the design of scalable quantum processors using various qubit technologies and different energetic regimes (cat-qubits, photons, electron spins, atoms, etc.). It can provide a consistent set of design specifications that keeps resource costs reasonable, and help to identify potential technological bottlenecks to achieving these specification.
It can help define roadmaps for the various research and industry teams working on these technologies, as it relies heavily on the development of efficient and integrated control electronics, with parametric amplifiers, signal multiplexing techniques, software engineering techniques and the like. The MNR methodology will be useful in prioritizing the coordinated development of all components in the full-stack quantum computer.
Developing this mindset will be vital in ensuring that quantum computing avoids dead-ends that await environmentally unsustainable technologies. Beyond the case of quantum computing, we also suggest to apply the MNR approach to optimizing the resource cost of other quantum technologies such as quantum communications and quantum sensing. 

\section*{Acknowledgements}
This work benefits from the support of the European Union Horizon 2020 research and innovation programme under the collaborative project QLSI grant number 951852, the ANR Research Collaborative Project ``QuRes" (Grant No. ANR-PRC-CES47-0019), the Merlion Project (grant no.~7.06.17), the ANR programme ``Investissements d'avenir" (ANR-15-IDEX-02), the Labex LANEF 
and the "Quantum Optical Technologies" project within the International Research Agendas programme of the
Foundation for Polish Science co-financed by the European Union's European Regional Development Fund.
The authors warmly thank O. Ezratty for his constant feedback and support, as well as O. Buisson, C. Gidney, O. Guia, B. Huard, T. Meunier,  V. Milchakov, L. Planat, M. Urdampilleta and P. Zimmermann  for useful discussions that greatly contributed to this work. HK Ng acknowledges the support of a Centre for Quantum Technologies (CQT) Fellowship. CQT is a Research Centre of Excellence funded by the Ministry of Education and the National Research Foundation of Singapore.


\clearpage
\appendix

\section{COUNTING QUBITS AND GATES FOR $k$ LEVELS OF CONCATENATED ERROR CORRECTION}
\label{Sect:Appendix-logical_to_physical}

\subsection{Fault-tolerant Clifford gates}
Figure \ref{Fig:steane_method} is the complete circuit diagram for any Clifford gate with one level of 7-qubit code for error correction, including the 
ancilla factory \cite{Steane1996Jul,Gottesman1997May,Steane1997Mar,Aliferis2006,Nielsen2011Jan}.
It shows that one level of error correction (i.e., one concatenation level)
replaces one qubit by 7 data qubits, and uses 28 ancilla qubits to detect errors \cite{Steane1997Mar,Aliferis2006}.
Gates acting on groups of 7 qubits are {\it transversal}, which means the relevant  gate is applied to each of the 7 qubits individually.  For example, a transversal cNOT between one group of 7 qubits and another group of 7 qubits involves a cNOT between qubit $i$ in the first group and qubit $i$ in the second group for all $i$ from 1 to 7 in parallel.

Figure \ref{Fig:steane_method} shows that the preparation and use of the ancillas take a significantly longer time than data qubit operations. The full evolution of the ancillas (including their preparation, verification, interaction with data qubits, and final measurement) takes 9 time-steps, while the data qubit operations take only 3 time-steps.
Hence, while 28 ancillas are being used in the current gate, an additional $2\times 28$ ancillas undergo the preparation and verification steps, to be ready in time for the next two gates. Thus, each additional level of concatenation in the error correction involves replacing one qubit by 7 data qubits and $3\times28$ ancillas, giving a total of 91 qubits. 

The logical ancillas must be verified to be sufficiently error-free for use, before they interact with the data qubits, and this is done by the verification part of the circuit in Fig.~\ref{Fig:steane_method}.
Each logical ancilla has a small chance of failing the verification, so the code must always prepare and verify a small percentage of extra logical ancillas at each clock-cycle (not shown in Fig.~\ref{Fig:steane_method}), to immediately replace any ancillas that fail verification.
We will show elsewhere \cite{our-article-in-prep} that this increases the resource consumption associated with ancillas by less than 2\%. 
This is small enough to neglect here, and so we simplify the
analysis by assuming that all ancillas pass verification.

The concatenated nature of the 7-qubit code scheme means that this counting is repeated $k$ times for $k$ levels of error correction. For $k$ levels of error correction, the number of physical qubits is 
\begin{eqnarray}
   Q = (91)^{k}\ Q_{\rm L} 
\end{eqnarray}
for $Q_{\rm L}$ logical qubits. This number is independent of the type of logical Clifford gates that are being implemented on the logical qubits.

A counting of gates in Fig.~\ref{Fig:steane_method}
gives the numbers of physical gates per logical gate in Table~\ref{table:FT_gate_list}.
Then, with $k$ concatenation levels, the number of physical gates done in parallel is related to the number of logical gates done in parallel, according to
\begin{subequations}
\label{Eq:N_physical_from_N_logical}
\begin{align}
\left(\!\begin{array}{c}
   N_{\rm 2qb} \\
   N_{\rm 1qb} \\
   N_{\rm Id}\\
   N_{\rm meas}
\end{array}\!\right) = A^k \, 
\left(\!\begin{array}{c}
   N_{\rm 2qb;L} \\
   N_{\rm 1qb;L} \\
   N_{\rm Id;L}\\
   N_{\rm meas;L}\\
\end{array}\!\right)  ,
\label{Eq:N_physical_from_N_logical-part1}
\end{align}
with 
\begin{align}
A=\frac{1}{3}\left(\begin{array}{cccc}
 135 & 64 & 64  & 0 \\
 56 & 35 & 28 & 0 \\
 58 & 29 & 36 & 0 \\
56 & 28 & 28 & 7
\end{array}\right),
\end{align}
\end{subequations}
where the elements of $A$ come from transposing Table.~\ref{table:FT_gate_list}.
The prefactor of $1/3$ in $A$ is because it takes three time-steps to perform each logical gate. Thus, the number of physical gates acting in parallel (averaged over those three time-steps) is $1/3$ the number of physical gates in a single level of concatenation. This $1/3$ appears at each level of concatenation, giving the prefactor in $A$.

\begin{table}[t]
\begin{tabular}{|r||c|c|c|c|}
	\cline{2-5}
 	\multicolumn{1}{c}{\ }& \multicolumn{4}{|c|}{\ N$^{\rm o}$ physical gates for given logical gate \ } \\
 	\cline{1-1}
 Logical   gate \ & \ 2qb gates \ & \ 1qb gates \ &\ Id gates \ & \ Meas \ \\
	\hline\hline
  Logical 2qb \ &   135 & 56 & 58 & 56 \\
  \hline
  Logical 1qb \ &  64 & 35 & 29 & 28 \\
  \hline
  Logical Id \ & 64 & 28 & 36 & 28\\
  \hline
  \ Logical Meas \ &  0 & 0 & 0 & 7\\
  \hline
\end{tabular}
\caption{
The number of physical gates in a given logical Clifford gate for one concatenation level of the 7-qubit code, as shown in Fig.~\ref{Fig:steane_method}.
Here ``1qb'' refers to any single-qubit gate in the set $\{ X,Y,Z,H,S \}$, and ``2qb'' refers to the two-qubit cNOT gate.
The numbers include the gates required to prepare and verify the logical ancillas. }
\label{table:FT_gate_list}
\end{table}

The matrix $A$ has two eigenvalues, $192/3=64$ and $7/3$. The larger one dominates Eq.~(\ref{Eq:N_physical_from_N_logical}).
For example, 
Eq.~(\ref{Eq:N_physical_from_N_logical}) gives the number of physical single-qubit gates as
\begin{eqnarray}
N_{\rm 1qb} \!&=&\! \frac{28\,(64)^k}{185}\Big(2N_{\rm 2qb;L}+N_{\rm 1qb;L}+ N_{\rm id;L}\Big)
\nonumber \\
& &\!\!- \frac{(7/3)^k}{185}\Big(56N_{\rm 2qb;L}-157N_{\rm 1qb;L}+28 N_{\rm id;L}\Big)\!.\qquad
\label{Eq:N1qb_from_N_logical}
\end{eqnarray}
This is well approximated by its first term, which comes solely from the eigenvalue 64.  The approximation is best for large $k$, but it works reasonably for order-of-magnitude calculations of $N_{\rm 1qb}$, $N_{\rm 2qb}$, $N_{\rm id}$, and $N_{\rm meas}$  for all non-zero $k$; the differences between approximate and exact results are less than 25\% for $k=1$, and less than 1\% for $k\geq2$.
This means that the  number of physical components in parallel is well approximated by a function of $(2N_{\rm 2qb;L}+N_{\rm 1qb;L}+ N_{\rm id;L})$, independent of the individual values of $N_{\rm 2qb;L}$, $N_{\rm 1qb;L}$ and $N_{\rm id;L}$. 
This is nice, because, for any time-step where no logical measurements are occurring, $(2N_{\rm 2qb;L}+N_{\rm 1qb;L}+ N_{\rm id;L})$ is equal to the number of logical qubits storing information at that time-step, irrespective of what (if any) gates are being performed on those logical qubits.

This is particularly simple in the context of quantum algorithms that can be approximated by {\it rectangular} circuits.  A rectangular circuit is one which uses the same number of qubits in each time-step, and does the logical measurements on these qubits at the end.  This is a reasonable approximation of the algorithm to break the RSA encryption in Ref.~\cite{Gidney2021Apr} (see Sec~\ref{sect:details-RSA} below).
In such a case, the number of logical qubits storing information in any given time-step is equal to the total number of logical qubits available in the computer, $Q_{\rm L}$. We can then write
\begin{eqnarray}
2N_{\rm 2qb;L}+N_{\rm 1qb;L}+ N_{\rm id;L} = Q_{\rm L}, 
\label{Eq:Logical-gate-sum=Q_L}
\end{eqnarray}
and the number of physical elements in parallel
in any time-step is 
\begin{eqnarray}
N_{\rm 2qb} \simeq \frac{64}{185}\left(64\right)^{\!k}Q_{\rm L},  &\hskip 4mm &
N_{\rm 1qb}\simeq \frac{28}{185}\left(64\right)^{\!k}Q_{\rm L}, 
\nonumber \\
N_{\rm id}\simeq \frac{29}{185}\left(64\right)^{\!k}Q_{\rm L},  &\hskip 4mm & \hskip -1.2mm
N_{\rm meas}\simeq \frac{28}{185}\left(64\right)^{\!k}Q_{\rm L}, \qquad
\label{Eq:approx_gate_numbers_from_QL}
\end{eqnarray}
where $\simeq$ indicates the approximations and assumptions made in the previous paragraphs. 

This gives us a useful intuitive rule for a quantum computer with error correction:  The number of physical gates in parallel after $k\geq 1$ levels of concatenations is about the same for any algorithm,  so long as the algorithm is using nearly all the logical qubits in the quantum computer in nearly all time-steps, with measurements of logical qubits remaining rare during the calculation \cite{physical-measurements}.
The number of each type of physical gates is then approximately given by Eq.~(\ref{Eq:approx_gate_numbers_from_QL}).

\subsection{Fault-tolerant $T$-gates}
\label{Sect:neglected_T_gates}

\red{The gate-set in Table~\ref{table:FT_gate_list} allows one to perform any Clifford operation. However, to perform an arbitrary quantum calculation (i.e., to perform an arbitrary unitary operation on the qubits), one also requires at least one non-Clifford gate in the gate-set. 
Without a non-Clifford gate in the gate-set, the calculations that one can perform on the quantum computer are limited to ones that can be efficiently simulated with classical computers.

We take the common choice to add a non-Clifford gate called the $T$-gate to the gate-set. The $T$-gate is a single-qubit rotation of $\pi/4$ around the $z$-axis of the Bloch sphere, i.e. $T \equiv \exp(-i \pi/8 \sigma_z)$. Then any non-Clifford operation (such as a Toffoli gate) is included in the circuit as a suitable combination of $T$-gates and Clifford gates. The $T$-gate's fault-tolerant implementation in the 7-qubit code scheme is done in a very different way from Clifford gates. It is done by making 
the logical qubit interact with ancillas in a so-called {\it magic state}.
These magic states have to be fault-tolerantly prepared beforehand
in the manner shown in Fig. 13 of Ref.~\cite{Aliferis2006}),
this circuit is more complicated than for Clifford gates. 
Each magic-state preparation also has a chance of failure, 
and must be verified before being inserted into the circuit. 
This means one must also prepare extra magic states to compensate for those that are discarded when they fail verification.
These considerations imply that a $T$-gate will be more costly than a Clifford gate.

However, $T$-gates are often rare in circuits of interest. The algorithm we consider here (see App.~\ref{sect:details-RSA}), and other algorithms in the literature \cite{Haner2020Apr,Gidney2021Apr,Zalka1998Jun,Fowler2012Sep,Scherer2017Jan}, have less than about one logical $T$-gate for every 70 logical Clifford gates (including identity gates is the count of  Clifford gates). 
It is reasonable to guess that a $T$-gate would not require 70 times more resources than a Clifford gate, so $T$-gates could be neglected when calculating power consumption, making our results in Sec.~\ref{Sect:fault-tolerant} are applicable to the type of algorithms that we consider here.  

However, to confirm that this guess is correct, a detailed calculation was necessary, to be presented in Ref.~\cite{our-article-in-prep}. There we compute the number of physical qubits and gates required by logical $T$-gates compared to the number required by logical Clifford gates. As the number of levels of concatenation $k$ increases,
the physical requirements for logical $T$-gates could grow differently than for Clifford gates, because the preparation of a logical magic state contains physical $T$-gates and physical Clifford gates, while logical Clifford gates only need physical Clifford gates. However,  Ref.~\cite{our-article-in-prep} will show that this is not the case. This can be seen by carefully considering the circuit in Fig. 13 of Ref.~\cite{Aliferis2006}, including all the Clifford gates necessary to prepare the logical states $|cat\rangle, |0\rangle$ and $|+\rangle$ that are required inputs into that circuit.
This shows that the magic-state preparation has a low number of physical $T$-gates compared to physical Clifford gates. A straight-forward calculation then shows that this means that the physical resources required by a $T$-gate has the same scaling with $k$ as we gave above for a Clifford gate, but with a different prefactor. The prefactor is about 5 times greater for a $T$-gate than for a Clifford gate for two reasons: (i) \mf{a logical $T$-gate implemented on a logical qubit needs the additional logical magic-state as an ancilla}, (ii) some extra magic-states are needed to replace those that fail verification. Thus at any value of $k$, a logical 
$T$-gate requires about 5 times more physical resources than a logical Clifford gate.  Thus for a circuit in which logical $T$-gates represent only $1/70$ of the total number of logical gates (including identity gates), it is a good approximation to neglect the $T$-gates when calculating power consumption.}

\begin{table*}[t]
\begin{tabular}{|l|c|l|}
\hline
\ {\bf Variables in Optimization} & \ {\bf Symbol} \ & \ {\bf Value}
\\
\hline \hline
\ Power consumption & $P_{\rm C}$ & \ To be minimized under constraint of 
given metric, ${\cal M}$. \ \  
\\
\hline
\ Qubit temperature & $T_{\rm qb}$ & \ To be found from minimization of $P_{\rm C}$. 
\\
\ Signal generation temperature & $T_{\rm gen}$ & \ To be found from minimization of $P_{\rm C}$. 
\\
\ Attenuation between  $T_{\rm gen}$ \& $T_{\rm qb}$ \ \ & $A$ & \ To be found from minimization of $P_{\rm C}$.
\\
\ Error correction's concatenation level \ \ & $k$ & \ Integer to be found from minimization of $P_{\rm C}$. 
\\
\hline
\end{tabular}
\vskip 3mm
\begin{tabular}{|l|c|l|}
\hline
\ {\bf Parameters} & \ {\bf Symbol} \ & \ {\bf Value}
\\
\hline 
\hline
\ Typical qubit frequency & $\omega_0/(2\pi)$ &  \ 6\,GHz (similar to Google Sycamore \cite{Google-Sycamore-datasheet,kjaergaard2020superconducting,krantz2019quantum})
\\
\hline
\ 1qb gate time & $\tau_{\rm 1qb}$ &  \ 25\,ns  (similar to  Google Sycamore \cite{Google-Sycamore-datasheet,kjaergaard2020superconducting,krantz2019quantum})
\\
\hline
\ 2qb gate time &  $\tau_{\rm 2qb}$ & \ 100\,ns (similar to cross-resonance scheme involving interaction  via 
\\
& & \ 
a bus \cite{chow2011simple,Sheldon2016Jun}, which allows implementation of the long-range 2qb gates 
\\
& & \ 
 necessary for concatenated error correction).
\\
\hline
\ Measurement time & $\tau_{\rm meas}$ &  \ 100\,ns (similar to scheme in Ref.~\cite{Jeffrey2014May})
\\
\hline
\ Timestep of quantum computer & $\tau_{\rm step}$ & \ 100\,ns (time of slowest gate)
\\
\hline
\ Type of error correction & -- & \ 7-qubit error correction code (concatenated error correction) \cite{Steane1996Jul,Gottesman1997May,Steane1997Mar,Aliferis2006,Nielsen2011Jan}
\\
\hline
\ Threshold for error correction & $\pthr$ & \ $2\times 10^{-5}$
\\
\hline
\ Error timescale  
&  $\gamma^{-1}$ &  \ from 3\,ms and 1\,s, see Fig.~\ref{Fig:P_C-map}a.
\\
\hskip 3mm $\equiv\,$ spontaneous emission rate 
& & 
\ \ \ The lowest $\gamma^{-1}$ here is {\it a few order of magnitude larger than Google}
\\
 \hskip 1.5cm into microwave line & & \  \ \  
 Sycamore \cite{kjaergaard2020superconducting,krantz2019quantum,Google-Sycamore-datasheet}. However,  smaller $\gamma^{-1}$ is not possible here, 
\\
\hskip 3mm $\sim\,$ decoherence time & & 
\ \ \ 
because it would put us above $p_{\rm thr}$, meaning the 7-qubit error correction  \  
\\
\hskip 1.5cm at $T_{\rm qb}=0$, $A\to \infty$& & 
\ \ \ 
code would fail completely.
\\
\hline
\ Errors per physical gate 
& $\perror$ &  \ Given in terms of $\gamma^{-1}$, $T_{\rm qb}$ and $A$ by Eq.~(\ref{Eq:eta_vs_T}).
\\
& &
\ At the smallest $\gamma^{-1}$ in Fig.~\ref{Fig:P_C-map} ($\gamma^{-1}=3$\,ms) this corresponds 
\\
& & \ to $\perror\simeq 0.4\pthr$ in the limit of $T\to0$ and $A\to \infty$. 
\\
\hline
\ Cryostat efficiency &  -- & \ Carnot efficiency at all temperatures
\\
& & \hskip 3mm (The state-of-art is 10\%-30\% of Carnot efficiency\cite{Parma2014}).   
\\
\hline \ 
N$^{\rm o}$ of refrigeration stages for  & $K$ & \ 5  
\\
\ control lines & & \ as sketched in Fig.~\ref{Fig:full-stack-stages}, with temperature \& attenuation in Eq.~(\ref{Eq:A_i_and_T_i}).
\\
\hline
\ Thermal conductivity & $\dot{q}_{\rm cond}$ & \ Coax  above $10\,$K \& superconducting microstrip below $10\,$K;
\\
\ of control lines  & & \ details in App.~\ref{Sect:conduction}
\\
\hline
\ Heat produced at $T_{\rm gen}$ by & $\dot{q}_{\rm gen}$ & 
\ Scenario A: 1\,mW per physical qubit; futuristic CMOS logic (see text)  
\\
\ signal generation \& readout & & \ Scenario B: 10\,$\mu$W per physical qubit 
\\
\ (demux-mux, DAC, amp \& ADC) & & \ Scenario C: 0.1\,$\mu$W per physical qubit; perhaps future SFQ logic (see text)
\\
\hline
\ Heat produced at $T_{\rm qb}$ & --- & \ smaller than other heat sources at $T_{\rm qb}$ (see text), so neglected. 
\\
\  by paramps  & & 
\\
\hline
\ Heat produced at 4\,K & $\dot{q}_{\rm para}$ & \ Scenario A:
$1\,\mu$W per physical qubit, value from Ref.~\cite{Malnou2021Oct}.    
\\
\ by paramps & & \ Scenario B: $10\,$nW per physical qubit. 
\\
\  & & \ Scenario C: $0.1\,$nW per physical qubit. 
\\
\hline
\ Heat produced at 70\,K stage  & $\dot{q}_{\rm hemt}$ & 
\ Scenario A:  50\,$\mu$W per physical qubit, value from Ref.~\cite{Malnou2021Oct}.
\\
\ by HEMT amps & & \ Scenario B: absent since $T_{\text{gen}}<70K$ (see text).
\\
\ & & \ Scenario C: absent since $T_{\text{gen}}<70K$ (see text).
\\
\hline
\ Heat conduction from $T_{\rm ext}$ to $T_{\rm gen}$ 
& -- & 
\ 
Absorbed into $\dot q_{\rm gen}$ (see text).
\\
\ (optical fiber, dc \& local osc.\,lines)\ & & 
\ 
heat generated at $T_{\rm gen}$, so it does not modify the values 
of $\dot q_{\rm gen}$.
\\
\hline
\ Joule heating in all lines
& -- &
\ The number and cross-section of lines in Fig.~\ref{Fig:full-stack-stages} is chosen to ensure Joule
\\
& & 
\  heating is less than other heat sources, and so can be neglected.
\\
\hline

\end{tabular}
\caption{Summary of variables and parameters used in our full-stack analysis of large-scale fault-tolerant quantum computing. These are the parameters used in Figs.~\ref{Fig:P_C-map}, \ref{Fig-RSA-quantum-versus-classical}, and \ref{Fig:gain_power_4K_fct_gamma}. Fig.~\ref{Fig:bad-refrigerator} also uses most of these variables and parameters, but it takes a lower cryogenic efficiency and adds a heat source at $T_{\rm qb}$ (see Sec.~\ref{App:varying_power_per_physical}). In all plots, we take the same scenario (A,B or C) for $\dot{q}_{\rm gen}$, $\dot{q}_{\rm para}$, and $\dot{q}_{\rm hemt}$; we do not mix scenarios. 
}
\label{table:parameters}
\end{table*}

\begin{figure}
         \includegraphics[width=0.95\columnwidth]{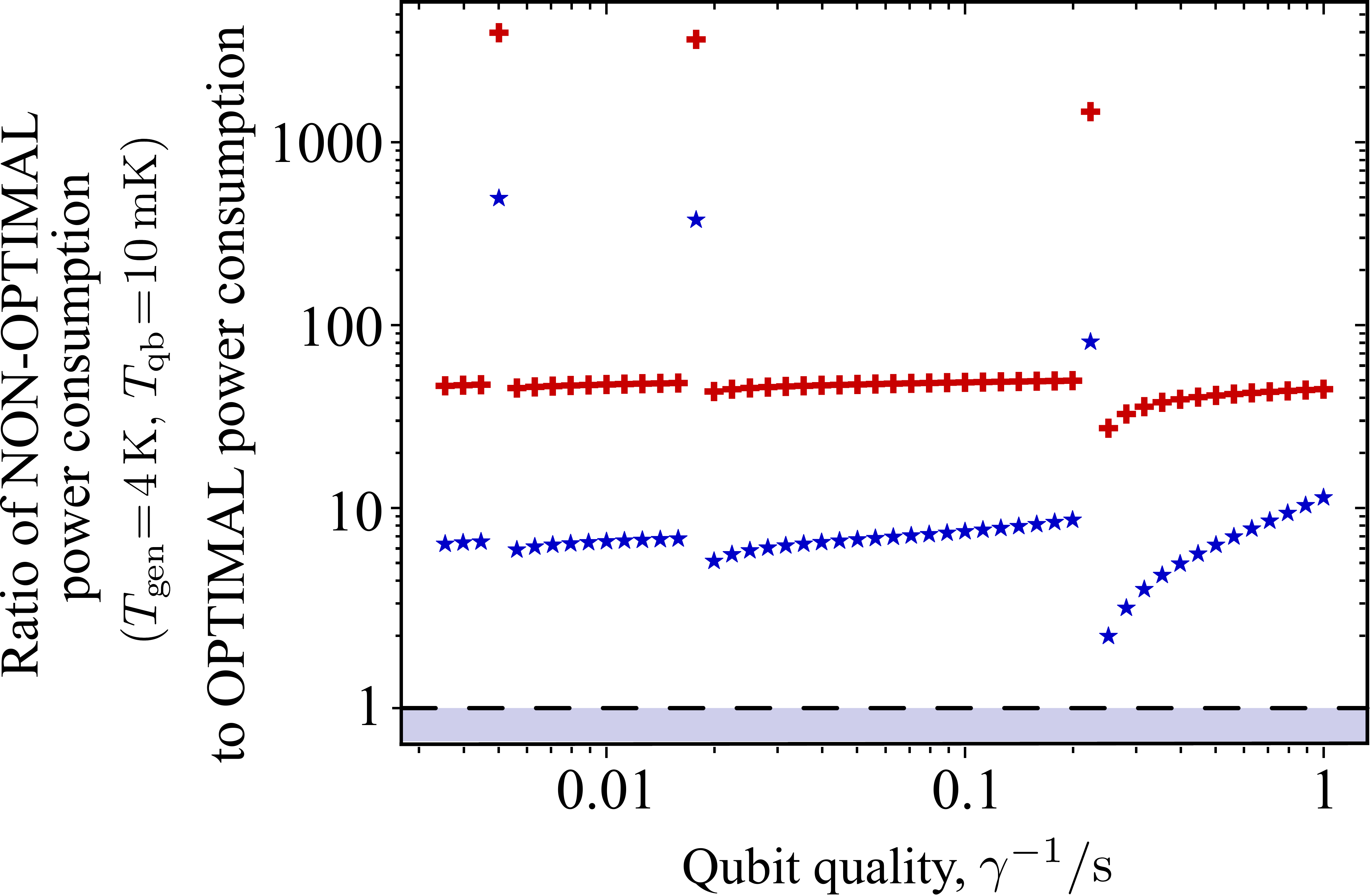}
         \caption{Power saved by optimization, compared to forcing the signal generation temperature $T_{\rm gen}= 4$\,K, and qubit temperature $T_{\rm qb}=10$\,mK, and only optimizing the attenuation, $A$, and the error correction's concatenation level, $k$.  All parameters are as in Fig.~\ref{Fig:P_C-map}. The different symbols are for different heat-generation at the signal generation stage; red crosses are for the current state of the art (Scenario A in table~\ref{table:parameters}) and  blue stars are $100$ times better (scenario B in table~\ref{table:parameters}). The discontinuities are when the fully optimized full-stack has a lower optimal $k$ than that given by the optimization in which we force $T_{\rm gen}= 4$\,K and $T_{\rm qb}=10$\,mK.
         }
         \label{Fig:gain_power_4K_fct_gamma}
\end{figure}

\section{PARAMETERS FOR THE FULL-STACK SUPERCONDUCTING QUANTUM COMPUTER}
\label{Sect:Appendix-hardware-parameters}


As stated in Sec.~\ref{Sect:optimizing-fault-tolerant},
we find the optimal value of four parameters:  the temperature of the signal generation $T_{\rm gen}$ (top stage in Fig.\ref{Fig:full-stack-stages}), the temperature of the qubits $T_{\rm qb}$, the total attenuation $A$ between $T_{\text{gen}}$ and $T_{\text{qb}}$, and the concatenation level $k$ for the error correction.
However, there are many more parameters in our model that are not optimized. For these, our philosophy is to take optimistic but realistic numbers of the current state of the art. When we are forced to make simplifications, 
we aim for a simplification that gets that contribution to power consumption within an order of magnitude of the correct result.
There is, however, one critical parameter for which we are vastly more optimistic than the state of the art:
We assume that the qubits and gates are at least {\it a few orders of magnitude better} than those in Google's current Sycamore chips.  The reason for this vastly optimistic assumption is that concatenated error correction scheme based on the 7-qubit code fails unless the 
error is below the threshold $\pthr= 2 \times 10^{-5}$.
While Google's current Sycamore chips typically have 2-qubit gate error probability of about 0.01, other recent works \cite{Wang2022Jan} suggest that
error probability per gate of $2\times 10^{-4}$ might soon be achievable (taking their $T_2$ with a gate time of 100\,ns).
Thus, we hope that error probabilities significantly below the threshold $\pthr$ should be achieved within a few years. 

We take the gate times for physical gates from Google's Sycamore chip \cite{Google-Sycamore-datasheet}, except that we take a longer 2-qubit gate time of $\tau_{\rm 2qb}=100\,$ns,
because we have in mind the long-range 2-qubit gates necessary for the 7-qubit code scheme. This can be done, for example, by making the qubits interact with each other through a bus via a cross-resonance technique \cite{chow2011simple,Sheldon2016Jun}. This makes the 2-qubit gate slower than in the Sycamore chip, and makes this the longest gate time in our modeling.
The computer's time-step (its clock cycle) is fixed by this gate time, and hence $\tau_{\rm step}\sim 100\,$ns in our model. 

The principal components of the full-stack model of a fault-tolerant quantum computer are sketched in Fig.~\ref{Fig:full-stack-stages}. They are explained in detail in the following subsections, but can be briefly summarized as follows:
\begin{itemize}
    \item A stage for the qubits at temperature $T_{\text{qb}}$, which also houses attenuation to remove thermal noise on the drive signals, and superconducting parametric amplifiers to boost the read-out signal.
    \item A stage containing superconducting parametric amplifiers \cite{Malnou2021Oct} at $T_{\text{Amp}}=4\,K$, to boost the read-out signal to a level above the noise at 70\,K.
    \item A stage containing amplifiers made from High Electron Mobility Transistors (HEMT) \cite{Malnou2021Oct} at $T_{\text{Amp}}=$70\,K, to boost the read-out signal to a level above the noise at $T_{\rm gen}$.
     \item A number of stages between $T_{\text{qb}}$ and $T_{\text{gen}}$, with attenuators on each stage to attenuate the thermal noise in the qubit driving signal. Their role is to evacuate heat at intermediate temperatures, to reduce the amount of heat to be evacuated at the lowest temperature stage. 
    \item A stage at temperature $T_{\text{gen}}$ which we call the signal generation stage. It contains the classical electronics consisting of demultiplex-multiplex chips (demux-mux), digital-to-analogue converters (DACs), and analogue-to-digital converters (ADCs).  The demultiplex part of the chip takes as inputs the list of gates to be performed at a given instant from the optical fibers (information that has been multiplexed in the room temperature computer). This is then sent to the appropriate DAC that generate the appropriate waveform from one it has stored in memory. The waveforms are then multiplied by the local oscillator and sent toward the qubits. The ADCs receive measurements on the qubits, coming through the chain of amplifiers (including one at $T_{\rm gen}$). The multiplex part of the chip takes the digital version of the read-out signal from the ADC, multiplex the data and sends it back up the optical fiber, for it to be analyzed by the classical computer at room temperature.
    \item Electronics at room temperature, including the generation of the local oscillator, and a classical computer. The role of the latter is to manage the algorithm on the logical level, to decode the syndromes from error correction and to digitally demodulate the readout signals. In App.~\ref{Sect:Appendix-classical-computer}, we argue that all the power consumption at room temperature can be neglected in comparison to the power dissipated by the stage at $T_{\text{gen}}$, at least for our Scenario A inspired by a futuristic view of CMOS electronics. To help understand the parameter dependences, we also neglect it in our scenarios B and C, but it can easily included using the information in App.~\ref{Sect:Appendix-classical-computer}.
\end{itemize}
At each stage, the cryogenics must evacuate the heat generated at that stage, and the heat conducted down cables from higher temperatures. We now explain the details of our full-stack model, and our motivations for neglecting the power consumption for some of the components.

\subsection{Attenuation on microwave lines}
\label{sect:Attenuation-multi-stage}
As is standard, we assume that the thermal photon contribution to the noise is reduced to an acceptable level by a chain of attenuators on the incoming microwave line (see Fig.~\ref{Fig:full-stack-stages}). These attenuators are kept cold by the cryogenics, so they thermalise the signal coming down the line from hotter temperatures, reducing the population of thermal photons.  
For a chain of $K$ cooling stages, with $K-1$ attenuators (Fig.~\ref{Fig:full-stack-stages} shows a case with $K=5$), 
the error probability of a physical qubit is 
\begin{eqnarray}
\perror = \frac{\gamma \tau_\text{step}}{2}\left( \frac{1}{2} +
n(T_1) 
+ \sum_{i=1}^{K-1} \frac{ n(T_{i+1}) - n(T_{i})}{\widetilde{A}_i}\right),
\label{Eq:eta_vs_T}
\end{eqnarray}
where $T_1=T_{\rm qb}$, 
and $n(T)= (\exp[\hbar\omega/(k_{\rm B} T)]-1)^{-1}$ is the Bose-Einstein function at the qubit frequency.
Here $A_i$ is the attenuation on stage $i$ at temperature $T_i$
(as sketched in Fig.~\ref{Fig:full-stack-stages}), and we define $\widetilde{A}_i$ to be the total attenuation between $T_i$ and the qubits, so that $\widetilde{A}_i = A_i \cdots A_2A_1$. The sum in Eq.~(\ref{Eq:eta_vs_T}) is due to thermal photons that leak through the attenuators from higher temperatures.

We see that the noise can always be reduced by increasing the attenuation, but this comes at the cost of greater power consumption. This makes the attenuation a crucial parameter in our optimization, as explained in App.~\ref{Sect:attenuaton-per-stage}.

\begin{figure}
\includegraphics[width=0.95\columnwidth]{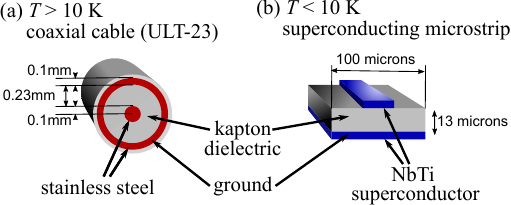}
\caption{
The types of control and readout lines considered in our modelling.  We need explicit models of these to calculate the heat that they will conduct between temperature stages
in the full-stack quantum computer. The coaxial cables are called ULT-23 with radius of order a millimetre. 
The microstrip lines are those discussed in Ref.~\cite{mcdermott2018quantum} and are much smaller than the coaxials cables (largest dimension is 0.1mm).  
}
\label{Fig-thermal-conduction}
\end{figure}

\subsection{Heat conducted by microwave lines}
\label{Sect:conduction}

We consider that the microwave signals are carried by coaxial cables above 10\,K and superconducting microstrip line below 10\,K, as sketched in Fig.~\ref{Fig-thermal-conduction}.
Unfortunately, these cables also conduct heat from higher to lower temperatures, and this heat must be evacuated by the cryogenics.  Thus, we must quantify the heat flow $\dot{q}_{\rm cond}$ from a higher temperature $T_2$ to a lower temperature $T_1$. For a cable of length $L$, made of both metal and Kapton (polyimide) dielectric, Fourier's law gives
\begin{eqnarray}
\dot{q}_{\rm cond} 
= \frac{1}{L} \int_{T_1}^{T_2} {\rm d} T \, \big[{\cal A}_{\rm kd}(T) \,\lambda_{\rm kd}(T) + {\cal A}_{\rm m}\,\lambda_{\rm m}(T) \big],
\label{Eq:q-cond-general}
\end{eqnarray}
where the Kapton dielectric has cross-sectional area  ${\cal A}_{\rm kd}(T)$ and thermal conductivity $\lambda_{\rm kd}(T)$ at temperature $T$,
while the metal has cross-sectional area 
${\cal A}_{\rm m}(T)$ and thermal conductivity $\lambda_{\rm m}(T)$ at temperature $T$.
Note that ${\cal A}_{\rm m}(T)$, is the total cross-section of stainless steel in the coaxial cable including the ground, or the total cross-section of the NbTi in the microstrip line including the ground. 

As the heat conduction is inversely proportional to the length $L$ of the cables between stages in the cryostat, it is good to have long microwave cables (coiled if necessary);
we take $L=$1\,m.
In the coaxial cable, the heat conduction is dominated by the stainless steel, so we can neglect the Kapton dielectric. In contrast, the heat conduction in the microstrip lines is dominated by the Kapton dielectric (because the NbTi is superconducting and so its thermal conductivity is similar to that of Kapton, but its cross-section is smaller).
Then, Eq.~(\ref{Eq:q-cond-general}) is reasonably well approximated by
\begin{eqnarray}
\dot{q}_{\rm cond} 
= \frac{1}{L} \int_{T_1}^{T_2} {\rm d} T \, {\cal A}(T)\, \lambda(T),
\label{Eq:q-cond}
\end{eqnarray}
where the relevant cross-sections are 
\begin{eqnarray}
{\cal A}(T) =\left\{ 
\begin{array}{ll}
 2.7 \times 10^{-7} {\rm m}^2 & \mbox{ for } T>10\,{\rm K}  
 \\
1.3\times 10^{-9} {\rm m}^2 & \mbox{ for } T< 10\,{\rm K}
\\
\end{array}\right.
\end{eqnarray}
The conductivity above $10\,$K is that of stainless steel, given by a fitting to experimental data in Ref.~\cite{Marquardt2002}, as
\begin{eqnarray}
\lambda (T > 10\,{\rm K}) &=& 10^{Z(T)}
\nonumber \\
& & \hskip -4mm \mbox{with } Z(T) \equiv \sum_{\alpha=0}^8 a_\alpha \left(\log_{10} T\right)^\alpha, 
\end{eqnarray}
where the fitting parameters are
$a_0=-1.4087$, 
$a_1=1.3982$, 
$a_2=0.2543$, 
$a_3=-0.6260$, 
$a_4=0.2334$, 
$a_5=0.4256$, 
$a_6=-0.4658$, 
$a_7=0.1650$, 
and $a_8=-0.0199$.
The conductivity below $10\,$K is that of the Kapton (polyimide) dielectric, which we model phenomenologically as 
\begin{eqnarray}
\lambda (T < 10\,{\rm K}) = \left\{ 
\begin{array}{ll}
4.6\, T^{\,0.56} & \mbox{ for } T< 4\,{\rm K}\\
3.0\, T^{\,0.98} & \mbox{ for } 4\,{\rm K}< T <10 \,{\rm K}
\end{array}\right.
\label{Eq:lambda_Kapton}
\end{eqnarray}
The form for $T<4$\,K is taken from experimental measurements on Kapton in the range $0.5\,{\rm K} < T <5\,{\rm K}$ \cite{Lawrence2000Mar}, which we assume can be extrapolated down to 0\,K, although we rarely need to extrapolate it below about 0.05\,K.
The form for $T>4$\,K is a simplification of the 
form extracted from experiments in the range $4-300$\,K
\cite{Marquardt2002}. This form slightly over estimates the conductivity in the range  $4-10$\,K compared to \cite{Marquardt2002}, and leads to a small discontinuity at 4\,K, which would be absent in a more accurate model.  However, a more realistic $T$-dependence would not change $\dot{q}_{\rm cond}$ by more than a few percent in the range $0-10$\,K, and we observe that $\dot{q}_{\rm cond}$ in this range makes only a small contribution to the total power consumption. Hence, for our purposes, Eqs.~(\ref{Eq:q-cond}-\ref{Eq:lambda_Kapton}) give a sufficiently accurate model of the heat conduction in the microwave lines.

\subsection{Control electronics for signal generation and read-out}
\label{Sect:sig_gen_amps}
The electronics on the signal generation stage (at $T_{\rm gen}$) has two jobs. Its first job is to generate the microwave signals driving the qubits. This is done by first demultiplexing (demux) the data sent down the optical fiber (which contains a few bits of information describing which gate has to be performed). Such information is then sent to the appropriate DAC which (i) reads the description of the waveform to generate from memory, (ii) generate the waveform, (iii) multiplies it with the local oscillator. The microwave signal is then sent toward the qubits through the coaxial cables. 

The second job of the electronics at this stage is to take the output of the measurements, digitize them (ADC), and  multiplex (mux) them to be sent up the optical fiber to the
traditional computer at room temperature.

We consider 3 scenarios to do the jobs of these control electronics (demux-muxs, DACs and ADCs), with increasing levels of energy efficiency.
Scenarios A and C are inspired by futuristic views of technologies that could do the job, while Scenario B is simply
a point half way between A and C, to better understand the parameter dependences.

Scenario A (the least energy efficient, red curves in Fig.~\ref{Fig:P_C-map}a) is a parameter regime that is a futuristic view for CMOS, for which we assume the power consumption $ \dot{q}_{\rm gen}=1\,$mW per qubit (assuming the quantum computer's time-step, $\tau_{\rm step}=100\,$ns).
This is a bit more than an order of magnitude better than current state-of-the-art CMOS \cite{Park2021Feb,frank2022cryo,Kang2022Aug}, but that state of the art is rapidly improving at present.  
We use Eq.~(\ref{Eq:bitrate-per-qubit}) for the heat dissipated to multiplex and demultiplex (Mux-Demux chip). Existing multiplexing consumes 0.8\,pJ/bit \cite{wade2018bandwidth}, giving a heat dissipation $\lesssim 0.5$\,mW. Thus, we take the optimistic view that the control electronics at $T_{\rm gen}$ will generate a total heat of $\dot{q}_{\rm gen}=$1\,mW per physical qubit.

Scenario B (intermediate energy efficiency, blue curves in Figs.~\ref{Fig:P_C-map}a) is taken to be halfway between scenarios A and C, so it is hundred times more energy efficient than scenario A with $\dot{q}_{\rm gen}= 10\,\mu$W. 

Scenario C (the most energy efficient, green curves in Figs.~\ref{Fig:P_C-map}a) is in a parameter regime inspired by single-flux quantum (SFQ) logic. This is a type of classical logic that is under development, based on superconducting circuits. Initial estimates suggest that it could be $10^4$ times more energy efficient than CMOS \cite{mcdermott2018quantum}, so we take  $\dot{q}_{\rm gen}= 100\,$nW. 

At various points in the following sections we will compare certain heating mechanisms at or near $T_{\rm gen}$ to the heat generated by the control-electronics in scenario A ($\dot{q}_{\rm gen}= 1\,$mW), as a way of justifying neglecting those heating mechanism.  We chose to also neglect those heat-mechanism when plotting results for scenario B and C (even when this may not be justified in a technology corresponding to these scenarios) to help us understand the parameter dependence of the mechanisms that we do {\it not} neglect. In practice, this means that one should not expect our results for scenario C to apply directly to SFQ logic.  We prefer to say that scenario C is an indication of the potential interest of SFQ logic,
while being modest about our lack of concrete knowledge of what an SFQ implementation would look like.  For example, would its implementation involve an optical fiber from SFQ logic to a classical computer at room-temperature, or would all the classical computing be done with SFQ logic at low-temperature? It is too early to tell, and this would need to be clarified before going beyond the very naive estimates made with our scenario C.

\subsection{Amplification stages}
\label{sect:amp-stages}
The qubit measurements require amplifiers. We first assume we can measure about 100 physical qubits with a single readout line. We get the factor of 100 as follows.
The readout time of 100\,ns means the readout on each physical qubit needs a bandwidth of about 0.01\,GHz. If we assume the qubit frequencies are spread over about 1\,GHz, then we can have about 100 qubits operating at different frequencies on a single readout line.
We assume the signal in each readout line is amplified by three amplifiers, one at the qubit temperture, one at 4\,K and one at 70\,K.  Here we take experimental numbers for the 4\,K and 70\,K amplifiers from Ref.~\cite{Malnou2021Oct}, while the parametric amplifiers (paramps) at $T_{\rm qb}$ is treated separately in Sec.~\ref{Sect:power_per_measurement} below \cite{amp-at-Tgen}.
They propose HEMT amplifiers at $T_{\rm hemt} =70$\,K, which cannot be turned off between measurements, so they are continually generating 5\,mW of heat generation per amplifier. With one amplifier for 100 physical qubits, this gives us $\dot{q}_{\rm hemt}= 50 \,\mu$W.
The job of this amplifier is to amplify the signal well above the noise at 300\,K.  However if the signal readout temperature $T_{\rm gen}$ is below 70\,K, it is then clear that this amplifier is unnecessary, and we can set $\dot{q}_{\rm hemt}$ to zero.  It turns out that our optimization places $T_{\rm gen}$ above 70\,K in our scenario A (so we keep $\dot{q}_{\rm hemt}=50\,\mu$W), but it places $T_{\rm gen}$ below 70\,K
in our scenarios B and C, for which $\dot{q}_{\rm hemt}=0$.

Ref.~\cite{Malnou2021Oct} proposes using superconducting paramps at $T_{\rm para} =4$\,K, which are powered by microwave pump signals sent from room temperature to 4\,K through a 20\,dB attenuator (to reduce the noise on the line). 
We estimate that these will need a driving power of order $10^{-6}$\,W to amplify the readout of 100 physical qubits. Thus, the 20\,dB attenuator will dissipate heat of order $10^{-4}$\, W for the 100 physical qubits.
This gives $\dot{q}_{\rm para}= 1 \,\mu$W.
For simplicity, we take the worst-case scenario, where the paramp microwave pump signal is always on, and we take all the attenuation on this microwave driving to be at $T_{\rm para}=4$\,K, so that $\dot{q}_{\rm para}$ is entirely dissipated at 4\,K. Clearly this can be improved significantly, by turning off the pump signal when no qubits are being measured, and by having the 20\,dB of attenuation being generated by a chain of attenuators at different temperatures. In practice, we take $\dot{q}_{\rm para}= 1 \,\mu$W for our scenario A. In the case of scenario C, the readout electronics are already at a temperature close to $4\,$K given the results of our optimisations, so it is likely that this amplifier at 4\,K will not be necessary. In our model, it is modelled by considering $\dot{q}_{\rm para}= 0.1 \,$nW (a negligible contribution compared to the heat dissipated by control electronics at $T_\text{gen}$ in scenario C). Finally, scenario B,
which is chosen to be a situation halfway between A and C, has $\dot{q}_{\rm para}= 10 \,$nW.

\subsection{Power consumption at room temperature}
\label{Sect:Appendix-classical-computer}

At room temperature we have different sources of power consumption. First, there is the generation of the local oscillator. We believe it reasonable to neglect it compared to the power consumed by the DACs and ADCs for our scenario A, given the information available in the literature \cite{LeGuevel2020Feb}. Then, there is electronics multiplexing and demultiplexing data coming from the optical fibers. We will show that this power consumption can be neglected in Sec.~\ref{Sect:sig_gen_amps}. Finally, there is a classical computer which has three main purposes. It has (i) to digitally demodulate the readout signals coming from the ADC (in order to interpret the state of the measured qubits from the readout signals), (ii) to decode the syndromes from error correction, and (iii) to manage the algorithm on the logical level. We argue here that the power consumption for all these can be neglected compared to the power consumption we take for the signal generation stage.

For (i), the digital demodulation, we consider $N_d=100$ discrete points in time to provide a good accuracy for the digitization of the readout signals \cite{Salathe2017Sep}. Two quadratures of the readout signal must be found to get the state of the qubit from the phase of this signal. This is done by multiplying the digitized signal once by $\cos$ and once by $\sin$ (for the two quadratures) \cite{krantz2019quantum}, requiring $\sim 2 N_d$ operations. Then, we multiply this quantity by the number of measurements per unit time,  $N_{\text{meas}}/\tau_{\text{step}}$.
This is further multipled by the energy required to perform each operation. We over-estimate this by taking the energy cost of a floating-point operation to be $q_{\text{Float}} \approx 0.85$ pJ \cite{Tung2020May}.  
Then, the power consumption \textit{per physical qubit} required for the demodulation is given by
\begin{align}
    \dot{q}_{\text{demodulation}}=2 N_d\  \frac{N_{\text{meas}}(k)}{Q_P(k)\tau_{\text{step}}} \ q_{\text{Float}}.
    \label{eq:qdot_demodulation}
\end{align}
This decreases with $k$ and is around $200 \mu$W for $k = 1$ (and hence negligible compared to the power required for DACs and ADCs in scenario A). To be more precise, in a heterodyne demodulation scheme, the readout signals are multiplied by a local oscillator before being filtered and digitized \cite{krantz2019quantum}, with the cost of digitization given by Eq.~\eqref{eq:qdot_demodulation}. 
We assume that this multiplication of signals with a local oscillator is being done by the ADCs at $T_{\rm gen}$ and that the energy needed for it is accounted for in the $\dot{q}_{\rm gen}=1$\,mW of scenario A.

For (ii), decoding the syndrome, we need the number of operations per unit time required to infer which error occurred. Multiplying that number by the energy cost of one operation will give the required power.
The recursive nature of concatenated code makes an exact calculation of the number of operations unnecessarily complicated for our goal here.
Instead, we assume that one operation has to be performed per physical qubit per time-step.  It gives a pessimistic cost per physical qubit to decode the syndrome:
\begin{align}
    \dot{q}_{\text{Syndrome}}\ =\ \frac{1}{\tau_{\text{step}}} \,q_{\text{Float}} \ \approx \ 8\, \mu{\rm W}.
\end{align}
This cost is several orders of magnitude lower than the power consumption of the electronics in our scenario A (1\,mW per physical qubit), so we neglect it in our calculations.
A more exact calculation of the number of operations necessary to decode a syndrome is sufficiently complicated that we have not rigorously shown this to be an upper-bound on $\dot{q}_{\text{Syndrome}}$, but we believe it is not far from such a bound.

For (iii), we considered it reasonable to neglect the total cost because the logical algorithm and its decomposition into a physical circuit comprising gates from the chosen gate-set can be pre-computed and stored in memory. Because memory storage is usually cheap in electronics, we consider it reasonable to neglect the associated cost. The only processing necessary is to add the gates required to correct any errors to this pre-computed circuit. As errors are rare (less than one every $10^5$ time-steps for each qubit) adding such error-correction gates will be equally rare,

and  this will require a tiny fraction of the processing required to decode the syndrome, so it can be safely neglected. 

\subsection{Power consumption per physical gate}
\label{Sect:attenuaton-per-stage}
The power consumption per physical gate is averaged over the time-step of the computer, $\tau_\text{step}$, assuming that the 2qb gates take one time-step, but the
1qb gates are faster (taking a time $\tau_\text{1qb} <\tau_\text{step}$). This gives
\begin{align}
P_\text{2qb}&= P_\pi \sum_{i=1}^{K}\frac{T_\text{ext}-T_i}{T_i} (\widetilde{A}_i-\widetilde{A}_{i-1}),
\label{Eq:P2qb}
\\
P_\text{1qb}&=\frac{\tau_{\text{1qb}}}{\tau_{\text{step}}} \ P_\text{2qb},
\label{Eq:P1qb}
\end{align}
where $\widetilde{A}_i$ is given below Eq.~(\ref{Eq:eta_vs_T}). Note that, as all signals arriving at stage 1 are eventually dissipated as heat in that stage, one must take $\widetilde{A}_0=0$ in the sum. The power supplied for 1qb and 2qb gates during the gate is the same, but 1qb gates are faster, so the power averaged over the time-step is smaller for 1qb gates than 2qb gates in Eqs.~(\ref{Eq:P2qb}) and (\ref{Eq:P1qb}). In our examples we take $\tau_\text{1qb}=\frac{1}{4}\tau_\text{step}$.

To keep the optimization tractable, we do not optimize the temperatures and attenuation at each refrigeration stage.  Instead we take the common rule of thumb, that stages should have equal attenuation and be regularly spaced in orders of magnitude of temperature between $T_{\text{gen}}$ and $T_{\text{qb}}$. In other words if we want a total attenuation of $A$, we take
\begin{eqnarray}
A_i = A^{1/(K-1)}, \qquad  T_i=T_{\rm qb} \left( \frac{T_{\rm gen}}{T_{\rm qb}}\right)^{(i-1)/(K-1)}
\label{Eq:A_i_and_T_i}
\end{eqnarray}
for $K$ stages of cooling. While this is not optimized, we suspect that the power consumption is not far from the optimal, because we observed in specific cases that increasing $K$ from 4 to 5 did not greatly reduce the power consumption. All plots in this work are for $K=5$, since this is typical of current cryostats.

\subsection{Power consumption per physical measurement}
\label{Sect:power_per_measurement}
To estimate the power consumption per measurement, $P_\text{meas}$, we note that it originates from the heat dissipated when performing the amplification of the signals coming from the qubits. The first amplification occurs using the superconducting parametric amplifiers at temperature $T_{\rm qb}$ in Fig.~\ref{Fig:full-stack-stages}. Being superconducting, these dissipate negligible heat, but they require microwave driving signals to be sent down through the attenuators. The microwave driving of the parametric amplifier must be about $100$ times its output signal; 
i.e., 100 $\times$ input signal $\times$ the amount of amplification.
For the measurement of a single-qubit state, the input signal is one photon during the measurement time, $\tau_{\rm meas}$, so the power being measured 
is $\hbar \omega_0/\tau_{\rm meas}\sim 10^{-17}$. This needs to be amplified by a factor of 100,
so the microwave driving signal for the parametric amplifier is about $10^{-13}\,$W per physical measurement. This is at least 40 times smaller than the microwave driving necessary for a single-qubit gate for the values of $\gamma$ that we consider, given by $P_\pi$ in Eq.~(\ref{P_pi}).
Thus the heat generated in the attenuators due to the driving of the parametric amplifiers can be neglected, compared to the heat generated in the attenuators due to the driving of the single-qubit and two-qubit gates. 

The other stages of amplification (at 4\,K, 70\,K and $T_{\rm gen}$) are
assuming to be always on and so dissipate heat constantly; see Sec.~ \ref{sect:amp-stages}. This means they  contribute to $P_\text{Q}$ (see Sec.~\ref{Sect:PQ}) rather than to $P_{\rm meas}$.  Thus, $P_{\rm meas}$ is at least 40 times smaller than $P_{\rm 1qb}$ and $P_{\rm 1qb}$, and we neglect it.

\subsection{Power consumption per physical qubit}
\label{Sect:PQ}

The heating proportional to the number of qubits (independent of the number of gates or measurements being performed) comes from thermal conduction in cables and the heat generated by electronics that cannot be switched off (amplifiers and signal generation and readout).
We define $\dot{q}_{\rm cond}(T_i,T_{i+1})$ as the heat conduction per physical qubit due to the cables between cryogenic stages at $T_{i+1}$ and $T_{i}$. It is given in Sec.~\ref{Sect:conduction} above.
We define $\dot q_{\rm gen}$ as the power consumed (and turned into heat) per physical qubit by the control electronics at $T_{\rm gen}$.
We define $\dot q_{\rm hemt}$ as the power consumed (and turned into heat) per physical qubit by the conventional HEMT amplifiers at $T_{\rm hemt}=70$\,K, and define $\dot q_{\rm para}$ as the power consumed (and turned into heat) per physical qubit by the superconducting parametric amplifiers at $T_{\rm para}=4$\,K. Values for $\dot q_{\rm gen}$,  $\dot q_{\rm hemt}$, and  $\dot q_{\rm para}$ are given in Apps.~\ref{Sect:sig_gen_amps} and \ref{sect:amp-stages}.
For $K$ stages of cryogenics between signal generation and qubits 
(as sketched in Fig.~\ref{Fig:full-stack-stages} for $K=5$), 
the power consumption per qubit is 
\begin{align}
P_\text{Q}&= 
\frac{T_\text{ext}}{T_{\rm gen}} \dot{q}_{\text{gen}}
+ \frac{T_\text{ext}}{T_{\text{hemt}}}  \dot{q}_{\text{hemt}} 
+\frac{T_\text{ext}}{T_{\text{para}}}  \dot{q}_{\text{para}} 
\notag \\
&+\sum_{i=1}^{K}\frac{T_\text{ext}-T_i}{T_i} \Big(\dot{q}_{\text{cond}}(T_{i},T_{i+1})-\dot{q}_{\text{cond}}(T_{i-1},T_i)\Big),\notag \\
\label{Eq:PQ}
\end{align}
where $T_1\equiv T_{\rm qb}$ and $T_K\equiv T_{\rm gen}$.
We take $K=5$ in \eqref{Eq:PQ} for the reason explained in App.~\ref{Sect:attenuaton-per-stage}.
Note that terms like $(T_\text{hemt}/T_{\rm gen}) \dot{q}_{\text{hemt}}$ are the sum of the power supplied to the HEMT amplifier which is dissipated as heat, $\dot{q}_{\text{hemt}}$, and the cryogenic power cost to remove that heat 
with Carnot efficiency, $((T_\text{hemt}-T_{\rm gen})/T_{\rm gen}) \dot{q}_{\text{hemt}}$.
Note also that, as there is no stage below stage 1, one must take $\dot{q}_{\text{cond}}(T_0,T_1)=0$ in the sum. 

We neglected the heat conduction between the laboratory and the stage at $T_K=T_{\text{gen}}$. This is because relatively few cables are required for the local oscillator and dc cables. In principle, the wires from the local oscillator and the dc sources need filtering to ensure thermal noise at $T_{\text{ext}}$ does not perturb the signal generation at $T_{\text{gen}}$. As each such wire carries a single specific frequency (rather than complex waveforms), we assume the filtering can be done by reflection rather than absorption, so we neglect heating due to such filtering. Hence, the main source of heat conduction at $T_{\text{gen}}$ is dominated by the optical fibers that allow the exchange of data with the laboratory. 
To show that such heat conduction can be neglected, we need the ratio of optical fibers to physical qubits. This is given by the ratio of an optical fiber's bit rate, $N_{\text{bitrate}}=400$ Gb/s, to the information exchanged per unit time (bit-rate) for a single physical qubit.  Most of the information exchanged between $T_{\text{ext}}$ and $T_{\text{gen}}$ is the digitized version of the readout signals going up the fiber from ADCs to the room-temperature computer. We take \cite{Salathe2017Sep} each readout to have its amplitude digitized with $N_{\text{encoded}}=14$ bits at $N_{\text{samples}}=100$ discrete points in time during a time-step, $\tau_{\text{step}}=100$ ns. 
There are $N_{\text{meas}}(k)/\tau_{\text{step}}$ measurements per unit time, so we arrive at  the following bit-rate per physical qubit,
\begin{align}
    \frac{N_{\text{encoded}} N_{\text{samples}}}{\tau_{\text{step}}} \frac{N_{\text{meas}}(k)}{Q(k)} \ \lesssim\ 1.5\,\text{Gb/s}.  
    \label{Eq:bitrate-per-qubit}
\end{align}
 Eqs.~(\ref{Eq:physical_vs_logical-qubits}) and (\ref{Eq:approx_gate_numbers_from_QL}) show that the ratio of measurements to physical qubits $N_{\text{meas}}(k)/Q(k)$ is maximal for $k=1$. Thus, we took $k=1$ to get the above over-estimate of the bit-rate. This gives one optical fiber for every $\sim 270$ physical qubits. 

The heat-conducted from $T_{\rm ext}$ to $T_{\rm gen}$ per physical qubit is thus $1/270$ times the heat conducted of an optical fiber, which is much less than 1\,mW, per physical qubit, giving a heat conducted down
to $T_{\rm gen}$ of much less than 4\,$\mu$W per physical qubit.
This is a significant over-estimate, because 1\,mW is the approximate heat conducted between 300\,K and 0\,K by a coaxial cable in App.~\ref{Sect:conduction}, 
when the temperature drop here is less than 300\,K to 0\,K, and 
optical fibers carry much less heat than coaxial cables (they are insulators rather than metallic). 
We thus neglect the heat conduction from $T_{\rm ext}$ to $T_{\rm gen}$, because it is negligible compared to the 1\,mW of heat dissipated by control electronics at $T_{\rm gen}$ in our scenario A.

\subsection{Number of qubits per microwave line}
Minimizing the number of cables going to the qubit stage minimizes the conduction of heat to the qubits from higher temperatures.   
This can be done by driving multiple qubits with a single microwave line, and reading out multiple qubits with a single readout line.  This is multiplexing at the level of qubits, but it should not be confused with the multiplexing elsewhere in this paper (which is multiplexing of data in an optical fiber).

The basic idea is to place superconducting qubits at slightly different frequencies, i.e., ensure that each qubit has a slightly different $\omega$. Then, one can have a single microwave line coupled to multiple qubits, and one can send different signals on resonance with different qubits at the same time, thereby performing different gates on different qubits at the same time. 

The fastest gate operations in our scheme take 25\,ns, and thus, the signal that performs a gate operation will have a bandwidth of about 40\,MHz. To ensure the signal only affects the desired qubit, all qubits coupled to a given microwave line must have their frequencies spaced by 40\,MHz. Assuming qubit frequencies that are spread over a range from 5.5 to 6.5\,GHz, this means a single microwave line can control about 25 qubits.

Similarly, we assume each measurement of a qubit takes about 100\,ns, following the scheme in Ref.~\cite{Jeffrey2014May}. Thus, the readout signal will have a bandwidth of about 10\,MHz. If we consider 100 qubits, and spread their frequencies in an intelligent way over the range from 5.5 to 6.5\,GHz, we can have them be driven by 4 microwaves lines, but read out with only one line, i.e., one amplification chain per 100 qubits.

\subsection{Efficiency of cryogenics}
\label{Sect:cryo-efficiency}

Here, we assume ideal (Carnot-efficient) cryogenics. As no cryogenics are ideal, real cryogenics will have larger power consumption than those we consider here.  The small-scale cryogenics
used in most research laboratories put flexibility before efficiency, and so often have very low efficiencies.  However, once the quantum computing hardware is fixed, and it is known how much heat will be evacuated at each temperature, then cryogenics engineers are good at optimizing efficiency.
Some of the most efficient designs for cooling to cryogenic temperatures are used at CERN, and they have efficiencies from 10\% to 30\% of Carnot efficiency \cite{Parma2014}.

Of course high cryogenic efficiency requires that there is minimal heat leaking from one stage of the cryogenics to another. Here, we assume that this heat leakage is strictly zero. In other words, if the cryostat was empty (no cables conducting heat into it, and no attenuators or amplifiers generating heat inside it), then it would require negligible power to keep it cold. This is clearly an idealization.

To be more realistic, we should take (i) the efficiency for heat extraction and (ii) the heat leakage into the cryostat from datasheets for the industrial state of the art.  To treat point (i) in the full-stack model will require replacing the factor of $(T_{\rm ext}-T_i)/T_i$ in Eqs.~(\ref{Eq:power_singlequbit}), (\ref{Eq:P2qb}), and (\ref{Eq:PQ}) with a realistic efficiency for heat evacuation at each $T_i$, where each $T_i$ is for a given temperature stage. To treat (ii) in the full-stack model, we could assume the magnitude of the heat-leakage between stages scales with the size of the cryostat, and so scales linearly with the number of physical qubits. If this is the case, we just need to know the heat-leakage between stages per physical qubit, and then include it in the $\dot{q}_{\rm cond}$ in Eq.~(\ref{Eq:PQ}).
If, in contrast, the heat leakage scales non-linearly with the number of physical qubits, it must be included as a new term in Eq.~(\ref{Eq:PQ}).

\section{ALGORITHMIC PARAMETERS}
\label{Sect:Appendix-algo-parameters}

\subsection{Comment on Fig.~\ref{Fig:P_C-map}}

We explain here how the data in Fig.~\ref{Fig:P_C-map} are calculated.
Rather than using Eq.~(\ref{Eq:power_versus_QL}), which is an approximation that fails for $k=0$, Fig~\ref{Fig:P_C-map} is a calculation based on the exact formula \eqref{Eq:N1qb_from_N_logical} (within the assumptions made in the model),
in which we assume only identity gates at the logical level so that $N_{\rm 1qb;L}=N_{\rm 2qb;L}=N_{\rm meas;l}=0$.

For $k=0$ this is less power consuming than an arbitrary calculation
that includes arbitrary gate operations,
because identity gates add no dynamic costs (unlike other gates). The plot hence only includes the static costs of the calculation but not the dynamic costs (see below Eq.~(\ref{Eq:power_generic_physical}) for the definition of static and dynamic). However, as we have found that most of the power consumption is static rather than dynamic for the parameters in Fig.~\ref{Fig:P_C-map}, it still gives reasonable order of magnitude estimate of $P_C$ for any calculation.

As soon as there is error correction, $k\geq 1$, so that many gates are necessary for the error correction, the power consumption of an arbitrary circuit is extremely similar to that of a circuit made only of logical identity gates.  For $k=1$, the dynamic part of the power consumption varies by much less than an order of magnitude between the case of a circuit solely composed of logical identity gates, or solely composed of non-identity gates (see Eq. \eqref{Eq:N1qb_from_N_logical}).  As most of the power consumption is static rather than dynamic, Fig.~\ref{Fig:P_C-map}'s result for a circuit with only logical identity gates at $k=1$ gives a fairly accurate estimate for an arbitrary circuit at $k=1$.  

For $k\geq 2$, the dynamic part of the power consumption will be the same, within a percent or so, for any computation. Thus, Fig.~\ref{Fig:P_C-map} gives a very accurate estimate for any calculation at $k\geq2$.

Of course, when we say Fig.~\ref{Fig:P_C-map} gives an accurate result for a given calculation, we mean accurate within the assumptions of the modelling. Specifically, we assume the absence of $T$-gates.
\red{App.~\ref{Sect:neglected_T_gates} considers $T$-gates and argues that they are rare enough in the circuit that we consider
(a protocol to crack RSA encryption), that they will not contribute significantly to the power consumption of such a circuit.  Thus we can apply the results in Fig.~\ref{Fig:P_C-map} to a circuit that is cracking RSA encryption.}

\subsection{Implementation of protocols to crack the RSA encryption}
\label{sect:details-RSA}

For our full-stack analysis of a fault-tolerant quantum computer in Sec.~\ref{Sect:fault-tolerant}, we need to know the number of logical qubits $Q_{\rm L}$ and logical depth $D_{\rm L}$ necessary for a typical calculation.
The best studied calculation is the cracking of RSA-$n$ encryption, so we use that as a benchmark. The current record for cracking the RSA encryption with a classical supercomputer is for RSA-829 \cite{Boudot2020Aug}. 

Here we take the quantum protocol to crack the RSA-$n$ encryption from 
Ref.~\cite{Gidney2021Apr}, which proposes
\begin{eqnarray}
Q_{\rm L} &=& 3n + 0.002n {\rm Log}_2[n], 
\nonumber\\
D_{\rm L} &=& 500n^2 + n^2{\rm Log}_2[n], 
\label{Eq:RSA-QL-DL}
\end{eqnarray}
for large $n$, where the algorithm has been decomposed using the gate-set comprising Clifford, Toffoli, and $T$-gates. Thus, to crack RSA-2048 ($n=2048$), which is considered uncrackable on current classical supercomputers, one requires $Q_{\rm L}=6175$ and $D_{\rm L}=2.1\times 10^9$, as in Fig.~\ref{Fig:P_C-map}a.
The symbols (star, triangle and square) in Fig.~\ref{Fig:P_C-map}b indicate $Q_{\rm L}$ and $D_{\rm L}$ for different $n$. \mf{It might also be worth noting that the estimate from \cite{Gidney2021Apr} is very similar to the early one from Zalka \cite{Zalka1998Jun}, where he found $Q_{\rm L}=5n$ and $D_{\rm L}=600n^2$.}

It is critical to chose the right implementation of the protocol, if one wishes to minimize power consumption. For example, another recent implementation of the protocol to crack RSA-$n$ encryption \cite{Haner2017} uses fewer logical qubits, at the cost of a larger logical depth. It has 
\begin{eqnarray}
Q_{\rm L} &=& 2n + 2, 
\nonumber\\
D_{\rm L} &=& 52n^3 +{\cal O}[n^2].
\label{Eq:RSA-QL-DL-less_good}
\end{eqnarray}
For RSA-2048, this gives $Q_{\rm L}=4098$ and $D_{\rm L}=4.4\times 10^{11}$, so that $Q_{\rm L}$ is about two-thirds of that in Eq.~(\ref{Eq:RSA-QL-DL}) but $D_{\rm L}$ is 200 times that in Eq.~(\ref{Eq:RSA-QL-DL}).
Our full-stack analysis shows that even though this uses
fewer qubits than in Eq.~(\ref{Eq:RSA-QL-DL}), it uses much more energy. 
In many parameter regimes, the extra depth with respect to the implementation in Eq.~(\ref{Eq:RSA-QL-DL}) means that more error correction is required, so there will be more physical qubits per logical qubit than for Eq.~(\ref{Eq:RSA-QL-DL}). This means the power consumption will be more than for Eq.~(\ref{Eq:RSA-QL-DL}), and the calculation takes longer to run. However, for the parameters in Fig.~\ref{Fig:P_C-map}b, it happens that the extra depth required by the implementation in Eq.~(\ref{Eq:RSA-QL-DL-less_good}) is not enough to require another concatenation level; see Eq.~(\ref{Eq:transitions-k_to_k+1}). Hence the power needs of both implementations are similar. However, as the implementation in Eq.~(\ref{Eq:RSA-QL-DL-less_good}) takes 200 times as long, the total energy cost of the computation is $200\times 2/3 = 133$ times that of Eq.~(\ref{Eq:RSA-QL-DL-less_good}). This takes its energy cost from the equivalent of about twenty car-tanks of gasoline \cite{gasoline} to the equivalent of about 2666 car-tanks of gasoline!

These results show how important research is on optimizing the implementation of algorithms. Even modest reductions of the prefactors in equations like Eq.~(\ref{Eq:RSA-QL-DL}) can have significant effects. In particular, the effect can be huge if it happens that this reduction takes the system into a regime in which the calculation can be done successfully with one fewer concatenation level. 

This is different from common situations in classical computing algorithms, where implementing the algorithm in a way that runs faster, does not make it consume less power. So the total energy consumption is reduced linearly with the increase in speed. Here, the fact that a faster quantum algorithm also requires less power (because it requires less error correction) means one will get a {\it better than linear} gain from any faster implementation of the algorithm.

There is also a {\it hardware} aspect of circuit implementation that is worth mentioning here; the results in Eq.~(\ref{Eq:RSA-QL-DL}) depend on which gates can be directly implemented by the hardware of the quantum computer.  
Thus it is worth briefly examining an example of how this could affect the power consumption and energy cost of a calculation.
Eq.~(\ref{Eq:RSA-QL-DL}) is based on the assumption that the hardware
supports a fault-tolerant gate set consisting of Clifford, Toffoli and $T$-gates.  
However, it could well be that the fault-tolerant Toffoli gates in a concatenated 7-qubit code must be built out of $T$-gates, something which takes about nine time-steps,
involving cNOT and $T$-gates \cite{Selinger2012Oct}. This is the point of view taken in Ref.~\cite{our-article-in-prep}. In this case, the fact the hardware cannot directly implement Toffoli gates would not change $Q_{\rm L}$, but it would increase $D_{\rm L}$. 
A careful estimate of how much $D_{\rm L}$ would increase by would require delving into details of the circuit to see how many Toffoli gates are done in parallel. We do not do that here. Instead, we look at the worst case and the best case. The worst case would be to start with Eq.~(\ref{Eq:RSA-QL-DL}) and assume there is no timestep in the logical algorithm without a Toffoli gate, and that when one replaces each Toffoli by a series of gates taking nine time-steps, all other qubits just wait for this series of gates to finish before continuing with the algorithm.  Then the logical depth, $D_{\rm L}$, would be nine times that in Eq.~(\ref{Eq:RSA-QL-DL}).  However, we believe it to be much too pessimistic.
Ref.~\cite{Gidney2021Apr} shows that only about one gate in every 5000 gates is a Toffoli gate; they say their ``Toffoli+T/2 count" is $0.3n^3 + 0.0005n^3 {\rm Log}_2[n]$, which is about 5000 times smaller than $Q_{\rm L}\times D_{\rm L}$. Thus the best case would be that the depth is only increased by 0.02\% (this would require that all logical qubits to do all Toffoli gates in parallel).  The true result will be somewhere between the two, although as Ref.~\cite{Gidney2021Apr} uses a lot of parallization,  we can hope it will be much better than the worst case. In any event, we expect that such an increase of $D_{\rm L}$ combined with the fact $T$-gates are a more demanding in physical resources (see App \ref{Sect:neglected_T_gates}) could increase our energy estimates by an order of magnitude, which would not change the qualitative conclusions in our Sec.~\ref{Sect:quantum_energy_advantage}.  
Conversely, the field of circuit optimization is very active, so new protocols with smaller $D_{\rm L}$ than Eq.~(\ref{Eq:RSA-QL-DL}) are likely to appear in coming years.
Thus for simplicity, we use Eq.~(\ref{Eq:RSA-QL-DL}) in our calculations here.

\subsection{First-order phase transition between concatenation levels}
\label{Sect:Appendix-first-order}

The transition from the $k$ to $k+1$ levels of concatenation in the error correction seen in Fig.~\ref{Fig:P_C-map}b is analogous to a first-order phase transition. To understand why, suppose one increases the depth of the calculation $D_{\rm L}$ for a given concatenation level $k$. To maintain the calculation's metric of ${\cal M}_0$, one must reduce the error probability per gate operation, $\perror$ in Eq.~(\ref{Eq:eta_vs_T}), by reducing $T_{\rm qb}$ and increasing $A$. Doing this will cause $P_{\rm C}$ to diverge at a finite value of $D_{\rm L}$, because $\perror$ remains finite when $T_{\rm qb}\to 0$ and $A\to\infty$, but $P_{\rm C}$ diverges. Shortly before this divergence occurs, this power consumption
becomes more than the power consumption if one allowed larger $\perror$, but added a concatenation level. Thus, the transition occurs when the minimum $P_{\rm C}$ for $k$ levels of error correction exceeds the minimum $P_{\rm C}$ 
for $(k+1)$ levels of error correction. 
This is analogous to a first-order phase transition, such as the liquid-gas transition, which occurs when the energies of two phases cross.

Eq.~(\ref{Eq:metric_logical}) tells us that the maximum calculation size 
---i.e.,~largest ${\cal N}_{\rm L}$, where ${\cal N}_{\rm L}$ is defined above Eq.~(\ref{Eq:metric_logical})---
for a given metric ${\cal M}$, concatenation level $k$, and error probability $\perror < \pthr$, is 
\begin{eqnarray}
{\cal N}_{\rm L} &=& \frac{ \ln[\mathcal{M}]}{\ln\left[\left(1-\pthr (\perror/\pthr)^{2^k}\right)\right]}
\nonumber
\\
&\simeq& \frac{\ln[1/{\cal M}]}{\pthr} \left(\frac{\pthr}{\perror}\right)^{2^k},
\label{Eq:maxN-1}
\end{eqnarray}
where the second line is a small $\pthr$ approximation,
using the fact that $\pthr = 2\times10^{-5} \ll 1$.

This means that the maximum possible calculation size
for a given $k$ is when $\perror$ in Eq.~(\ref{Eq:eta_vs_T}) takes its minimum value, $\gamma\tau_{\rm step}/4$
(the value it takes when $n_{\rm noise}=0$ because $T_{\rm qb}\to0$ and $A\to \infty$). Thus, the maximum possible ${\cal N}_{\rm L}$ for a given $k$ is found by replacing $\perror$ by $\gamma\tau_{\rm step}/4$ in Eq.~(\ref{Eq:maxN-1}).
This is the value of ${\cal N}_{\rm L}$ at which the power consumption for $k$ levels of error correction diverges. Following the argument above, 
it will be energetically favourable to switch from $k$ to $k+1$ levels of error correction as ${\cal N}_{\rm L}$ approaches this value from below.

Thus, as we assume ${\cal N}_{\rm L}=Q_{\rm L}\times D_{\rm L}$,
the transition from $k$ to $k+1$ levels of error correction must occur for 
\begin{eqnarray}
Q_{\rm L}\times D_{\rm L}
\ &\lesssim&\ \frac{\ln[1/{\cal M}]}{\pthr} \left(\frac{4\pthr}{\gamma\tau_{\rm step}}\right)^{2^k}.
\label{Eq:transitions-k_to_k+1}
\end{eqnarray}
Now, in general, the power consumption per qubit has a term that depends on $T_{\rm qb}$
(and diverges as $T_{\rm qb}\to 0$), and another term that is $T_{\rm qb}$-independent. The latter term contains such things as the power consumption of the electronics at $T_{\rm gen}$, and the resources required to evacuate the heat conducted down wires at temperatures above 10\,K.
We observe that when the $T_{\rm qb}$-independent term becomes larger larger, then the transition moves toward the line defined by the equality in Eq.~(\ref{Eq:transitions-k_to_k+1}).
Indeed, for the parameters considered in this work (summarized in Table~\ref{table:parameters}), it is a reasonable approximation to say that the transitions occur at  
\begin{eqnarray}
Q_{\rm L}\times D_{\rm L}
\ &=&\ \frac{\ln[1/{\cal M}]}{\pthr} \left(\frac{4\pthr}{\gamma\tau_{\rm step}}\right)^{2^k}.
\label{Eq:transitions-k_to_k+1-specialcase}
\end{eqnarray}
Indeed, deviations from this approximation are hardly noticeable, if one  superimposes Eq.~(\ref{Eq:transitions-k_to_k+1-specialcase}) on the log-log plot in Fig.~\ref{Fig:P_C-map}b.

We note that Eq.~(\ref{Eq:transitions-k_to_k+1-specialcase}) can also be arrived at from the conventional wisdom
that error correction is so 
expensive that, heuristically, one should always
adjust other control parameters (qubit temperature, etc) to minimize the amount of error correction (in this case the concatenation level $k$). However, we emphasise that while this conventional wisdom works for the parameters given in Table~\ref{table:parameters}, it can fail drastically for slightly different parameters (such as the less ideal parameters in App.~\ref{App:varying_power_per_physical}).
Thus, while Eq.~(\ref{Eq:transitions-k_to_k+1-specialcase}) helps us give a simple interpretation of this aspect of the full-stack modelling,
it would be dangerous to rely on it without {\it first} performing the full-stack modelling. 
The reason is that to know whether Eq.~(\ref{Eq:transitions-k_to_k+1-specialcase}) is a good approximation or not, one needs to know the relative strengths of the $T_{\rm qb}$-dependent and $T_{\rm qb}$-independent terms in the power consumption.  This is typically information that is only accessible after one has optimized the full-stack quantum computer (and thereby already found the transitions). Thus, we see Eq.~(\ref{Eq:transitions-k_to_k+1-specialcase}) as a way to help understanding the optimization's result {\it a posteriori}, rather than for making a prediction {\it a priori}.

Typically, Eq.~(\ref{Eq:transitions-k_to_k+1-specialcase}) fails when the $T_{\rm qb}$-dependent term in the power consumption is more significant than for the parameters given in Table~\ref{table:parameters}.  We give a plausible example of this below in App.~\ref{App:varying_power_per_physical} with Fig.~\ref{Fig:bad-refrigerator}.
There, we see that the position of the transition from $k=3$ to $k=2$ depends on unexpected control parameters, such as the power consumption of the control electronics. Hence, there is no simple way to guess the optimal value of $k$ for a given algorithm size 
without performing the full-stack optimization.

\section{Classical energy efficiency}
\label{Sect:classical-RSA}
To compare quantum computers to classical computers, we need the energy and computational duration required by the classical supercomputer to crack an RSA-$n$ private key, where $n$ is the bit-size of the private key. To proceed, we can consider the asymptotic estimation of the number of operations required by the classical algorithm called the General Number Field sieve (GNFS), which is the best known classical algorithm to factorise a number into prime factors \cite{Boudot2020Aug}.  The number of operations required to crack the RSA-$n$ encryption is 
\begin{align} 
N_{\text{GNFS}}(n)=& \exp \Bigg[\left(\frac{64 n\ln(2)}{9} \, \Big[\ln\big(n\ln(2)\big)\Big]^2\,\right)^{\! 1/3}
\nonumber \\
& \qquad\qquad \times
\big[1+o(1)\big]\Bigg].
\end{align}
We consider $1+o(1) \approx 1$; while it could hide a possibly large constant, to our knowledge, there is no better estimate available today. From this total number of operations, we extrapolate the total energy $E_{\text{GNFS}}$ and time $t_{\text{GNFS}}$ this classical algorithm would take on the classical supercomputer JUWELS Module 1, which was used in the state-of-the-art cracking of the RSA encryption \cite{Boudot2020Aug}).
This gives $E_{\text{GNFS}}(n)=E_{\text{GNFS}}(830) N_{\text{GNFS}}(n)/N_{\text{GNFS}}(830)$, where $E_{\text{GNFS}}(830) \approx 1$\,TJ. 
For $t_{\text{GNFS}}$, we assume that the calculation can be fully parallelized on all cores of the JUWELS Module 1, so $t_{\text{GNFS}}(830) \approx $ 8-9 days. This underestimates $t_{\text{GNFS}}$, as some steps in the algorithm cannot be parallelized in this way; see \cite{Boudot2020Aug}. Then $t_{\text{GNFS}}(n)=t_{\text{GNFS}}(830) N_{\text{GNFS}}(n)/N_{\text{GNFS}}(830)$.
We define the energy efficiency as $\eta=n/E(n)$, and this gives the black curve in Fig.~\ref{Fig-RSA-quantum-versus-classical}.

\section{A different situation in which $P_{\rm C}$ is dominated by heat-production at $T_{\rm qb}$}
\label{App:varying_power_per_physical}

\begin{figure*}
         \includegraphics[width=0.9\textwidth]{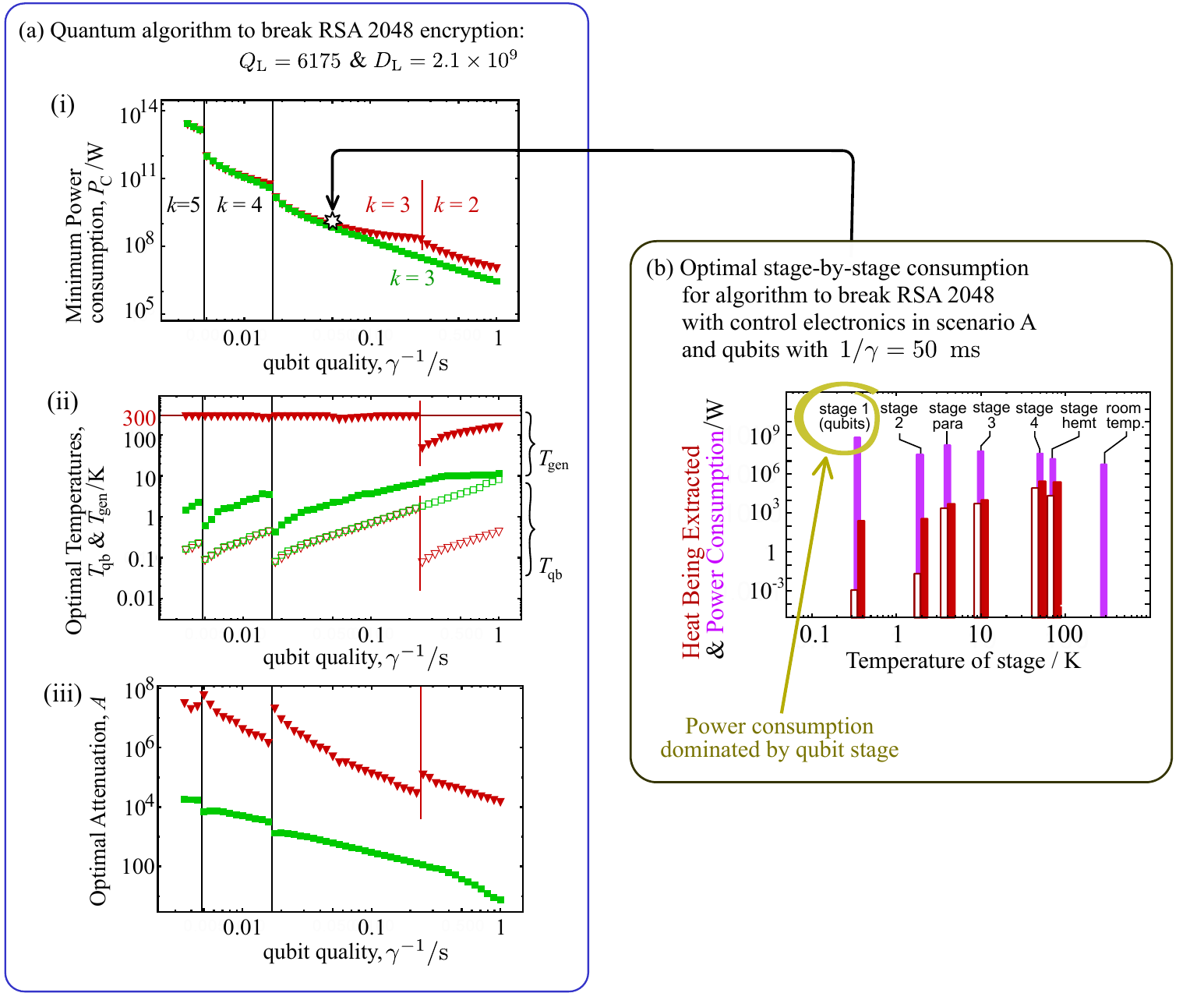}
         \caption{(a) Minimization of power consumption with less efficient cryogenics and more heat sources at $T_{\rm qb}$ compared to Fig.~\ref{Fig:P_C-map}  (see Sec.~\ref{App:varying_power_per_physical}). Plots show (i) the power consumption after minimization, with (ii) and (iii) showing the temperatures and attenuation.  
         As in  Fig.~\ref{Fig:P_C-map}, the red curves are for control electronics in scenario A, and the green curves are for scenario C.
         (b) The power consumption stage-by-stage at the point given by the star on the red curve in (a). Here, the power consumption is dominated by the cryogenic stage containing the qubits at $T_{\rm qb}$, unlike in 
         Fig.~\ref{Fig:P_C-map}(c).
         This is typical for the parameter regime that we explore here (not just at the star). It is the reason that the power consumption depends strongly on qubit temperature. This in turn means that the optimal amount of error correction can depend on the power consumption of the control electronics. We see that the transition from $k=3$ to $k=2$ is absent for the green curve (occurring outside the plot at $\gamma^{-1}>1\,$s). This is very different from Fig.~\ref{Fig:P_C-map}a, where all transitions are almost independent of the power consumption of the electronics. 
         }
         \label{Fig:bad-refrigerator}
\end{figure*}

For the parameters taken throughout this work (summarized in Table~\ref{table:parameters}), the power consumption of the cryogenic stage at temperature $T_{\rm qb}$ is a tiny contribution to the overall power consumption; see e.g.~Fig.~\ref{Fig:P_C-map}c.
Many of the our observations followed from this fact. 

In this appendix, we look at a full-stack model in the {\it opposite} situation, where the power consumption is dominated by what happens at temperature $T_{\rm qb}$.  This will be a common situation if the cryogenics are less ideal than we assumed above, and if extra heat must be dissipated at the qubit temperature for any reason.  
The model we take here differs from that elsewhere in this article in two ways:
\begin{itemize}
\item[(i)] We replace the Carnot efficiency with the efficiency for typical small-scale cryostats, given by a phenomenological formula constructed by fitting data in Ref.~\cite{irds}.
This formula says that the power required to remove heat at a rate $\dot{Q}$ from a cryogenic stage at $T$ is $3.24 \times 10^5 \times \dot{Q} (1-T/T_{\text{ext}})/T^2$, for $T_{\rm ext}=300\,$K.  This diverges much faster than Carnot-efficient refrigeration at small $T$, and so the low-temperature stages will contribute a much bigger proportion of the cryogenic power consumption.  
\item[(ii)] We add an additional heat-load of $50 nW$ per physical qubit at the stage  at $T_{\text{qb}}$.  We took this from Ref.~\cite{krinner2019engineering}, which estimated such an additional heat load in their technology due their flux biasing of each physical qubit to bring it to its operational sweet-spot  (the flux bias being necessary to compensate for intrinsic magnetic fields).
\end{itemize}
These two modifications greatly increase the power consumption of the part of the cryogenics that extracts heat at $T_{\rm qb}$ with respect to other contributions to the power consumption.

Fig.~\ref{Fig:bad-refrigerator} shows results for the same parameters as in Fig.~\ref{Fig:P_C-map}a.  Of course, it is no surprise that 
the power consumption is much bigger now the cryogenics is less efficient and there is extra heat to extract at $T_{\rm qb}$. What is interesting is that now the power consumption varies more with $\gamma^{-1}$ within a given concatenation level of error correction, than at the jumps between levels of error correction.
This means that it is so expensive to make the qubits colder that it is sometimes more competitive to perform an additional level of concatenation. 
As a consequence, there is no easy way to see what is the best 
level of concatenation for a given set of parameters. 
In Fig.~\ref{Fig:P_C-map}, the places where the concatenation level changed were the same for all curves, and it was fairly well approximated by Eq.~(\ref{Eq:transitions-k_to_k+1-specialcase}).  Here, 
the optimal concatenation level completely changes between the red and green curves in the region $\gamma^{-1} > 0.25\,$s, simply because the control electronics have different power consumption for the red and green curves. 
This neatly shows the inter-dependence of technologies in the quantum computer. 
When the power consumption per physical qubit is low enough (green curve), adding more error correction (more physical qubits per logical qubit) costs less power than cooling the physical qubits further.
Thus, the region of the green curve with $\gamma^{-1} > 0.25\,$s is a regime in which it is better to have hotter physical qubits (more  errors per physical qubit) and compensate with more error correction (i.e., $k=3$ rather than $k=2$). This is counter to the conventional wisdom outlined in App.~\ref{Sect:Appendix-first-order}.
It shows that, in general, only a full-stack optimization will find the optimal working conditions for a quantum computer.

This also gives us a clear example in which our MNR methodology allows us to reduce the power consumption by more than three orders of magnitudes. Consider the power consumption for SFQ electronics (green curve) in  Fig.~\ref{Fig:bad-refrigerator}a. Suppose we had qubits  with $\gamma^{-1}=1\,$s, but we applied plausible but non-optimal values of attenuation, qubit temperature, and signal generation, such as those that would be optimal if $\gamma^{-1}=0.02\,$s.  Then, the power consumption would be about 10\,GW (the same as if $\gamma^{-1}=0.02\,$s), when the
optimization tells us that a power consumption of only about 2\,MW is necessary for  $\gamma^{-1}=1$s.  This would be a saving of more than three orders of magnitude, in a regime of high power consumption.
Such an energy saving is possible because our MNR methodology links the fundamental noise model of the qubit to the classical hardware, electronics, cryogeny, etc.

\bibliography{Energetics_FTQC-References}

\begin{thebibliography}{129}
\bibitem [{att()}]{attenuation}%
  \BibitemOpen
  \href@noop {} {}\bibinfo {note} {{Unlike some works, we take positive dB
  values for attenuation. An attenuator with $A_{\rm dB}=20dB$ has $A=100$,
  corresponding to the attenuator's output being 100 time smaller than the
  input.}}\BibitemShut {Stop}%
\bibitem [{gas()}]{gasoline}%
  \BibitemOpen
  \href@noop {} {}\bibinfo {note} {{Gasoline has about 0.03\,GJ of energy per
  litre and a typical car-tank contains about 60 litres.}}\BibitemShut {Stop}%
\end{thebibliography}

\end{document}